\documentclass[11pt]{article}

\newcommand{\Dc}{ {\mathcal{D}} }

\newcommand{\Lc}{ {\mathcal{L}} }

\newcommand{\Nc}{ {\mathcal{N}} }

\newcommand{\Pc}{ {\mathcal{P}} }

\newcommand{\bsx}{\boldsymbol{x}}

\newcommand{\bsp}{\boldsymbol{p}}

\newcommand{\bsmu}{\boldsymbol{\mu}}

\usepackage{graphicx}
\usepackage{epstopdf, epsfig}
\usepackage[colorlinks]{hyperref}
\usepackage{amsfonts}
\usepackage{amssymb}
\usepackage{mathrsfs}
\usepackage[T1]{fontenc}
\usepackage{mathtools}
\usepackage{amsmath}
\usepackage{dcolumn}
\usepackage{bm}
\usepackage[export]{adjustbox}
\usepackage{wrapfig}
\usepackage{lipsum}
\usepackage{authblk}
\usepackage{makecell}
\usepackage[font=small,labelfont=bf]{caption}
\allowdisplaybreaks
\usepackage[titletoc,toc,title]{appendix}	
\DeclareMathOperator*{\argmin}{arg\,min}

\usepackage[
style=phys,
sorting=none,
citestyle=numeric-comp,
backend=biber,
maxbibnames=3, 
minbibnames=3, 
eprint=true,
]{biblatex}

\usepackage{tikz}
\usetikzlibrary{arrows.meta,positioning,fit,backgrounds}
\usepackage[utf8]{inputenc}
\usepackage{newunicodechar}
\newunicodechar{ʼ}{'}

\usepackage[
a4paper,
margin=1in,          
headheight=15pt,     
footskip=0.5in       
]{geometry}

\usepackage{hyperref}
\hypersetup{
	colorlinks=true,
	linkcolor=red,
	filecolor=magenta,      
	urlcolor=cyan,
	citecolor = cyan,
}

\usepackage{blindtext} 
\usepackage{booktabs}
\usepackage{makecell}
\usepackage{siunitx}

\usepackage{algorithm}
\usepackage{algpseudocode} 
 
\title{\vspace{-10mm} Physics-Informed Kolmogorov-Arnold Networks\\ for Grad-Shafranov Tokamak Equilibria} 
\author[1,2]{D. A. Kaltsas \thanks{kaltsas.d.a@gmail.com}}
\author[1]{A. Kuiroukidis}
\author[3]{J. Liu}
\author[2]{L. Magafas}
\author[1]{G. N. Throumoulopoulos}
\affil[1]{Department of Physics, University of Ioannina, GR 451 10 Ioannina, Greece}
\affil[2]{Department of Informatics, Democritus University of Thrace, GR 654 04 Kavala, Greece}
\affil[3]{Weihai Institute for Interdisciplinary Research, Shandong University, Weihai, 264209, China}

\date{}

\begin{document}
\maketitle

 \vspace{-4mm}
\begin{abstract}
We employ equation-driven, physics-constrained deep learning to solve the fixed-boundary Grad-Shafranov (GS) equilibrium problem, constructing axisymmetric magnetohydrodynamic equilibria with tokamak-relevant characteristics. Equilibria across linear (Solov'ev) and nonlinear profile functions are constructed using Physics-Informed Kolmogorov-Arnold Networks (KANs) that approximate GS solutions while satisfying appropriate boundary conditions. A highly nonlinear pressure profile recreating high-confinement mode phenomenology, such as pressure pedestals and significant bootstrap current components, is also considered. To enable efficient convergence, guided training schemes are employed, specifically homotopy-based continuation curriculum learning and transfer learning via pretrained networks. While computing nonlinear equilibria employing standard Multi-Layer Perceptrons under unguided physics-informed training remains an elusive or computationally inefficient task, our framework overcomes this limitation. Specifically, we demonstrate that the combination of three key elements, namely KAN architecture, guided training schemes, and the self-scaled Broyden optimization method, enables stable, efficient, and accurate equilibrium computation with simultaneous profile parameter identification in view of equilibrium constraints.

\end{abstract}


\section{Introduction}
\label{Sec_I}
Magnetohydrodynamic (MHD) equilibrium is a fundamental concept in the context of magnetic fusion research since it is necessary for stability and transport calculations and is crucial for designing fusion devices and experimental scenarios. Serving as the initial background state for plasma dynamics, the characteristics of the equilibrium state affect various processes, such as macroscopic MHD instabilities that may lead to an abrupt loss of confinement, and microscopic instabilities that drive turbulence and enhanced transport, thereby leading to the degradation of energy and particle confinement. Such losses must be minimized to achieve steady energy production by fusion reactions in magnetic confinement devices. Consequently, calculating the underlying equilibrium state is an essential prerequisite to assess and optimize a magnetic configuration in terms of confinement properties. Equilibrium calculations allow us to determine the confining magnetic field architecture and other physical features in regions within the plasma which are not accessible by measurements in magnetic fusion devices. Typically, the equilibrium configuration in tokamaks and spherical tokamaks (as well as other axisymmetric configurations), is determined by solving the static Grad-Shafranov (GS) equation \cite{Grad1958,Grad1958b,Shafranov1958,Shafranov1960,Shafranov1966} which governs ideal magnetohydrodynamic (MHD) force balance in the absence of flow.

Beyond the static case, generalized versions of the GS equation have been developed to incorporate additional physical effects. These include anisotropic pressure \cite{Mercier1961,Grad1967}, toroidal flow (e.g., \cite{Green1973,Maschke1980,Throumoulopoulos1989}), flows of arbitrary direction (e.g., \cite{Tasso1998,Hameiri1983,Guazzotto2004}), Hall-drift and electron inertial effects (e.g., \cite{Throumoulopoulos2006,Kaltsas2018}), and energetic particle effects (e.g., \cite{Kaltsas2024}). Such generalizations are physically motivated by auxiliary heating techniques, such as neutral beam injection  and radio frequency waves, which frequently induce pressure anisotropy and generate substantial populations of energetic particles. Furthermore, these methods, alongside intrinsic self-organization processes, can drive significant macroscopic plasma rotation \cite{Rice2016}. This rotation is crucial for plasma confinement, as it is well established that plasma flow and flow shear play a significant role in improving confinement by suppressing radial turbulence and inducing internal or external transport barriers, thereby facilitating the transition to the high-confinement mode (H-mode) \cite{Terry2000,Liang2020}. While these effects are important and require generalized versions of the GS equation to accurately determine equilibria, the present work focuses on the static GS equation as a fundamental and essential starting point for further extensions in the future.

To determine the plasma equilibrium in tokamaks, one must construct solutions to the GS equation subject to appropriate constraints and boundary conditions. This can be approached as a forward equilibrium problem where the plasma profile functions and the boundary conditions (such as the position and  shape of the Last Closed Flux Surface (LCFS), in a fixed-boundary setup) are prescribed a priori. Another, more experiment-oriented approach is the inverse or equilibrium reconstruction approach, where experimental measurements, including magnetic diagnostics (e.g., Mirnov coils, flux loops), motional Stark effect data, and Thomson scattering profiles, are utilized to reconstruct the equilibrium profiles and the plasma shape. Although analytic solutions to the GS equation are invaluable for qualitatively studying plasma equilibrium properties, incorporating complex boundaries and experimental constraints requires dedicated numerical codes. Traditional forward solvers (such as HELENA and CHEASE) and reconstruction codes (such as EFIT) construct numerical solutions for the poloidal magnetic flux function on a discrete grid, rendering them fundamentally mesh-dependent. Furthermore, conventional algorithms often face convergence issues when highly nonlinear profiles are introduced, such as those featuring steep gradients characteristic of H-mode pedestals. Lastly, conventional numerical approaches lack parametric flexibility, meaning that even minor changes in the physical parameters or boundary conditions requires the entire numerical procedure to be reinvoked. These shortcomings can be mitigated by utilizing physics-informed machine learning methods, which offer continuous, differentiable, and computationally efficient neural network alternatives to conventional grid-based numerical solutions.

In this paper, we utilize physics-informed machine learning (PIML) to construct axisymmetric plasma equilibria in a forward manner, using an equation-driven, physics-constrained machine learning method. This approach implements the physics-informed neural network (PINN) framework, which has flourished in recent years by utilizing deep neural networks (DNNs) to approximate solutions to partial differential equations (PDEs) \cite{Raissi2019,Karniadakis2021,Lagaris1998}. In contrast to conventional data-driven machine learning methods employed in fields like image recognition or time-series prediction, PINNs solve physics-relevant boundary and initial value problems with limited or even no data \cite{Lagaris1998,Raissi2019,Sirignano2018,Karniadakis2021}. The training process is largely data-free, as the loss function minimized during training consists of the PDE residual evaluated at collocation points selected via Monte Carlo sampling, alongside penalty terms that enforce boundary conditions and additional physical and data constraints. Consequently, the physics-informed terms in the loss function can significantly enhance the accuracy and fidelity of machine learning models when training data is scarce and enable prediction even when experimental data are absent. For a comprehensive overview of PINNs across scientific computing, the reader is referred to \cite{Cuomo2022}. The use of PINNs to approximate solutions to PDEs like the GS equation offers several key advantages: it provides continuous, differentiable solutions that eliminate the need for a discrete computational grid, rendering the method entirely mesh-free; enables the model to be efficiently adapted or retrained for new parametric values; allows for the straightforward handling of complex geometries, bypassing the intricate mesh-generation and mapping techniques required by traditional solvers; and facilitates the seamless integration of physical constraints and experimental data.

Neural networks have evolved from early applications in direct axisymmetric equilibrium calculations \cite{vanMilligen1995} into robust PIML frameworks for magnetic confinement. Modern PIML algorithms enable the solution of the GS equation under realistic constraints, encompassing real-time magnetic signal processing \cite{Joung2019, Joung2023}, alongside extensions to 3D equilibria \cite{Merlo2021}, axisymmetric equilibria with flows \cite{Kaltsas2022}, flexible geometry and profile parameterizations \cite{Jang2024}, and physically constrained profile mappings \cite{Zhang2024}. To meet real-time control demands, physics-informed surrogate models and specialized architectures, such as deep ensembles and extreme learning machines, have been deployed to accelerate legacy solvers (e.g., EFIT and LIUQE), enabling fast plasma shape control, boundary reconstruction, and robust uncertainty quantification \cite{Lao2022, Madireddy2024, Fiorenza2025, Grandin2026}. Furthermore, recent studies demonstrate the efficacy of PINNs in tackling complex multi-diagnostic inverse problems, incorporating nonlinear polarimetry and bolometry diagnostics, while establishing formal guidelines for architecture and loss-weight optimization \cite{Rossi2023, Rutigliano2025, Rutigliano2026}. Beyond laboratory fusion devices, the scope of PINNs has also expanded to astrophysical plasma modeling, including solar coronal equilibria and magnetic reconnection dynamics \cite{Baty2024}.

Despite their advantages though, PINNs often exhibit lower accuracy and higher computational overhead compared to traditional numerical algorithms, especially in low-dimensional problems. This accuracy and efficiency deficit primarily originates from optimization bottlenecks, including vanishing or exploding gradients and the slow training convergence characteristic of deep architectures required for high expressivity. To mitigate these limitations, alternative network architectures with enhanced expressivity (e.g., \cite{Liu2025}) and advanced second-order, curvature-aware optimization methods (e.g., \cite{Kiyani2025,Jnini2026}) have recently been explored. In this work, to improve PINN performance, we leverage Kolmogorov–Arnold Networks (KANs), introduced by \cite{Liu2025} and further explored as Physics-Informed KANs (PIKANs) (e.g., in \cite{Toscano2025}), as a promising, more expressive alternative to traditional multi-layer perceptrons (MLPs). Additionally, to accelerate convergence, we employ the self-scaled Broyden (SSBroyden) optimization algorithm \cite{Urban2025,Kiyani2025,Jnini2026}, which is an advanced second-order quasi-Newton optimizer. These two new elements are combined with guided learning techniques that enable convergence in cases with nonlinear plasma equilibrium profiles, namely transfer learning via the utilization of pretrained networks and a curriculum learning strategy \cite{Bengio2009,Soviany2022}. This strategy is inspired by parameter continuation methods \cite{Pathak2025,Ko2023}, which gradually change the optimization problem from an easier setup to the more challenging problem of interest by continuously varying a homotopy parameter, enabling the network to learn progressively more complex plasma equilibria starting from the simplest possible solution. The combination of these three elements, i.e., KANs, second-order quasi-Newton optimizer, and guided training, enables the stable, efficient, and accurate determination of tokamak equilibria with nonlinear profiles.

This paper is organized as follows: In Sec.~\ref{Sec_II}, we present the static GS equation and define the forward equilibrium problem, introducing standard equilibrium profile parameterization and a specific pressure profile function that reproduce pressure pedestals. In Sec.~\ref{Sec_III}, we describe the PIKAN algorithm, introducing the basic characteristics of the KAN architecture and defining the optimization problem. In Sec.~\ref{Sec_IV}, auxiliary algorithms used in this work are presented, namely the enforcement of physical constraints for profile parameter optimization, the magnetic axis tracking algorithm, transfer learning, and homotopy-based curriculum learning. In Sec.~\ref{Sec_V}, we present the resulting neural network equilibria, covering three distinct classes: Solov'ev equilibria, nonlinear equilibria with typical low confinement mode (L-mode) profiles, and H-mode equilibria with pressure pedestals. Finally, in Sec.~\ref{Sec_VI}, we summarize the main conclusions of this study.

\section{Grad-Shafranov equilibrium}
\label{Sec_II}
\subsection{The Grad-Shafranov equation}
The static GS equation is derived from the steady-state ideal MHD equations assuming axisymmetry and the absence of plasma flow. In the cylindrical coordinate system ($R,\phi,Z$), axisymmetry translates to the invariance of physical quantities with respect to the toroidal angle $\phi$. Under these assumptions, the standard, static GS equation takes the form \cite{Grad1958,Grad1958b,Shafranov1958,Shafranov1960,Shafranov1966}:
\begin{equation}
	\Delta^* \Psi + \mu_0 R^2 \frac{d P(\Psi)}{d\Psi} + F(\Psi) \frac{dF(\Psi)}{d\Psi} = 0\,, \label{gse_1}
\end{equation}
where $\Psi(R,Z)$ is the poloidal magnetic flux function (representing the poloidal magnetic flux per radian); $P(\Psi)$ is the plasma pressure, which is constant on magnetic surfaces labeled by $\Psi(R,Z) = \text{const.}$; and $F(\Psi)$ is the poloidal current stream function that is constant on magnetic surfaces, related to the toroidal magnetic field via $B_\phi = F/R$. The poloidal component of the magnetic field is given by $\bm{B}_p = \nabla \Psi \times \nabla\phi$. In Eq.~\eqref{gse_1}, $\mu_0$ is the magnetic permeability of free space, and $\Delta^*$ is the Shafranov operator, which is defined as follows:
\begin{equation}
	\Delta^* = R\frac{\partial}{\partial R}\left(\frac{1}{R}\frac{\partial}{\partial R}\right)+\frac{\partial^2}{\partial Z^2}\,.
\end{equation}

To non-dimensionalize the GS equation, we introduce the following normalized quantities:
\begin{align}
	r &= R/R_0,, \quad z = Z/R_0\,, \quad \psi = \Psi/(R_0^2 B_0)\,, \notag \\
	\tilde{P} &= \mu_0 P / B_0^2\,, \quad \tilde{B} = B / B_0\,, \quad \tilde{F} = F/(R_0 B_0)\,, \label{nondimensional}
\end{align}
where $R_0$ is the characteristic length and $B_0$ is the characteristic magnetic field strength. For tokamak equilibria, $R_0$ is identified as the major radius of the torus, and $B_0$ as the vacuum magnetic field at a distance $R = R_0$ from the axis of symmetry.

Substituting the non-dimensional quantities given by \eqref{nondimensional} into Eq.~\eqref{gse_1} and omitting the tildes for notational simplicity, we obtain the dimensionless GS equation:
\begin{equation}
	\Delta^*\psi(r,z) + r^2 \frac{dP}{d\psi} + \frac{dH}{d\psi} = 0\,, \label{gs_norm}
\end{equation}
where $H(\psi) := F^2(\psi)/2$. The forward, fixed-boundary equilibrium problem consists of solving Eq.~\eqref{gs_norm} within a domain $\Dc$ for prescribed free functions $P(\psi)$ and $H(\psi)$, subject to the Dirichlet boundary condition $\psi = \psi_b$ on the domain boundary $\partial\Dc$. This boundary can be described mathematically either by analytically parameterized curves $(r_b(t), z_b(t))$ or via discrete coordinate pairs $(r_b, z_b)$ extracted from experimental data or free-boundary equilibrium codes.

\subsection{Equlibrium parametrization}
\subsubsection{Typical equilibrium parametrization} 
In this work, we employ a standard parameterization widely used in equilibrium studies and codes (e.g., \cite{Luxon1982,Jeon2015,freegs,Faugeras2022,Amorisco2024}):
\begin{align}
	\frac{dH}{d\psi} = H_1 \left(1-\psi_n^{a_m}\right)^{a_n}\,, \quad \frac{dP}{d\psi} = P_1 \left(1-\psi_n^{b_m}\right)^{b_n}\,, \label{freegs_ansatz}
\end{align}
where $H_1$, $P_1$, $a_m$, $a_n$, $b_m$, and $b_n$ are free parameters that can be tuned to satisfy specific equilibrium constraints and determine the shape of the toroidal current density profile, which is given by:
\begin{equation}
	J_\phi(r,z) = -\frac{\Delta^*\psi}{r} = P_1 r \left(1-\psi_{n}^{b_m}\right)^{b_n} + \frac{H_1}{r}\left(1-\psi_{n}^{a_m}\right)^{a_n} \,. \label{J_phi}
\end{equation}
In Eqs.~\eqref{freegs_ansatz} and~\eqref{J_phi}, $\psi_n$ denotes the normalized poloidal flux function, defined as
\begin{equation}
	\psi_n = \frac{\psi - \psi_a}{\psi_b - \psi_a}\,, \label{psi_n}
\end{equation}
where $\psi_a$ and $\psi_b$ are the values of the magnetic flux function $\psi$ on the magnetic axis and on the boundary, respectively. Consequently, $\psi_n \in [0,1]$, with $\psi_n = 0$ on the magnetic axis and $\psi_n = 1$ on the boundary.

\begin{figure}[!htb]
	\centering
	\includegraphics[scale=0.35]{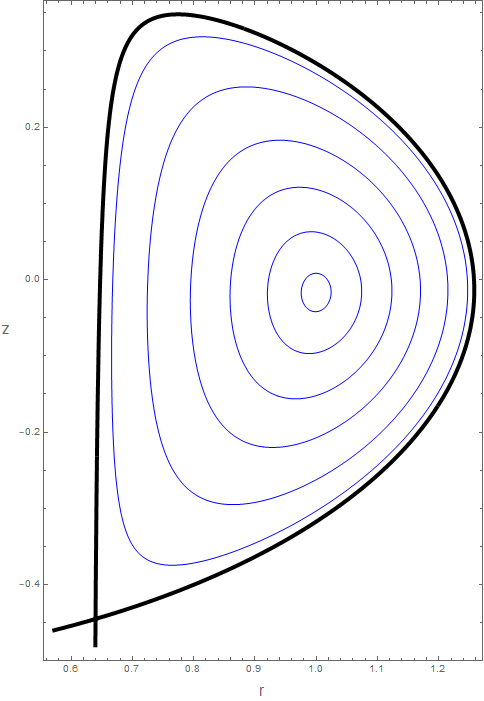}
	\caption{Magnetic surfaces of the Solov'ev equilibrium described by solution \eqref{Solovev_analytic}. The equilibrium features a separatrix (black curve) characterized by a lower hyperbolic X-point for $\gamma>0$. \label{fig_Solovev_analytic}}
\end{figure}

The simplest type of equilibria with non-vanishing pressure and self-consistent magnetic field is given by Solov'ev-type solutions \cite{Solovev1968}, which are obtained by assuming linear profiles for the functions $H(\psi)$ and $P(\psi)$, i.e., for $a_n = b_n = 0$. In this case, we obtain $dH/d\psi = H_1$ and $dP/d\psi = P_1$, rendering the toroidal current density a monotonic function of $r$. Although this configuration is of limited practical significance for fusion devices, the simplicity of the Solov'ev-type solutions compared with other analytic ones have made them a standard benchmark for testing numerical equilibrium solvers. For example, in a previous study \cite{Kaltsas2022}, we used a generalized Solov'ev solution describing a shaped equilibrium with incompressible flows of arbitrary direction to assess the accuracy of our neural network equilibrium models. However, the parameters in the Solov'ev solution were numerically determined by fitting a prescribed boundary, introducing errors into the analytical solution itself. To avoid such errors, one can employ the simplest Solov'ev-type equilibrium that spontaneously forms an LCFS  with single or double magnetic nulls (X-points). Setting $a_n = b_n = 0$ in the profiles given by Eq.~\eqref{freegs_ansatz}, we obtain the following analytical solution to Eq.~\eqref{gs_norm}:
\begin{equation}
	\tilde{\psi} = z^2 \left(r^2 - \frac{H_1}{2}\right) - \frac{1}{4}\left(1+\frac{P_1}{2}\right) (r^2 - 1)^2 + \gamma r^2 z\,, \label{Solovev_analytic}
\end{equation}
where $\gamma$ is a parameter that adjusts the up-down asymmetry of the configuration with respect to the  plane $z = 0$. For $\gamma = 0$, we obtain the standard double-null, up-down symmetric Solov'ev equilibrium, whose characteristics are described in detail in \cite{Kaltsas2026b}. For $\gamma \neq 0$, the equilibrium configuration possesses an LCFS with a single X-point and is thus up-down asymmetric, as illustrated in Fig.~\ref{fig_Solovev_analytic}. This corresponds to a lower X-point for $\gamma > 0$ and an upper one for $\gamma < 0$.

\subsubsection{Equilibria with pressure pedestal}
Strong auxiliary heating applied in tokamak plasmas can lead to a spontaneous transition from L-mode to H-mode, during which radial turbulent transport is substantially reduced and energy confinement is enhanced \cite{Wagner2007}. The edge profiles of pressure, particle density, and temperature in H-mode feature steep gradients, whereas the profiles in the plasma core retain their general L-mode shape, thus forming a characteristic pedestal structure. The reduction in turbulent transport is associated with the formation of an edge transport barrier closely linked to sheared plasma flows \cite{Burrell1997}. Although velocity shear is crucial for transport reduction, L–H transitions can also be achieved in slowly rotating plasmas \cite{McKee2009}, rendering dynamic pressure contributions small compared to thermal pressure in such cases. Consequently, it is relevant to model equilibria with a pressure pedestal even within the static framework, adopting a specialized profile that can phenomenologically capture steep edge gradients \cite{Pataki2013}:
\begin{equation}
	P(\psi) = P_1(P_2 + P_3\psi^2)\left( 1 - e^{-\psi^2/\nu}\right)\,, \label{pres_prof_ped}
\end{equation}
where $\nu$ is a parameter governing the steepness of the pressure profile toward the edge, $P_1$ scales the overall magnitude of the pressure, and $P_2$ and $P_3$ are shape parameters that determine the pedestal height relative to the core pressure. Note that Eq.~\eqref{pres_prof_ped} is expressed in terms of the poloidal flux function $\psi$.

\section{The PIKAN algrithm}
\label{Sec_III}
\subsection{PINNs}
\label{subsec_pinns}
Here we summarize the PINN algorithm for solving boundary value problems (BVPs) such as the fixed-boundary Grad-Shafranov equilibrium; however, for more detailed expositions, we refer the reader to \cite{Raissi2019, Yu2022, Blechschmidt2021}. Let us assume a BVP of the form:
\begin{align}
	F\left(\bsx, \ldots, \frac{\partial u}{\partial x_i}, \ldots, \frac{\partial^2 u}{\partial x_i \partial x_j}, \ldots; \bsmu \right) &= 0\,, \quad \bsx \in \mathcal{D} \subseteq \mathbb{R}^d\,, \label{gen_pde}\\
	u(\bsx) &= u_b\,, \quad \bsx \in \partial \mathcal{D}\,, \label{gen_bcs}
\end{align}
where $\mathcal{D}$ denotes the spatial domain of the BVP, $\bsx = (x_1, x_2, \dots, x_d)^\top$ is the vector of independent variables, and $\bsmu$ represents the physical parameters governing the system. The PINN framework approximates the exact solution $u(\bsx)$ via a parameterized neural network $u_{\text{net}}(\bsx, \bsp)$, where $\bsp$ denotes the set of trainable parameters. In standard MLPs, these parameters consist of the weights and biases governing the affine transformations between consecutive network layers. In addition, individual neurons transform their inputs via a fixed non-linear mapping, known as the activation function. Typical choices include the hyperbolic tangent ($\tanh$), the sigmoid, or the sigmoid linear unit ($\text{SiLU}$). Through this alternating composition of affine transformations and non-linear activation functions, the MLP achieves the expressive capacity necessary to approximate arbitrary smooth functions, as guaranteed by the Universal Approximation Theorem, making it well-suited for solving BVPs such as the one formulated above. The PINN algorithm relies on the minimization of the PDE residual, which is evaluated without introducing discretization errors because the network output $u_{\text{net}}$ is differentiated with respect to the spatial coordinates $\bsx$ via automatic differentiation \cite{Baydin2018}. This enables the exact computation (up to machine precision) of arbitrary-order differential operators required by the governing equations. This is a significant advantage of physics-informed frameworks over conventional grid-based numerical methods, which inevitably introduce discretization errors.

Training \(u_{\text{net}}\) to approximate the solution \(u\) of the BVP \eqref{gen_pde}--\eqref{gen_bcs} involves minimizing a composite loss function that accounts for both the PDE residual \eqref{gen_pde} and the boundary conditions \eqref{gen_bcs}.
 Accordingly, the overall loss function takes the general form
\begin{equation}
	\mathcal{L}(\bsp) = w_\Dc \Lc_{\Dc} + w_{\partial \Dc} \Lc_{\partial\Dc} \,, \label{gen_loss}
\end{equation}
where $w_\Dc$ and $w_{\partial \Dc}$ are weighting parameters balancing the individual loss components. The terms $\Lc_\Dc$ and $\Lc_{\partial \Dc}$ represent the mean squared error of the differential equation residual over a set $\Pc_{\Dc}$ of $N_{\Dc}$ interior collocation points, and that of the boundary condition over a set $\Pc_{\partial \Dc}$ of $N_{\partial \Dc}$ boundary points, respectively:
\begin{align}
	\Lc_{\Dc} &= \frac{1}{N_{\Dc}} \sum_{\bsx \in \Pc_{\Dc}}\left| F\left(\bsx, \ldots, \frac{\partial u_{\text{net}}}{\partial x_i},\ldots, \frac{\partial^2 u_{\text{net}}}{\partial x_i \partial x_j},\ldots; \boldsymbol{\mu} \right)\right|^2\,, \label{L_i_1} \\
	\Lc_{\partial \Dc} &= \frac{1}{N_{\partial\Dc}}\sum_{\bsx \in \Pc_{\partial \Dc}}\big|u_{\text{net}}(\bsx;\bsp)-u_b \big|^2\,. \label{L_b_1}
\end{align}

Additionally, inverse or semi-supervised frameworks can be adopted, wherein experimental/observational data or auxiliary physical constraints are integrated into the loss function as regularization terms to estimate the system parameters $\bsmu$. The external data or physical constraints $C_i$ are enforced over point sets $\Pc_{C_i}$ containing $N_{C_i}$ points via:
\begin{equation}
	\Lc_{C_i} = \frac{1}{N_{C_i}} \sum_{\bsx \in \Pc_{C_i}}\big|C_i(u_{\text{net}}(\bsx;\bsp);\bsmu)\big|^2\,. \label{L_ci_1}
\end{equation}
Then the additional loss term $\Lc_{C} = \sum_i w_{\text{\textsc{c}}_i} \Lc_{C_i}$ can be added to the total loss \eqref{gen_loss}.

The training of $u_{\text{net}}$ thus reduces to identifying the optimal set of network parameters $\bsp^*$ (and, if applicable, the physical parameters $\bsmu^*$) that minimize the loss function \eqref{gen_loss}:
\begin{equation}
	(\bsp^*, \bsmu^*) = \argmin_{(\bsp,\bsmu)} \Lc(\bsp,\bsmu)\,. \label{optimization_1}
\end{equation}
This minimization is performed using gradient-based algorithms that update the parameters iteratively over a predetermined number of epochs or until a convergence tolerance is reached. In this work, we employ the AdamW optimizer \cite{Loshchilov2017} followed by a second-order self-scaled Broyden optimizer \cite{Urban2025} to optimize the network parameters while $\bsmu$ are held fixed. The gradient  of the loss function $\mathcal{L}$ with respect to the parameter vector $\bsp$ is computed via automatic differentiation, specifically leveraging the \texttt{autograd} engine in PyTorch, to drive the iterative updates toward the optimal configuration. The optimization of the physical parameters $\bsmu$ is performed separately via an optimization method detailed in section \ref{Sec_IV}.

\subsection{KANs}
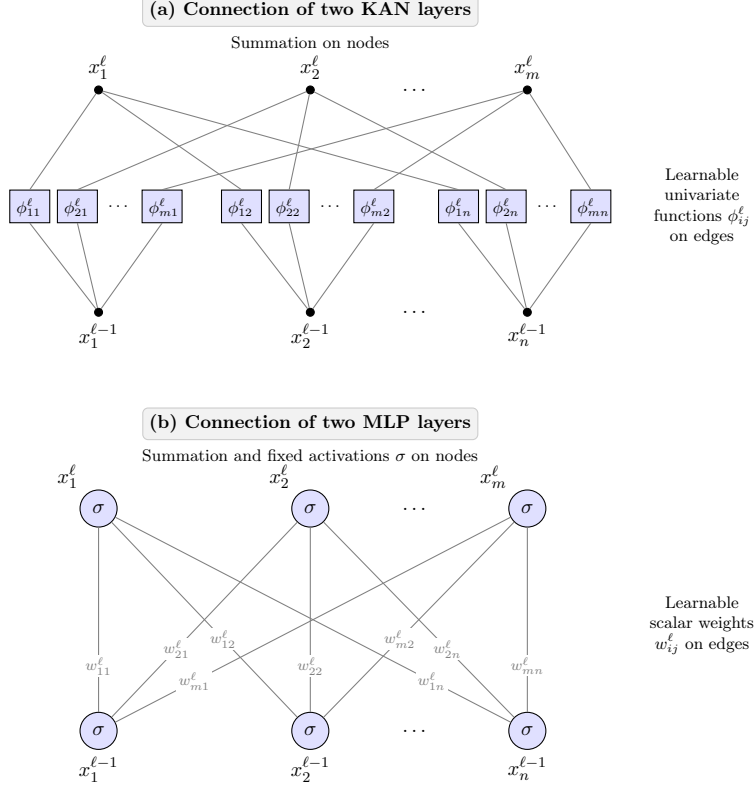
\begin{figure}[!htbp]
	\centering
	\begin{tikzpicture}[scale=0.7, transform shape]
		\useasboundingbox (0.0, -0.8) rectangle (16.0, 6.2);
		
		\node[font=\bfseries\small, fill=gray!10, draw=gray!40, rounded corners=2pt, inner sep=3.5pt] at (6.4, 5.7) {(a) Connection of two KAN layers};
		
		\node[circle, draw, fill=black, inner sep=1.5pt, label=below:$x^{\ell-1}_1$] (x1) at (2.4,0) {};
		\node[circle, draw, fill=black, inner sep=1.5pt, label=below:$x^{\ell-1}_2$] (x2) at (6.4,0) {};
		\node[font=\normalsize] at (8.4,0) {$\dots$};
		\node[circle, draw, fill=black, inner sep=1.5pt, label=below:$x^{\ell-1}_n$] (xn) at (10.5,0) {};
		
		\node[rectangle, draw, text width=0.5cm, text height=0.3cm, fill=blue!12, font=\footnotesize] (phi11) at (1.1, 2) {$\phi^{\ell}_{11}$};
		\node[rectangle, draw, text width=0.5cm, text height=0.3cm, fill=blue!12, font=\footnotesize] (phi12) at (2.0, 2) {$\phi^{\ell}_{21}$};
		\node[font=\footnotesize] at (2.8, 2) {$\dots$};
		\node[rectangle, draw, text width=0.5cm, text height=0.3cm, fill=blue!12, font=\footnotesize] (phi1n) at (3.6, 2) {$\phi^{\ell}_{m1}$};
		
		\draw[thin, gray] (x1) -- (phi11.south);
		\draw[thin, gray] (x1) -- (phi12.south);
		\draw[thin, gray] (x1) -- (phi1n.south);
		
		\node[rectangle, draw, text width=0.5cm, text height=0.3cm, fill=blue!12, font=\footnotesize] (phi21) at (5.1, 2) {$\phi^{\ell}_{12}$};
		\node[rectangle, draw, text width=0.5cm, text height=0.3cm, fill=blue!12, font=\footnotesize] (phi22) at (6.0, 2) {$\phi^{\ell}_{22}$};
		\node[font=\footnotesize] at (6.8, 2) {$\dots$};
		\node[rectangle, draw, text width=0.5cm, text height=0.3cm, fill=blue!12, font=\footnotesize] (phi2n) at (7.6, 2) {$\phi^{\ell}_{m2}$};
		
		\draw[thin, gray] (x2) -- (phi21.south);
		\draw[thin, gray] (x2) -- (phi22.south);
		\draw[thin, gray] (x2) -- (phi2n.south);
		
		\node[rectangle, draw, text width=0.5cm, text height=0.3cm, fill=blue!12, font=\footnotesize] (phin1) at (9.2, 2) {$\phi^{\ell}_{1n}$};
		\node[rectangle, draw, text width=0.5cm, text height=0.3cm, fill=blue!12, font=\footnotesize] (phin2) at (10.1, 2) {$\phi^{\ell}_{2n}$};
		\node[font=\footnotesize] at (10.9, 2) {$\dots$};
		\node[rectangle, draw, text width=0.5cm, text height=0.3cm, fill=blue!12, font=\footnotesize] (phinn) at (11.7, 2) {$\phi^{\ell}_{mn}$};
		
		\draw[thin, gray] (xn) -- (phin1.south);
		\draw[thin, gray] (xn) -- (phin2.south);
		\draw[thin, gray] (xn) -- (phinn.south);
		
		\node[circle, draw, fill=black, inner sep=1.5pt, label=above:$x^{\ell}_1$] (y1) at (2.4, 4.2) {};
		\node[circle, draw, fill=black, inner sep=1.5pt, label=above:$x^{\ell}_2$] (y2) at (6.4, 4.2) {};
		\node[font=\normalsize] at (8.4, 4.2) {$\dots$};
		\node[circle, draw, fill=black, inner sep=1.5pt, label=above:$x^{\ell}_m$] (y3) at (10.5, 4.2) {};
		
		\draw[thin, gray] (phi11.north) -- (y1);
		\draw[thin, gray] (phi21.north) -- (y1);
		\draw[thin, gray] (phin1.north) -- (y1);
		
		\draw[thin, gray] (phi12.north) -- (y2);
		\draw[thin, gray] (phi22.north) -- (y2);
		\draw[thin, gray] (phin2.north) -- (y2);
		
		\draw[thin, gray] (phi1n.north) -- (y3);
		\draw[thin, gray] (phi2n.north) -- (y3);
		\draw[thin, gray] (phinn.north) -- (y3);
		
		\node[font=\footnotesize, align=center] (lbl_edge) at (13.8, 2.0) {Learnable\\ univariate\\ functions $\phi^{\ell}_{ij}$\\ on edges};
		\node[font=\footnotesize, align=center] (lbl_node) at (6.4, 5.1) {Summation on nodes};
	\end{tikzpicture}
	
	\vspace{0.6cm}
	
	\begin{tikzpicture}[scale=0.7, transform shape]
		\useasboundingbox (0.0, -0.8) rectangle (16.0, 6.2);
		
		\node[font=\bfseries\small, fill=gray!10, draw=gray!40, rounded corners=2pt, inner sep=3.5pt] at (6.4, 5.8) {(b) Connection of two MLP layers};
		
		\node[circle, draw, minimum size=0.7cm, fill=blue!12, label=below:$x^{\ell-1}_1$] (x1) at (2.4, 0) {$\sigma$};
		\node[circle, draw, minimum size=0.7cm, fill=blue!12, label=below:$x^{\ell-1}_2$] (x2) at (6.4, 0) {$\sigma$};
		\node[font=\normalsize] at (8.4, 0) {$\dots$};
		\node[circle, draw, minimum size=0.7cm, fill=blue!12, label=below:$x^{\ell-1}_n$] (xn) at (10.5, 0) {$\sigma$};
		
		\node[circle, draw, minimum size=0.7cm, fill=blue!12, label=north west:$x^{\ell}_1$] (y1) at (2.4, 4.2) {$\sigma$};
		\node[circle, draw, minimum size=0.7cm, fill=blue!12, label=north west:$x^{\ell}_2$] (y2) at (6.4, 4.2) {$\sigma$};
		\node[font=\normalsize] at (8.4, 4.2) {$\dots$};
		\node[circle, draw, minimum size=0.7cm, fill=blue!12, label=north west:$x^{\ell}_m$] (y3) at (10.5, 4.2) {$\sigma$};
		
		\draw[thin, gray] (x1) -- node[pos=0.25, fill=white, inner sep=1pt, font=\scriptsize] {$w^{\ell}_{11}$} (y1);
		\draw[thin, gray] (x1) -- node[pos=0.35, fill=white, inner sep=1pt, font=\scriptsize] {$w^{\ell}_{21}$} (y2);
		\draw[thin, gray] (x1) -- node[pos=0.20, fill=white, inner sep=1pt, font=\scriptsize] {$w^{\ell}_{m1}$} (y3);
		
		\draw[thin, gray] (x2) -- node[pos=0.40, fill=white, inner sep=1pt, font=\scriptsize] {$w^{\ell}_{12}$} (y1);
		\draw[thin, gray] (x2) -- node[pos=0.25, fill=white, inner sep=1pt, font=\scriptsize] {$w^{\ell}_{22}$} (y2);
		\draw[thin, gray] (x2) -- node[pos=0.40, fill=white, inner sep=1pt, font=\scriptsize] {$w^{\ell}_{m2}$} (y3);
		
		\draw[thin, gray] (xn) -- node[pos=0.20, fill=white, inner sep=1pt, font=\scriptsize] {$w^{\ell}_{1n}$} (y1);
		\draw[thin, gray] (xn) -- node[pos=0.35, fill=white, inner sep=1pt, font=\scriptsize] {$w^{\ell}_{2n}$} (y2);
		\draw[thin, gray] (xn) -- node[pos=0.25, fill=white, inner sep=1pt, font=\scriptsize] {$w^{\ell}_{mn}$} (y3);
		
		\node[font=\footnotesize, align=center] (lbl_edge) at (13.8, 2.0) {Learnable\\ scalar weights\\ $w^{\ell}_{ij}$ on edges};
		\node[font=\footnotesize, align=center] (lbl_node) at (6.4, 5.2) {Summation and fixed activations $\sigma$ on nodes};
	\end{tikzpicture}
	
	\caption{Structural comparison between (a) a KAN layer with edge-bound learnable activation functions $\phi^\ell_{ij}$ and (b) a standard MLP layer with linear weighted edges $w^\ell_{ij}$ and node-level non-linear activations $\sigma$.}
	\label{fig:kan_vs_mlp}
\end{figure}

KANs \cite{Liu2025} form a class of neural network models inspired by the Kolmogorov--Arnold representation theorem (KART), which asserts that any continuous multivariate function can be represented as a finite composition of univariate functions:
\begin{equation}
	f(\bsx)=f(x_1,\dots, x_d ) = \sum_{q=1}^{2n+1}\Phi_q\left(\sum_{p=1}^d \phi_{q,p}(x_p)\right)\,. \label{KART}
\end{equation}
Inspired by the structure of \eqref{KART}, a KAN layer maps the input variables $(x_1,\dots, x_d)$ through learnable univariate functions $\phi_{ij}$, residing on the network edges and connecting the $j$-th input variable to the $i$-th layer node (or neuron). Each univariate function is constructed as a finite sum of suitable basis functions with learnable weights. The layer then computes the summation of these edge outputs at the network nodes, as shown in Fig.~\ref{fig:kan_vs_mlp}a.

This represents a fundamental paradigm shift compared to standard MLPs, which alternate affine transformations with fixed nonlinear activation functions. In an MLP layer, the connection between the $j$-th  input variable and the $i$-th neuron of the layer is parametrized by a learnable scalar weight $w^{\ell}_{ij}$. At the nodes, the weighted inputs are aggregated and  transformed via a fixed nonlinear activation function $\sigma$, as illustrated in Fig.~\ref{fig:kan_vs_mlp}b.  Essentially, while MLPs execute linear operations on the edges and apply non-linear activations at the nodes, KANs replace scalar edge weights with learnable univariate functions $\phi^\ell_{ij}$, parameterized as finite sums of basis functions, and reduce node operations to simple summation.

While \eqref{KART} essentially represents a two-layer network, Liu et al.~\cite{Liu2025} extended this concept to arbitrary depths by stacking multiple KAN layers (see Fig.~\ref{fig:kan_vs_mlp}). Mathematically, a KAN layer $\ell$ with $m$ nodes receiving $n$ inputs from layer $\ell-1$ is represented by a function matrix $\boldsymbol{\Phi}^{\ell}$ containing the learnable univariate edge functions:
\begin{equation}
	\boldsymbol{\Phi}^{\ell} = 
	\begin{pmatrix}
		\phi^{\ell}_{1,1} & \phi^{\ell}_{1,2} & \dots & \phi^{\ell}_{1,n} \\
		\phi^{\ell}_{2,1} & \phi^{\ell}_{2,2} & \dots & \phi^{\ell}_{2,n} \\
		\vdots & \vdots & \ddots & \vdots \\
		\phi^{\ell}_{m,1} & \phi^{\ell}_{m,2} & \dots & \phi^{\ell}_{m,n}
	\end{pmatrix}.
	\label{eq:kan_layer_matrix}
\end{equation}
The forward pass through layer $\ell$ is expressed in matrix notation as:
\begin{equation}
	\bsx^{\ell} = \boldsymbol{\Phi}^{\ell}\bsx^{\ell-1},
\end{equation}
i.e., each component of the vector $\bsx^\ell = (x^\ell_1, \dots, x^\ell_m)^\top$ is computed via node-level summation over the outputs of the evaluated edge functions:
\begin{equation}
	x^\ell_i = \sum_{j=1}^{n} \phi^\ell_{i,j}\left(x^{\ell-1}_j\right), \quad i = 1, \dots, m.
\end{equation}

A FastKAN architecture with $L_{\text{\textsc{kan}}}$ hidden layers is
specified by the sequence of layer widths
$(n_0,n_1,\ldots,n_{L_{\text{\textsc{kan}}}},
n_{L_{\text{\textsc{kan}}}+1})$, where $n_0$ and
$n_{L_{\text{\textsc{kan}}}+1}$ denote the input and output dimensions,
respectively. For networks with a uniform hidden width $n_1 = \dots = n_{L-1} = n$ (as is the case in this work), the architecture can be succinctly denoted by $(L, n)$. Consequently, an $L$-layer KAN is expressed as the composition:
\begin{equation}
	\mathrm{KAN}(\bsx)
	=
	\left(
	\boldsymbol{\Phi}^{L_{\text{\textsc{kan}}}}
	\circ \cdots \circ
	\boldsymbol{\Phi}^{0}
	\right)(\bsx).
\end{equation}
Thus, for example, the architecture
$[2,16,16,16,1]$ contains three hidden layers of width 16 and four
successive FastKAN transformations.

In regular KANs \cite{Liu2025}, each activation function $\phi$ is parametrized by a linear combination of k-th order B-splines and a SiLU function as follows:
\begin{equation}
	\phi(x) = w_\sigma x \sigma(x) + spline(x)\,,
\end{equation}
where 
\begin{equation} 
spline(x) =\sum_{i=1}^{G+k-1} w_i B_i(x)\,.
\end{equation}
Here, $G$ is the grid size, $B_i(x)$ are the  B-spline basis functions of degree-$k$, and $w_\sigma, w_i$ are trainable parameters. To bypass the computational overhead of evaluating B-splines, Li et al.~\cite{Li2024} introduced FastKAN, replacing the B-spline expansion with Radial Basis Functions (RBFs) using Gaussian kernels, i.e.   $spline(x)$ is replaced by a linear combination of the form:   
\begin{eqnarray}
	y(x) = \sum_{i=1}^G w_i \varphi(\|x-x_i\|)\,,
\end{eqnarray}
where $\varphi(r)= exp(-r^2/(2h^2))$ where \(h\) denotes the  characteristic width of the Gaussian radial basis functions. This formulation significantly accelerates model evaluation and simplifies GPU implementations without compromising accuracy. In this work we employ the FastKAN network using the models published in \cite{fastkan}.

Overall, KANs provide greater expressivity and flexibility, enabling KANs to capture complex nonlinear behaviors with significantly fewer parameters. Furthermore, KANs exhibit enhanced interpretability through the direct inspection of learned edge functions, alongside adaptability in continual learning, where the localized nature of the basis functions mitigates catastrophic forgetting of previously learned tasks.

\section{Solving the fixed-boundary GS equilibrium problem via neural network optimization and guided training}
\label{Sec_IV}

\subsection{The optimization problem}
\label{gs_optimization_problem}
Returning to the Grad--Shafranov equation, the PDE residual contributing to $\Lc_{\Dc}$ in Eq.~\eqref{gen_loss} corresponds to the left-hand side of the normalized GS equation \eqref{gs_norm}, while the boundary term $\Lc_{\partial\Dc}$ enforces the Dirichlet boundary condition $\psi = 0$ on $\partial\Dc$. The spatial domain is two-dimensional ($d=2$), with coordinates $(x_1, x_2) = (r, z)$. 

To  identify the physical profile parameters $\bsmu = (P_1, H_1)^\top$ appearing in Eqs.~\eqref{freegs_ansatz} and \eqref{pres_prof_ped}, two physical constraints are introduced. Rather than incorporating parameter estimation directly into the primary optimization problem \eqref{optimization_1}, we adopt a decoupled, alternating optimization scheme. Specifically, the flux network $\psi_{\text{net}}$ is trained for a fixed number of epochs while holding $\bsmu$ constant. Subsequently, $\bsmu$ is updated via gradient descent for a few iterations using exclusively the constraint loss $\Lc_C = w_{\text{\textsc{c}}_1} \Lc_{C_1} + w_{\text{\textsc{c}}_2} \Lc_{C_2}$ while keeping the weights of $\psi_{\text{net}}$ frozen. This alternating procedure is repeated until the total number of network optimization epochs is reached, simultaneously determining the flux function $\psi$ and the physical parameters $\bsmu$.

The constraints $C_1$ and $C_2$ are imposed to enforce predefined target values for the toroidal beta $\beta_t=2 \mu_0 \langle P \rangle /B_0^2 $ and the total toroidal plasma current $I_t$, as done in \cite{Kaltsas2022}.   Utilizing Monte Carlo quadrature over $N_C$ interior collocation points uniformly sampled within the plasma cross section the discrete estimates of these physical quantities are evaluated as:
	\begin{align}
		\hat{\beta}_t &= 2 \, \frac{\sum_{i=1}^{N_C} P(r_i, z_i) r_i}{\sum_{i=1}^{N_C} r_i} \,, \label{eq:beta_mc} \\
		\hat{I}_t &= \frac{S_p}{N_C} \sum_{i=1}^{N_C} J_\phi(r_i, z_i) \,, \label{eq:it_mc}
	\end{align}
where $S_p$ denotes the cross-sectional area enclosed by the plasma boundary $\partial\Dc$. Consequently, the individual constraint loss terms driving the update of $\bsmu$ are formulated as the squared errors relative to their target values:
\begin{align}
	\Lc_{C_1} = \big| \hat{\beta}_t - \beta_{t0} \big|^2 \,, \quad
	\Lc_{C_2} = \big| \hat{I}_t - I_{t0} \big|^2 \,, \label{Lc_1_2}
\end{align}
where $\beta_{t0}$ and $I_{t0}$ represent the target values for the toroidal beta and the total toroidal plasma current, respectively. 

Having defined the physical constraints, we reformulate the overall optimization procedure described in \ref{subsec_pinns}. During the network optimization phase, the network parameters $\bsp$ are updated by solving
\begin{equation}
	\bsp^* = \operatorname*{argmin}_{\bsp} \Lc(\bsp)\,, \label{optimization_2}
\end{equation}
with the AdamW and SSBroyden optimizers while holding the physical parameters $\bsmu$ fixed. Conversely, during the parameter update phase, the optimal physical parameter vector $\bsmu = (P_1, H_1)^\top$ is obtained by minimizing the total constraint loss $\Lc_C(\bsmu) = \sum_{i=1}^2 w_{\text{\textsc{c}}_i} \Lc_{C_i}(\bsmu)$, i.e.
	$\bsmu^* = \operatorname*{argmin}_{\bsmu} \Lc_C(\bsmu)\,,$
which is solved iteratively via standard gradient descent:
\begin{equation}
	\bsmu^{(k+1)} = \bsmu^{(k)} - \eta \nabla_{\bsmu} \Lc_C\left(\bsmu^{(k)}\right)\,,
\end{equation}
where network parameters $\bsp$ are held constant, and $\eta > 0$ denotes the gradient descent step size.

Alternatively, instead of the constraint involving the toroidal current, we can impose the target value of the safety factor on axis through 
\begin{equation}
	\Lc_{C_3} = \left| \hat{q}_a - q_{a0} \right|^2 \,, \label{eq:LC3}
\end{equation}
where
\begin{equation}
	\hat{q}_a = \left[ \frac{F(\psi)}{r} \left( \frac{\partial^2 \psi}{\partial r^2} \frac{\partial^2 \psi}{\partial z^2} \right)^{-1/2} \right]_{(r_a, z_a)} \,, \label{eq:qa_analytic}
\end{equation}
is an analytic formula for the safety factor on the magnetic axis \cite{Freidberg2014}.
The coordinates of the magnetic axis $(r_a, z_a)$ correspond to the O-point location, defined by the condition $|\nabla \psi|_{(r_a, z_a)} = 0$ with $\det(\mathbf{H})|_{(r_a,z_a)} > 0$, where $\mathbf{H}$ is the $2\times2$ Hessian matrix of the flux function. This position is evaluated at each iteration using a damped Newton--Raphson algorithm with adaptive Hessian regularization. Starting from an initial guess $\mathbf{x}^{(0)} = (r_0, z_0)^\top$, the position vector $\mathbf{x}^{(k)} = (r^{(k)}, z^{(k)})^\top$ is updated according to
\begin{equation}
	\mathbf{x}^{(k+1)} = \mathbf{x}^{(k)} - \alpha \, \mathbf{H}_{\text{reg}}^{-1}(\mathbf{x}^{(k)}) \, \nabla \psi(\mathbf{x}^{(k)}) \,, \label{find_axis}
\end{equation}
where $\alpha \in (0,1)$ is a damping parameter ensuring step stability, and $\nabla \psi$ and $\mathbf{H} = \nabla \nabla \psi$ are evaluated via automatic differentiation. To guarantee numerical robustness, if $\mathbf{H}$ is well-conditioned, we set $\mathbf{H}_{\text{reg}} = \mathbf{H}$; otherwise, if it becomes singular or ill-conditioned, a regularized matrix $\mathbf{H}_{\text{reg}} = \mathbf{H} + \epsilon \mathbf{I}$ is employed, where $\epsilon \ll 1$ is a small parameter and $\mathbf{I}$ is the $2\times2$ identity matrix.

\subsection{Guided learning: pretrained networks and homotopy-based curriculum learning} 
\label{subsec_transfer_and_curriculum_learning}
It turned out that the successful and efficient training of simple MLPs and KANs that approximate the solution of the GS equation with highly nonlinear terms requires the exploitation of guided training techniques to facilitate the convergence of the optimization problem. Here, we explore transfer and curriculum learning methods for this task.

Transfer learning refers to the technique where a network trained on a baseline task is reused as the initialization state for a target task, leveraging pre-learned representations to accelerate convergence and avoid optimization pitfalls. Here, we employ transfer learning by initializing our networks with the parameters of pretrained models that approximate simpler equilibrium solutions, namely Solov\'ev equilibria, with the same boundary conditions. Consequently, the networks start training while already satisfying the boundary conditions, which significantly improves their subsequent learning trajectory and convergence rate.

On the other hand, curriculum learning refers to a training paradigm where the model is gradually exposed to tasks of increasing complexity, trying to imitate a structured learning process. Thus, the network starts from an easier optimization problem and is progressively introduced to more demanding tasks, e.g. to GS equations with gradually increased nonlinearity. This helps the network optimize its loss landscape navigation and find global or deeper local minima more efficiently.

Here, we implement a homotopy-based curriculum learning strategy based on a homotopy continuation of the profile functions in the GS equation. Specifically, we define a continuous deformation between the baseline Solov'ev profiles and the target nonlinear  profiles using a homotopy parameter $\lambda \in [0, 1]$:
\begin{align*}
	\frac{dP}{d\psi} &= \lambda \left(\frac{dP}{d\psi}\right)_{\text{nonlinear}} + (1-\lambda)\left(\frac{dP}{d\psi}\right)_{\text{Solovev}} \\
	\frac{dH}{d\psi} &= \lambda \left(\frac{dH}{d\psi}\right)_{\text{nonlinear}} + (1-\lambda)\left(\frac{dH}{d\psi}\right)_{\text{Solovev}}
\end{align*}
As $\lambda$ steps from $0$ to $1$, the network smoothly adapts to the increasing nonlinearity of the plasma equilibrium.

In this work, the neural equilibria are trained using both strategies alongside the baseline unguided method for comparison. While both methods exhibit comparable performance, the homotopy-based method proves more robust, as it ensures convergence across all nonlinear equilibrium scenarios considered in section \ref{Sec_V}. Additionally, it offers the practical advantage of eliminating the need to store pretrained models.

\subsection{Training workflow}

The training framework accommodates three operational modes: a standard unguided baseline and two guided strategies (transfer learning and homotopy continuation). All three share a common two-stage optimization structure: an initial AdamW phase (Phase 1) followed by SSBroyden quasi-Newton refinement (Phase 2) to further minimize \eqref{gen_loss}, but differ in parameter initialization and nonlinearity scheduling:

\begin{enumerate}
	\item \textbf{Standard unguided training:} Network parameters $\bsp$ are initialized randomly via Kaiming uniform weights for the linear base transformation and a small-variance truncated normal distribution for the RBF coefficients \cite{Li2024}. Optimization directly targets the full nonlinear GS equation ($\lambda = 1$ fixed throughout) via the standard two-stage scheme  without domain transfer learning or continuation mechanisms.
	
	\item \textbf{Pretrained transfer-learning approach:} The KAN network is initialized with parameters $\bsp_0$ from a pretrained baseline model satisfying a simpler Solov'ev equilibrium under identical boundary conditions, ensuring $\psi_{\text{net}}|_{\partial\mathcal{D}} = 0$ at the first epoch. Training then proceeds directly through the standard two-stage scheme.
	
	\item \textbf{Homotopy continuation approach:} Parameters $\bsp$ are initialized randomly as in the baseline case. However, during the first stage, AdamW operates within a continuation scheme where a homotopy parameter $\lambda \in [0, 1]$ gradually introduces the GS nonlinearity. Once the full nonlinear profile is attained ($\lambda = 1$), the second stage applies SSBroyden refinement.
\end{enumerate}

Importantly, transfer learning and homotopy continuation are not applied simultaneously, but they are two alternative guided strategies that significantly improve convergence for the nonlinear optimization problem. Transfer learning provides a favorable initialization by starting from a network that has already learned a simpler Solov'ev equilibrium with the same boundary conditions, thereby placing the optimization closer to a physically admissible solution. On the other hand, homotopy continuation, constructs a continuous optimization path from the linear Solov'ev problem to the target nonlinear equilibrium, allowing the network to adapt progressively as the profile nonlinearity is introduced. In both approaches, referred to here as guided methods, AdamW performs the initial optimization and adaptation, while the subsequent SSBroyden phase provides a second-order refinement of the network parameters. As demonstrated by the results presented in Section \ref{Sec_V}, guided training plays a crucial role in achieving convergence for the nonlinear equilibria considered here, for which direct unguided optimization may fail to converge or may converge inefficiently. In addition, the subsequent SSBroyden stage further refines the network solution, reducing the loss terms beyond the levels attained during the first-order AdamW phase. The benefit of SSBroyden is also evident in the linear Solov'ev case, for which no guided training strategy is required, where the quasi-Newton phase substantially improves the final loss minimization, as shown in Section \ref{Sec_V}. Therefore, SSBroyden acts as a complementary, strategy-independent refinement stage that further improves the accuracy of the converged solution.

\begin{figure}[!htb]
	\centering
	\includegraphics[scale=0.7]{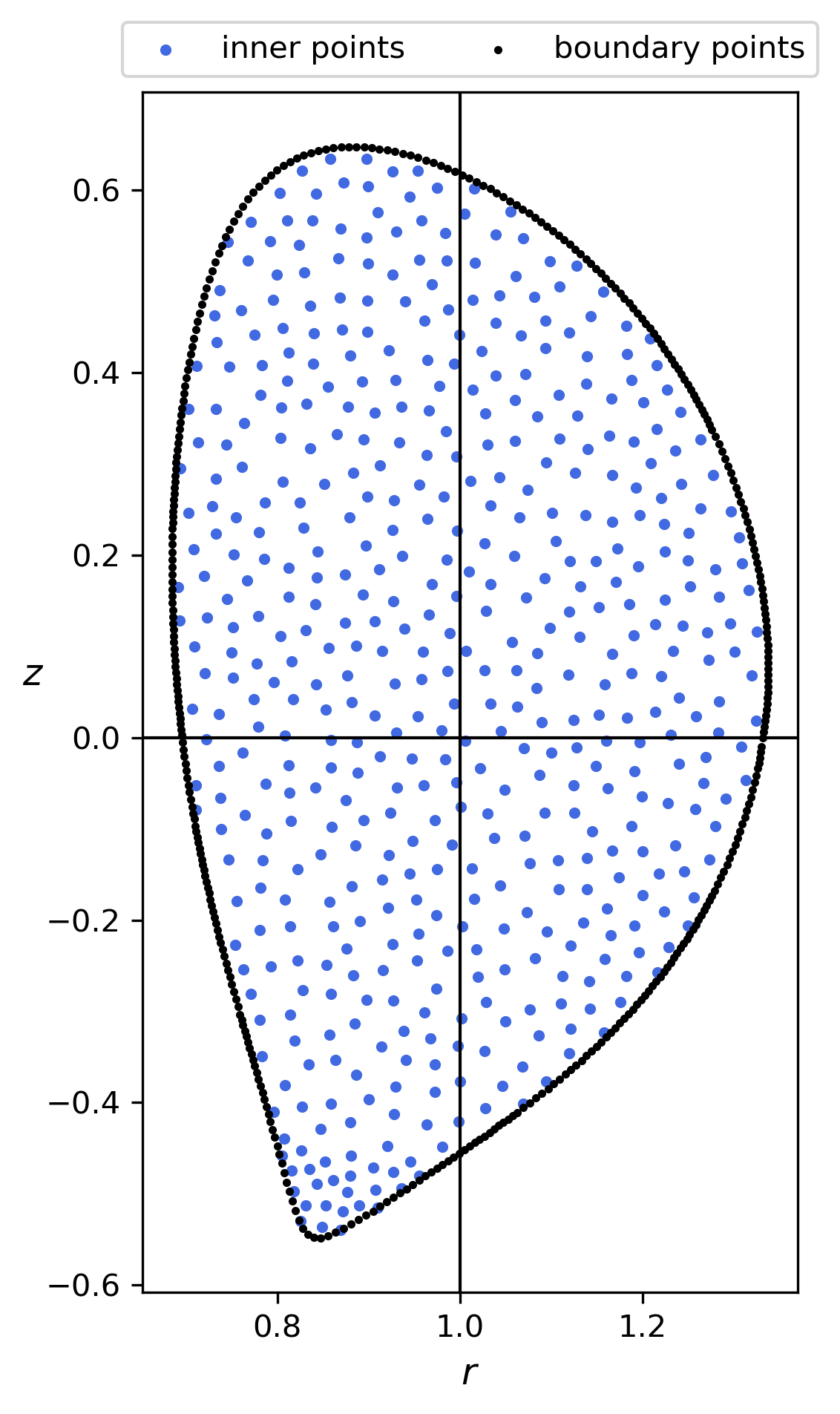}
	\caption{Typical spatial distribution of 390 boundary points and 500 internal collocation points generated by the method described in Appendix~\ref{appendix}, with a minimum inter-point distance of $d_{\text{min}} = 0.03$ (except for a small region near the X-point where the minimum distance is reduced to $0.018$).}
	\label{fig:point_cloud}
\end{figure}

Additionally, the equilibrium calculation algorithm employs the following procedures:
\begin{itemize}
\item \textbf{Collocation point sampling:} In this work, we refine our previous point cloud generation framework \cite{Kaltsas2022} by introducing several enhancements that achieve a balanced and uniform distribution of collocation points across the computational domain. The algorithm incorporates four main procedures:
(i)~\emph{Geometry-aware initial sampling:} The initial radial sampling is scaled by the local boundary distance from the geometric center to eliminate point density imbalances induced by vertical elongation.
(ii)~\emph{Clustering elimination:} Clumped point clusters are dissolved by enforcing a minimum distance criterion ($d_{\text{min}}$) between adjacent collocation points, efficiently evaluated via $k$-d tree spatial queries.
(iii)~\emph{Void filling via refinement:} Under-sampled regions are iteratively populated using an artificial potential energy-based scheme that identifies local spatial gaps.
(iv)~\emph{X-point  refinement:} High-curvature points along the separatrix (e.g. X-points) are automatically detected, increasing the local point density inside the domain adjacent to the boundary.

In the numerical experiments presented herein, the minimum  point distance thresholds for internal and boundary points are set to $d_{\text{min}}^{(i)} = 0.03$ and $d_{\text{min}}^{(b)} = 0.006$, respectively ($d_{\text{min}}^{(b)} = 0.2 \, d_{\text{min}}^{(i)}$). For the specific boundary adopted from \cite{Poulipoulis2021}, this choice of parameters yields a point cloud comprising approximately $500$ interior collocation points and $390$ boundary points, as illustrated in Fig.~\ref{fig:point_cloud}. The point cloud can be regenerated multiple times during training to avoid overfitting on particular point data. A more detailed description of the point cloud generation algorithm is provided in Appendix~\ref{appendix}.

Additionally, to prevent spatial overfitting to a static grid, interior collocation points are periodically regenerated every $M_{\text{reg}}$ epochs.
	\item \textbf{Dynamic axis tracking:} At every epoch, the magnetic axis location $(r_a, z_a)$ and its corresponding flux value $\psi_a$ are computed dynamically using the damped Newton root-finding scheme \eqref{find_axis}. This step ensures that the  normalized flux $\psi_n$ is computated accurately at each step.

\item \textbf{Optional parameter discovery:} When physical parameter discovery is active, the network update pauses every $M_{\bsp}$ epochs of primary optimization on $\bsp$. Subsequently, $M_{\bsmu}$ gradient descent steps are executed to update the physical profile parameters $\bsmu = (P_1, H_1)^\top$ using exclusively the constraint loss $\Lc_C$ as described in \ref{gs_optimization_problem}. Parameter discovery can be applied during Phase 1, Phase 2, or both. To allow the KAN representation to fully stabilize upon discovered parameters, parameter updates are restricted to a predefined cutoff fraction of optimization stage, remaining frozen thereafter.
 
\end{itemize}

All KAN architectures employed in this work are fully connected FastKANs (hence, hereafter the terms KAN and FastKAN are used interchangeably) configured with $G=8$ grid points for the RBF representation. Each network has two input variables corresponding to the spatial coordinates
$(r,z)$, one scalar output $\psi$, and
$L_{\text{\textsc{kan}}}$ hidden layers of uniform width
$n_{\text{\textsc{kan}}}$. Each learnable edge function is represented
using $G$ Gaussian RBFs defined on the grid interval
$[g_{\min},g_{\max}]$. We benchmark the FastKAN models against standard fully connected MLPs with $L_{\text{\textsc{mlp}}}$ hidden layers and $n_{\text{\textsc{mlp}}}$ neurons per layer. The training is performed backpropagating through the network graphs. All differential operators and derivatives in the loss are computed with automatic differentiation using PyTorch \texttt{autograd} engine. Network optimization is carried out using   the AdamW optimizer  operating with a static learning rate or a cosine annealing scheduler (\texttt{CosineAnnealingLR}) and the SSBroyden optimizer with a static learning rate. Refer to Table~\ref{tab:architectures_hyperparameters} for a detailed summary of the various architectures and hyperparameters used in the subsequent numerical experiments.

\begin{table}[!htb]
	\centering
	\footnotesize
	{\setlength{\tabcolsep}{5pt}
		\begin{tabular}{llll}
			\toprule
			\textbf{Parameter / Feature} & \textbf{Solov'ev} & \textbf{Nonlinear} & \textbf{Pedestal} \\
			\midrule
			\multicolumn{4}{l}{\textbf{Network Architecture \& Complexity}} \\
			FastKAN hidden-layer architecture
			$(L_{\text{\textsc{kan}}},n_{\text{\textsc{kan}}})$& $(2,16)$ & $(2,16)$ & $(3,16)$ \\
			Basis function type & \multicolumn{3}{c}{\texttt{Gaussian RBF}} \\
			Grid size ($G$) & $8$ & $8$ & $8$ \\
			MLP$_1$ hidden-layer architecture $(L_{\text{\textsc{mlp}}},n_{\text{\textsc{mlp}}})$ &  $(2,50)$ & $(2,50)$ & $ (3,49) $\\
			MLP$_2$ hidden-layer architecture $(L_{\text{\textsc{mlp}}},n_{\text{\textsc{mlp}}})$  & $(4,30)$ & $(4,30)$ & $(4,40)$\\
			\textit{Number of model parameters} & & & \\
			\quad FastKAN & $2,769$ & $2,769$ & $5,089$ \\
			\quad MLP$_1$ & $2,751$ & $2,751$ & $ 5,097 $ \\
			\quad MLP$_2$ & $2,911$ & $2,911$ & $5,081$ \\
			\midrule
			\multicolumn{4}{l}{\textbf{Optimization}} \\
			\quad $w_\Dc$      & 0.1 & 1.0 & 0.1 \\
			\quad $w_{\partial \Dc}^{(1)}$ / $w_{\partial \Dc}^{(2)}$ & 10 / 100 & 10 / 100 & 10 / 100\\
			\textit{Phase 1: AdamW (Global Exploration)} & & & \\
			\quad Max iterations (epochs) & $500$ & $500$ & $1000$ \\
			\quad Initial Learning rate & $5\times 10^{-4}$ & $5\times 10^{-4}$ & $5\times 10^{-4}$ \\
			\quad LR scheduler & \multicolumn{3}{c}{\texttt{CosineAnnealingLR}} \\ 
			\textit{Phase 2: SS-Broyden (Refinement)} & & & \\
			\quad Max iterations & $500$ & $500$ & $1000$ \\
			\quad Learning rate & $0.5$ & $0.3$ & $0.5$ \\
			\quad LR scheduler & \multicolumn{3}{c}{\texttt{None}} \\
			\textit{Physical Parameter Optimization} & & & \\
			\quad Opt. Frequency / Steps & $30$ / $10$ & $30$ / $10$ & $30$ / $10$ \\
			\quad Step size ($\eta$) & $0.1$ & $0.1$ & $0.001$ \\
			\bottomrule
		\end{tabular}
	}
	\caption{FastKAN and MLP network architecture specifications, trainable parameter counts, and two-stage optimization hyperparameters for the Solov'ev, typical non-linear, and pressure pedestal equilibrium profiles. For all test cases, the domain point cloud comprises $\sim 500$ interior points ($d_{\text{min}}^{(i)} = 0.03$) and $\sim 390$ boundary points ($d_{\text{min}}^{(b)} = 0.006$).}
	\label{tab:architectures_hyperparameters}
\end{table}

\section{Numerical experiments}
\label{Sec_V}
\subsection{Solovev equilibrium}
\label{subsec_Solovev_equilibrium}
The Solov'ev equilibrium is obtained by solving Eq.~\eqref{gs_norm} with profile functions specified in \eqref{freegs_ansatz}. Setting $a_n = b_n = 0$ reduces the Grad--Shafranov (GS) equation to the linear inhomogeneous PDE:
\begin{equation}
	\frac{\partial^2 \psi}{\partial r^2} - \frac{1}{r} \frac{\partial \psi}{\partial r} + \frac{\partial^2 \psi}{\partial z^2} + H_1 + P_1 r^2 = 0\,, \label{gs-Solovev}
\end{equation}
which admits analytical solutions constructed via a linear combination of a homogeneous solution and a particular inhomogeneous solution.

\subsubsection{Solov'ev equilibria with ITER-relevant boundary}
Although closed-form Solov'ev solutions such as \eqref{Solovev_analytic} exist in the literature, tailoring equilibria to customized boundary geometries still requires numerical fitting procedures. 

Owing to its linearity, the corresponding homogeneous PDE admits solutions expressed via truncated series expansions with free coefficients, which are determined by enforcing boundary conditions suitable for fusion-relevant plasma cross-sections (see, e.g., \cite{Throumoulopoulos2012, Kaltsas2014}). However, evaluating these coefficients introduces numerical ill-conditioning: as higher-order terms are retained in the series expansion, minor numerical inaccuracies can severely compromise boundary accuracy.  

Consequently, employing neural network surrogate models is well justified even for this quasi-analytical linear case. Furthermore, neural network Solov'ev solutions serve as an effective initialization for surrogate models subsequently trained to solve  nonlinear GS equations, directly supporting the transfer learning framework outlined in Section \ref{subsec_transfer_and_curriculum_learning}.

In this context, we apply the proposed PIKAN framework detailed in Sections~\ref{Sec_III} and~\ref{Sec_IV} to construct  equilibria featuring an ITER-like boundary with a lower single-null (X-point), approximating the configuration used in \cite{Poulipoulis2021}. The primary geometric and target physical parameters for this equilibrium configuration are summarized in Table~\ref{tab:geometric_target_parameters}. 

For this particular equilibrium class, the PIKAN algorithm is executed in its standard unguided configuration without transfer learning. Furthermore, homotopy continuation is redundant in this setup, as the target configuration is itself a Solov'ev equilibrium, which serves as the baseline reference state for our continuation scheme. The KAN architecture and corresponding training hyperparameters used in this and subsequent experiments are summarized in Table~\ref{tab:architectures_hyperparameters}.

\begin{table}[!htb]
	\centering
	\small
	\setlength{\tabcolsep}{14pt}
	\begin{tabular}{lc}
		\toprule
		\textbf{Parameter} & \textbf{Value} \\
		\midrule
		\multicolumn{2}{l}{\textit{Geometric Parameters (Boundary)}} \\
		\addlinespace[2pt]
		Inverse aspect ratio $\epsilon$ & $0.33$ \\
		Upper elongation $\kappa_u$ & $1.92$ \\
		Lower elongation $\kappa_d$ & $1.62$ \\
		Upper triangularity $\delta_u$ & $0.36$ \\
		Lower triangularity $\delta_d$ & $0.48$ \\
		\addlinespace
		\multicolumn{2}{l}{\textit{Target Physical Quantities}} \\
		\addlinespace[2pt]
		Toroidal plasma beta $\beta_t$ & $0.03$ \\
		Toroidal plasma current $I_t$ $[10^7\, A]$ & $1.20$ \\
		\bottomrule
	\end{tabular}
	\caption{Geometric  and target physical parameters for the ITER-like equilibrium configuration.}
	\label{tab:geometric_target_parameters}
\end{table}

\begin{figure}[!htb]
	\centering
	\includegraphics[scale=0.6]{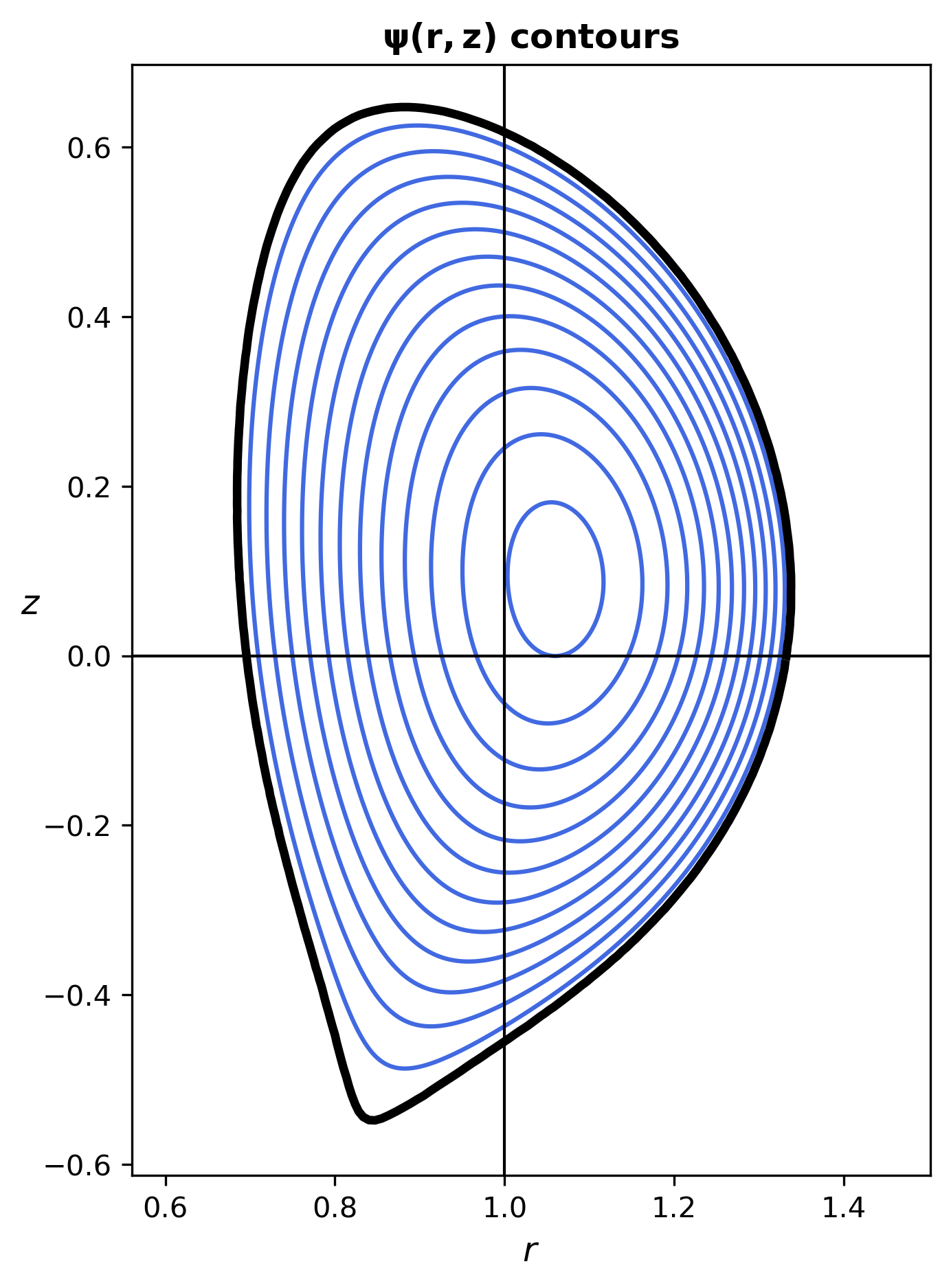}
	\includegraphics[scale=0.6]{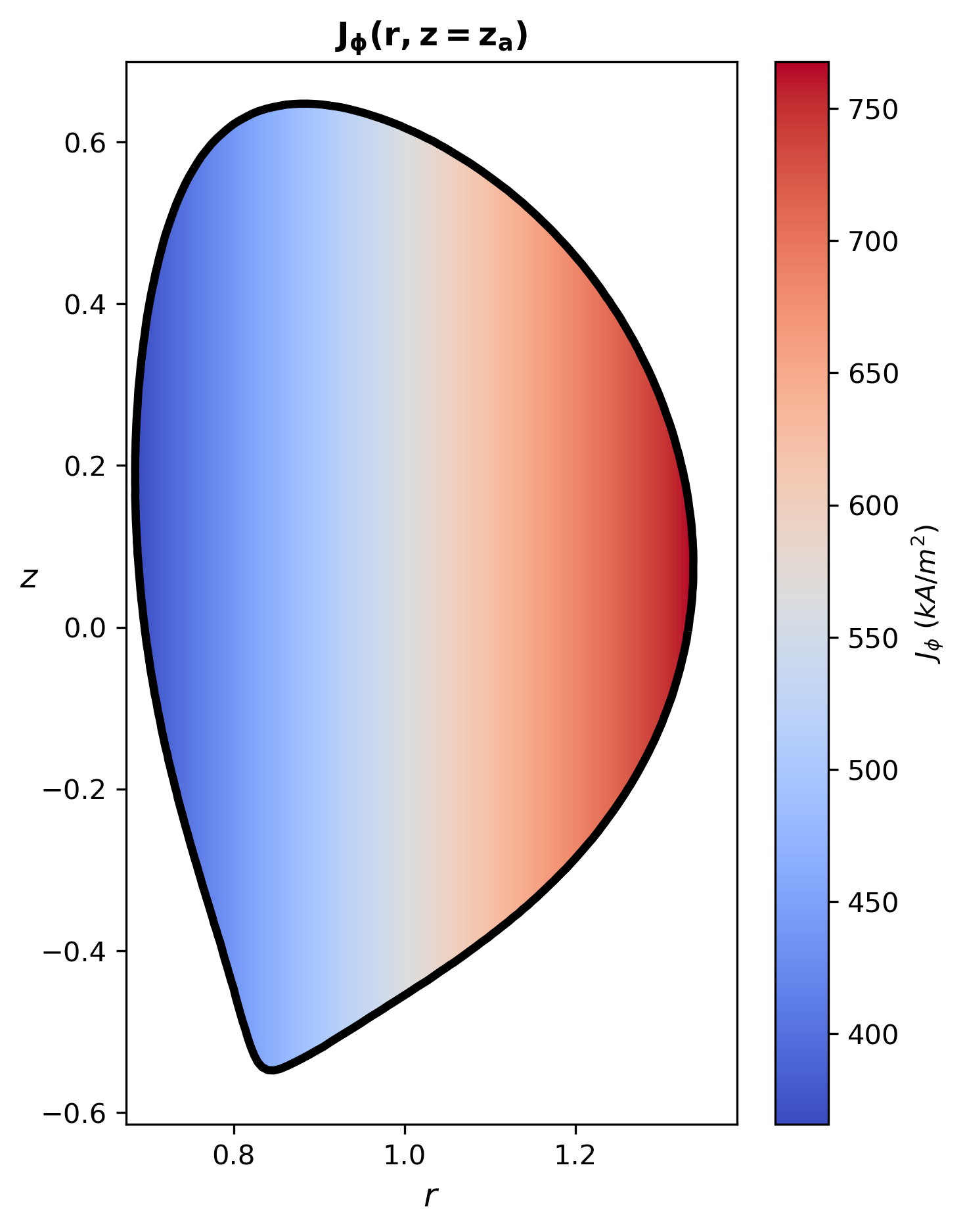}
\caption{(Left panel) Poloidal cross-section of the magnetic flux surfaces for the Solov'ev KAN equilibrium. (Right panel) The corresponding 2D toroidal current density distribution across the poloidal cross-section. Note that this equilibrium, characterized by a monotonic \(J_\phi\) profile, is a linear model that serves as a benchmark test case rather than a physically realistic tokamak equilibrium.}

	\label{fig:Solovev_contours_and_Jt}
\end{figure}

\begin{figure}[!htb]
	\centering
	\includegraphics[scale=0.4]{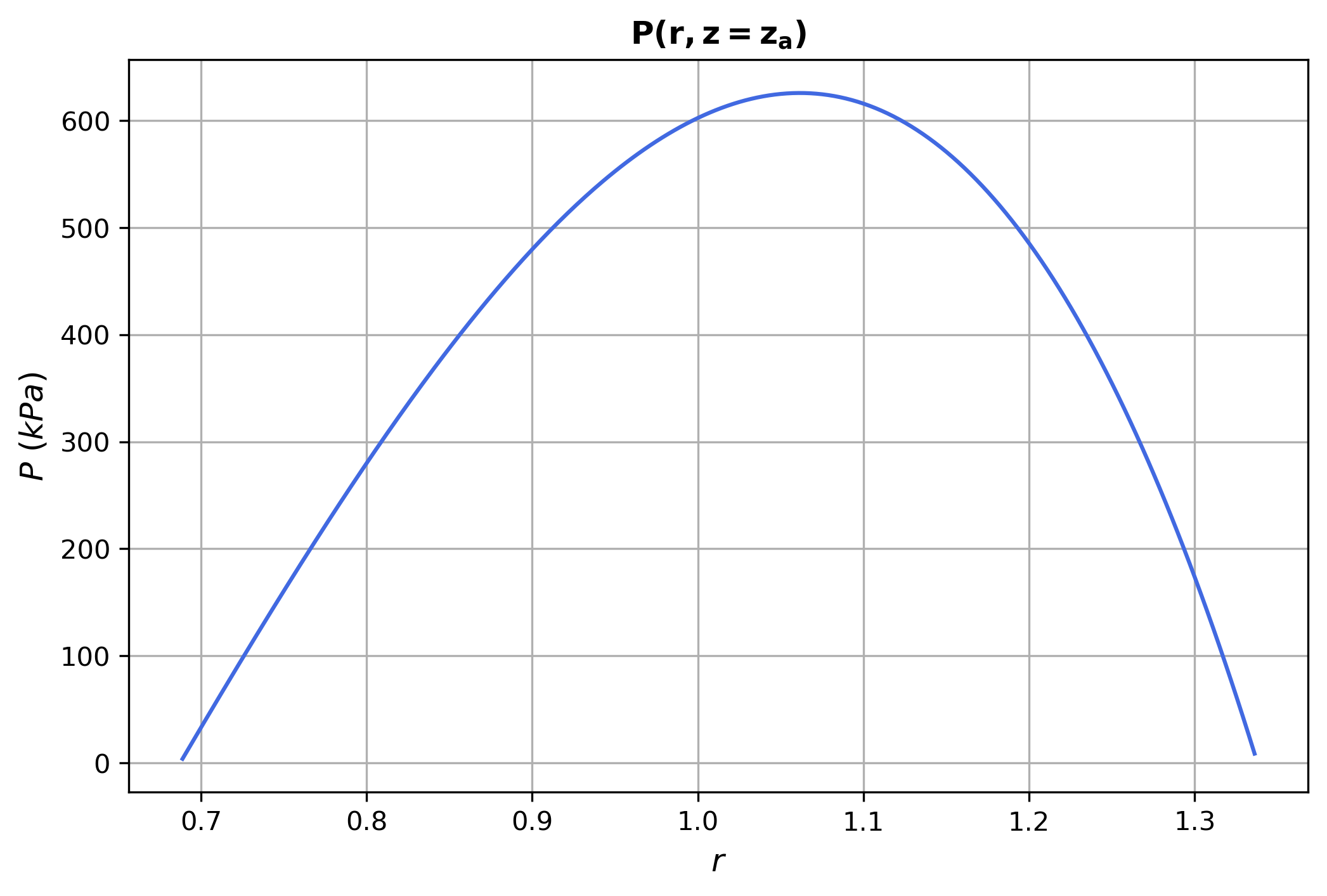}
	\includegraphics[scale=0.4]{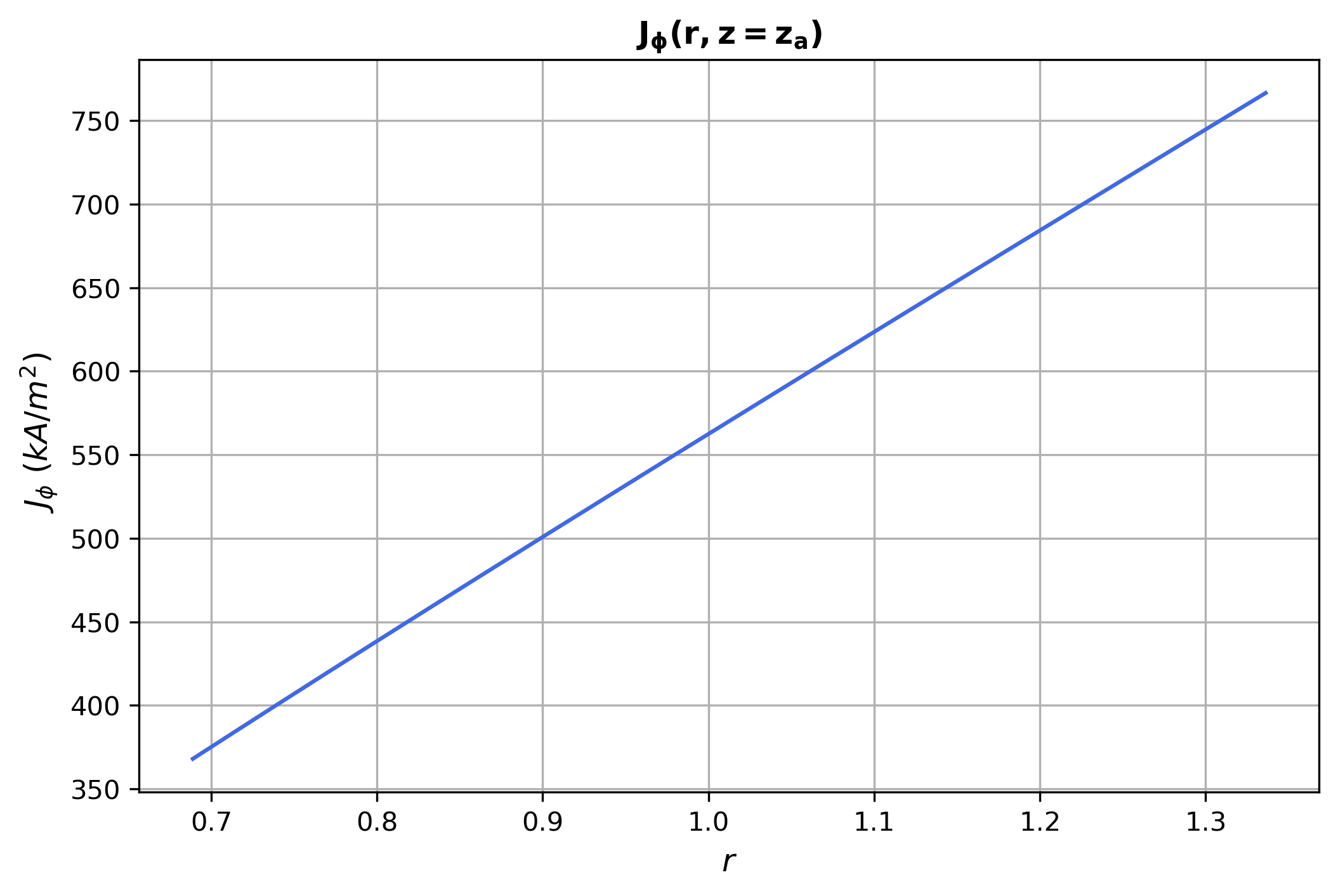}
	\includegraphics[scale=0.4]{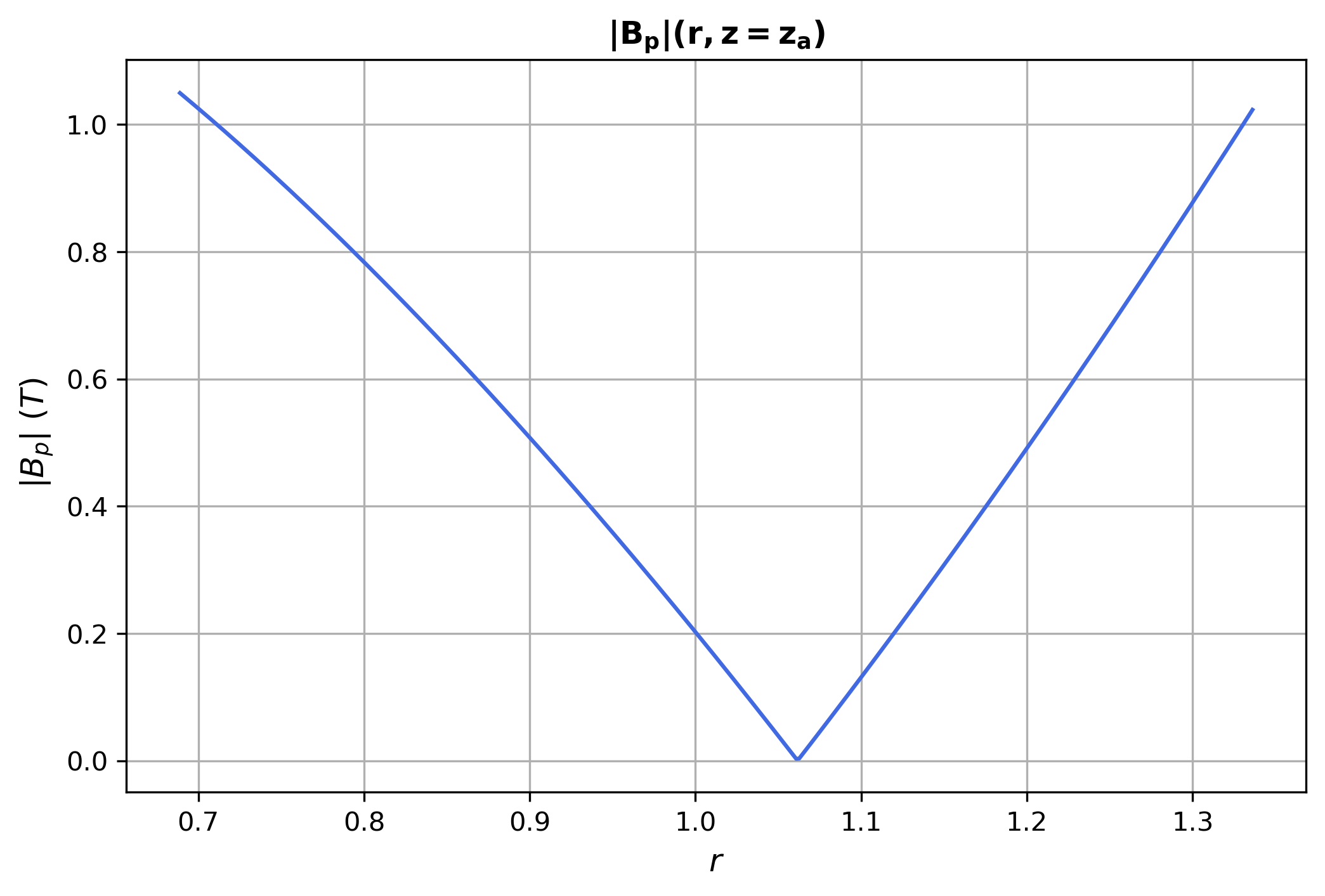}
	\includegraphics[scale=0.4]{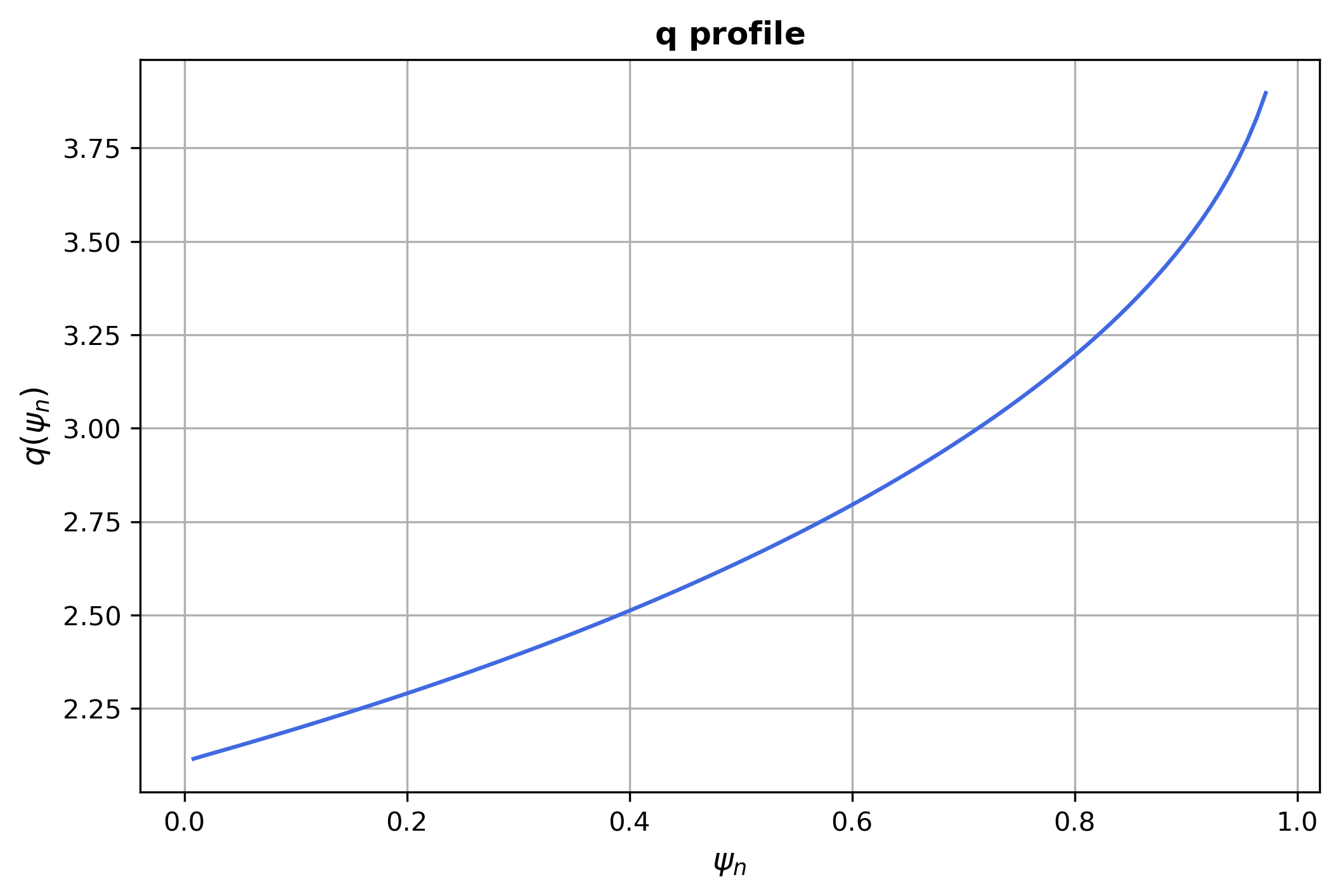}
	\caption{Profiles of the plasma pressure $P(r, z_a)$ (top-left), the toroidal current density $J_\phi(r, z_a)$ (top-right), and the modulus of the poloidal magnetic field $B_p(r, z_a)$ (bottom-left) on the midplane $z = z_a$, alongside the safety factor profile $q(\psi_n)$ as a function of the normalized poloidal flux $\psi_n$ (bottom-right) for the Solov'ev equilibrium.}
	\label{fig:Solovev_profiles}
\end{figure}

\begin{figure}[!htb]
	\centering
	\includegraphics[scale=0.5]{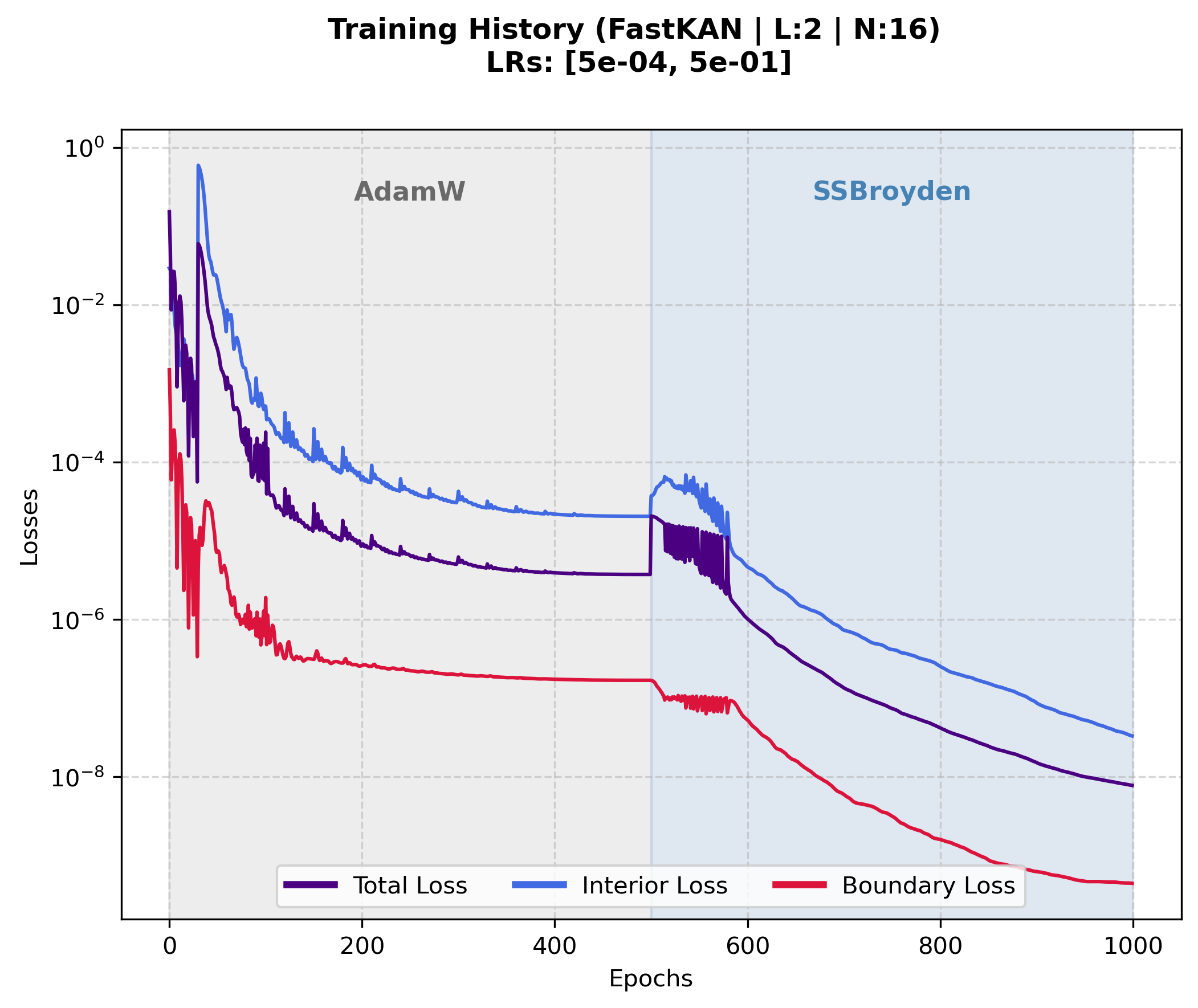}

	\caption{The training history of the KAN network in the Solov'ev case with discoverable profile parameters.} \label{fig:Solovev_train_hist}
\end{figure}

The reconstructed free parameters $(P_1, H_1)$, alongside key figures of merit, equilibrium characteristics, and the GPU runtime, are summarized in the first row of Table~\ref{tab:mlp_kan_comparison_Solovev} while the training history, representing the dynamic evolution of the loss terms during training, is illustrated in Fig.~\ref{fig:Solovev_train_hist}.  

The poloidal cross-section of the magnetic flux surfaces $\psi(r,z)$ revealing the magnetic topology and the 2D spatial distribution of the toroidal current density $J_\phi$ are depicted in Fig.~\ref{fig:Solovev_contours_and_Jt}. Furthermore, Fig.~\ref{fig:Solovev_profiles} illustrates profiles of the plasma pressure $P$, the toroidal current density $J_\phi$, and the modulus of the poloidal magnetic field $B_p=|\nabla \psi|/r$ along the plane $z=z_a$, together with the safety factor profile $q(\psi_n)$ plotted against the normalized poloidal flux $\psi_n$. The safety factor profile is evaluated as a function of $\psi$ via the line integral:
\begin{equation}
	q(\psi) = \frac{F(\psi)}{2\pi}\oint \frac{\mathrm{d}\ell}{r^2 B_p} \,,
\end{equation}
evaluated on a set of flux surfaces. Here $\mathrm{d}\ell$ represents the arc length element along the flux surface contour in the poloidal plane.

\subsubsection{Comparison with MLPs}
To evaluate the efficacy of the KAN solver, its accuracy and convergence
properties are benchmarked against standard MLPs with approximately matched
numbers of trainable parameters. Here,
$L_{\text{\textsc{kan}}}$ and $L_{\text{\textsc{mlp}}}$ denote the
numbers of hidden layers in the respective architectures. Note, however, that an individual KAN layer
is not functionally equivalent to an MLP layer: KANs employ learnable
univariate functions on the edges, whereas MLPs use scalar edge weights
followed by fixed activation functions placed at the nodes. We therefore use the total number of trainable parameters as the primary measure of model-size comparability.

\begin{figure}[!htb]
	\centering 
	\includegraphics[scale=0.5]{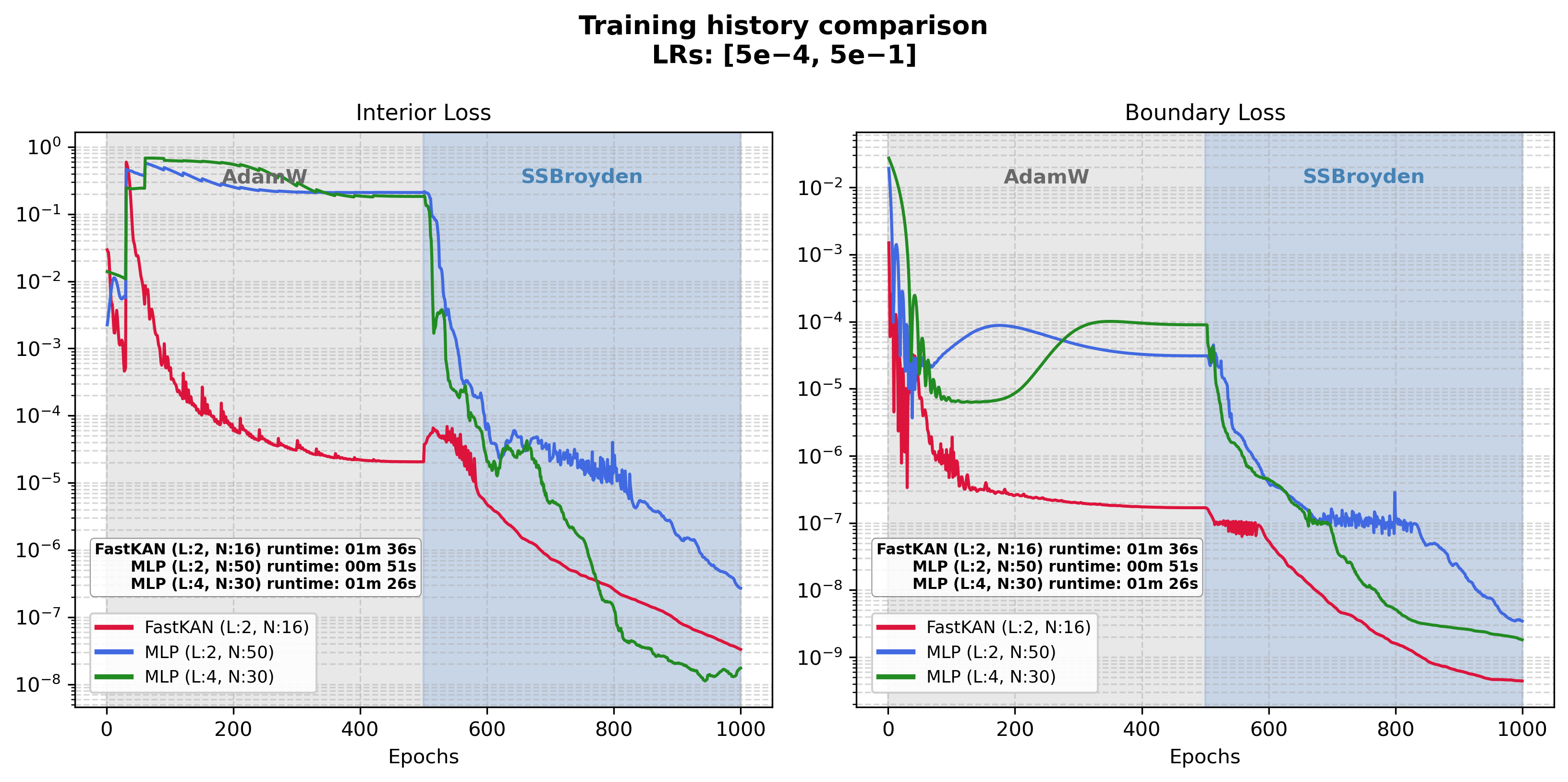}
\caption{Comparison of training histories among the Solov'ev FastKAN (red), $\text{MLP}_1$ (blue), and $\text{MLP}_2$ (green) models with profile parameter discovery: (left) interior loss term minimization dynamics and (right) corresponding boundary loss dynamics.}
\label{fig:fastkan_vs_mlp_discovery}
\end{figure}

For a FastKAN architecture with two input dimensions $(r,z)$ and a single
scalar output $\psi$, the total number of trainable parameters
$\Nc_{\text{\textsc{kan}}}$ is given by
\begin{equation}
	\Nc_{\text{\textsc{kan}}}
	=
	(L_{\text{\textsc{kan}}}-1)n_{\text{\textsc{kan}}}^2(G+1)
	+
	n_{\text{\textsc{kan}}}
	(3G + L_{\text{\textsc{kan}}} + 3)
	+ 1\,,
\end{equation}
where $G$ denotes the FastKAN grid size, i.e., the number of Gaussian RBF
basis functions associated with each learnable edge function. The corresponding parameter count for the MLP architecture is given by:
\begin{equation}
	\Nc_{\text{\textsc{mlp}}} = (L_{\text{\textsc{mlp}}}-1)n_{\text{\textsc{mlp}}}^2 + (L_{\text{\textsc{mlp}}}+3)n_{\text{\textsc{mlp}}} + 1\,.
\end{equation}
Because each edge in a KAN performs a univariate functional fit using $G$ localized RBF grid points, FastKAN achieves high functional expressivity with relatively compact hidden dimensions. Consequently, matching the parameter count in standard MLPs requires scaling either the network width $n_{\text{\textsc{mlp}}}$ (model MLP$_1$) or depth  $L_{\text{\textsc{mlp}}}$ (model MLP$_2$). The former increases the linear feature combination capacity per layer, whereas the latter introduces deeper hierarchies of non-linear activation functions.
\begin{table}[!htb]
	\centering
	\vspace{0.2cm}
	\footnotesize
	\setlength{\tabcolsep}{2.5pt} 
	\begin{tabular}{l c c c c c c c c}
		\toprule
		\makecell{\textbf{Model} \\ \textbf{Method}} & $(P_1, H_1)$ & $u_a$ & $\beta_t$ & $I_t \; [10^7\text{ A}]$ & $q_a$ & $(r_a, z_a)$ & \makecell{\textbf{Losses} \\ $\mathcal{L}_{\mathcal{D}} / \mathcal{L}_{\partial \mathcal{D}}$} & \textbf{Runtime} \\
		\midrule
		\makecell{\textbf{KAN} \\ \textbf{Unguided}}   & $(0.916, -0.040)$ & $0.034$ & $0.029$ & $1.207$ & $2.108$ & $(1.062, 0.090)$ & \makecell{$3.3 \times 10^{-8}$ \\ $4.4 \times 10^{-10}$} & 01m 50s \\ \addlinespace
		\makecell{\textbf{MLP$_1$} \\ \textbf{Unguided}} & $(1.067, -0.178)$ & $0.035$ & $0.034$ & $1.210$ & $1.999$ & $(1.067, 0.090)$ & \makecell{$2.7 \times 10^{-7}$ \\ $3.4 \times 10^{-9}$} & 00m 51s \\ \addlinespace
		\makecell{\textbf{MLP$_2$} \\ \textbf{Unguided}} & $(0.904, 0.012)$  & $0.036$ & $0.030$ & $1.266$ & $2.046$ & $(1.059, 0.091)$ & \makecell{$1.7 \times 10^{-8}$ \\ $1.8 \times 10^{-9}$} & 01m 26s \\
		\bottomrule
	\end{tabular}
	\caption{Comparative  results between the reference FastKAN and two parameter-matched MLP models scaling either the network width (MLP$_1$) or the depth (MLP$_2$) for the Solov'ev case. }
	\label{tab:mlp_kan_comparison_Solovev}
\end{table}
Table~\ref{tab:mlp_kan_comparison_Solovev} summarizes the architectures and parameter counts of the three models being compared (KAN, MLP$_1$ and MLP$_2$), along with the corresponding reconstructed values of the trainable physical parameters, equilibrium figures of merit, achieved minimum losses, and computational runtimes. The convergence history for each architecture is depicted in Fig.~\ref{fig:fastkan_vs_mlp_discovery}, where the internal PDE loss $\mathcal{L}_{\mathcal{D}}$ and boundary loss $\mathcal{L}_{\partial \mathcal{D}}$ of the FastKAN model are compared against the MLP models. The FastKAN model converges more rapidly and minimizes the loss terms more effectively than the MLP models during the AdamW stage (where parameter reconstruction takes place); however, during the SSBroyden phase the deeper MLP$_2$ architecture minimizes the PDE loss more effectively. In addition the KAN model exhibits a higher overall GPU execution time\footnote{All networks were trained using double precision ($\texttt{float64}$) on an NVIDIA RTX 5070 GPU.}. It should be noted that each architecture may benefit from specific learning rate configurations. Therefore, under optimal hyperparameter tuning, MLPs could potentially outperform  the KAN model in both accuracy and efficiency; however, such an investigation was not pursued in this study.

\subsubsection{Benchmark against analytic solution}
In this subsection, we benchmark the accuracy of the previously used lightweight KAN model (featuring two hidden layers and 16 nodes per layer) against the known Solov'ev equilibrium given by Eq.~\eqref{Solovev_analytic}. Here, we aim to assess the accuracy of the KAN approximation for fixed values of $P_1$ and $H_1$.

To perform this benchmark, we construct an equilibrium using Eq.~\eqref{Solovev_analytic} by a priori fixing the parametric values $(P_1, H_1, \gamma) = (-3.28, 0.8, 0.02)$ to obtain a configuration with the desired geometric characteristics (e.g., fusion-relevant inverse aspect ratio, elongation, and the presence of a single lower X-point). We then extract the coordinates of the boundary points, i.e. the points along the spontaneously formed separatrix that encloses the magnetic flux surfaces degenerating toward the magnetic axis located near $(r,z) = (1,0)$. These points serve as boundary data to generate a spatial cloud of collocation points for training the KAN model with frozen physical parameters.

\begin{figure}[!htb]
	\centering
	\includegraphics[scale=0.5]{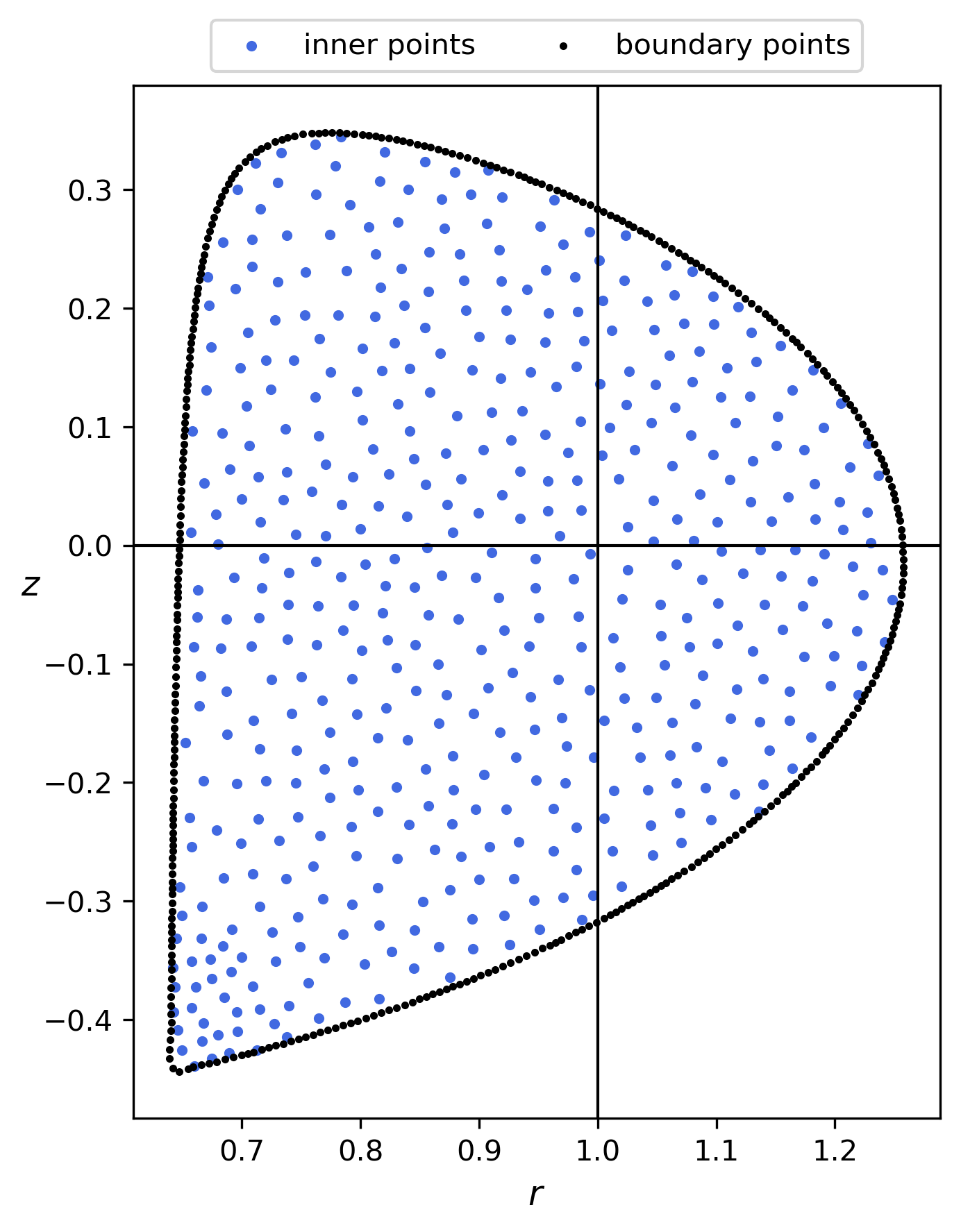}%
	\includegraphics[scale=0.5]{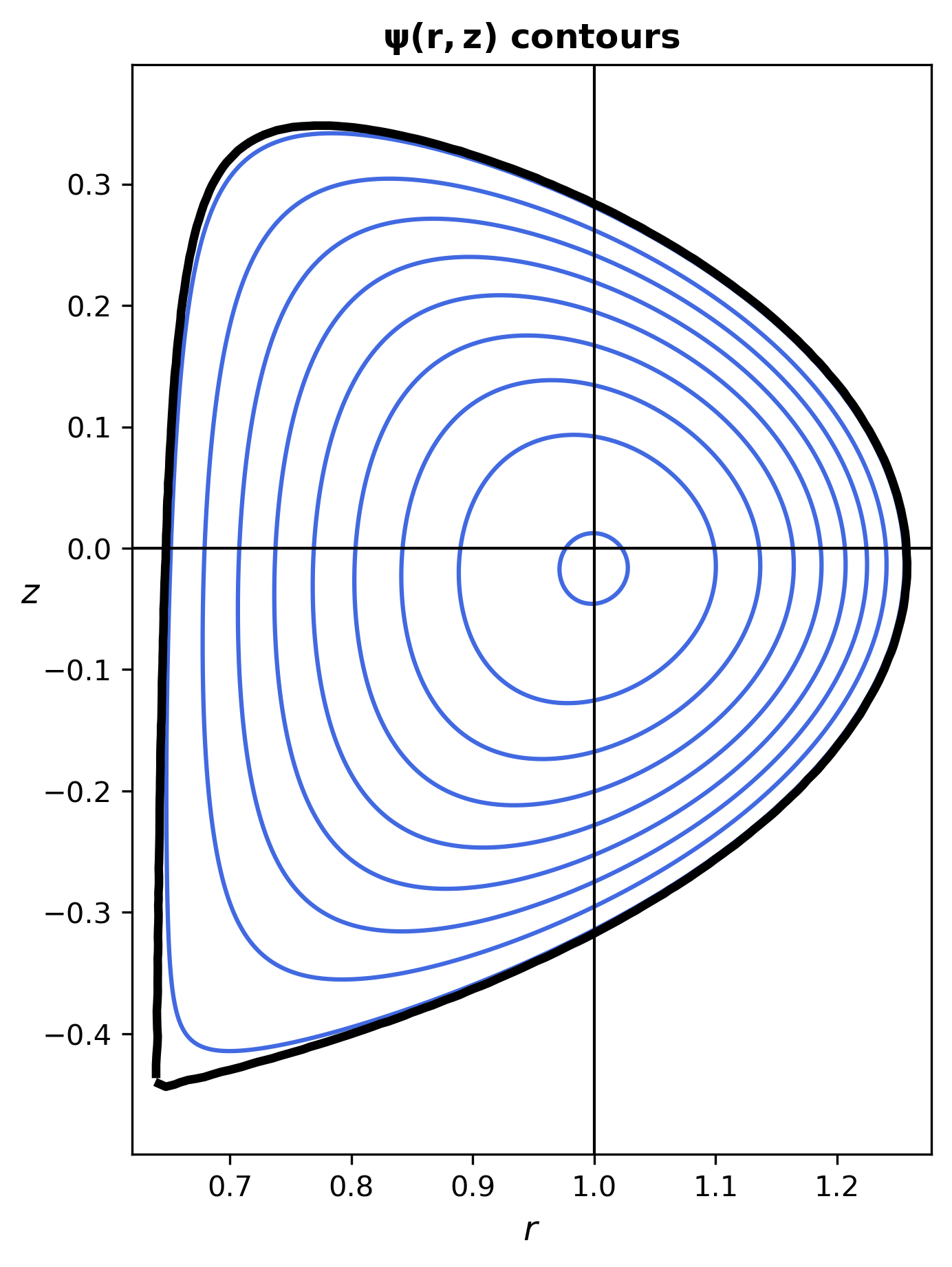}
	\caption{(Left) Distribution of internal and boundary collocation points generated for the benchmark test. (Right) Magnetic flux surfaces on the poloidal plane constructed from the FastKAN solution.}
	\label{fig:point_cloud_and_contours_bench}
\end{figure}

In Fig.~\ref{fig:point_cloud_and_contours_bench}, we show the cloud of collocation points alongside the computed magnetic flux surfaces on the poloidal plane. We construct this equilibrium utilizing two different architectures, namely the FastKAN and the MLP$_1$ models detailed in Table~\ref{tab:architectures_hyperparameters}, and we assess their relative accuracy and computational efficiency. 

\begin{figure}[!htb]
	\centering
	\includegraphics[scale=0.5]{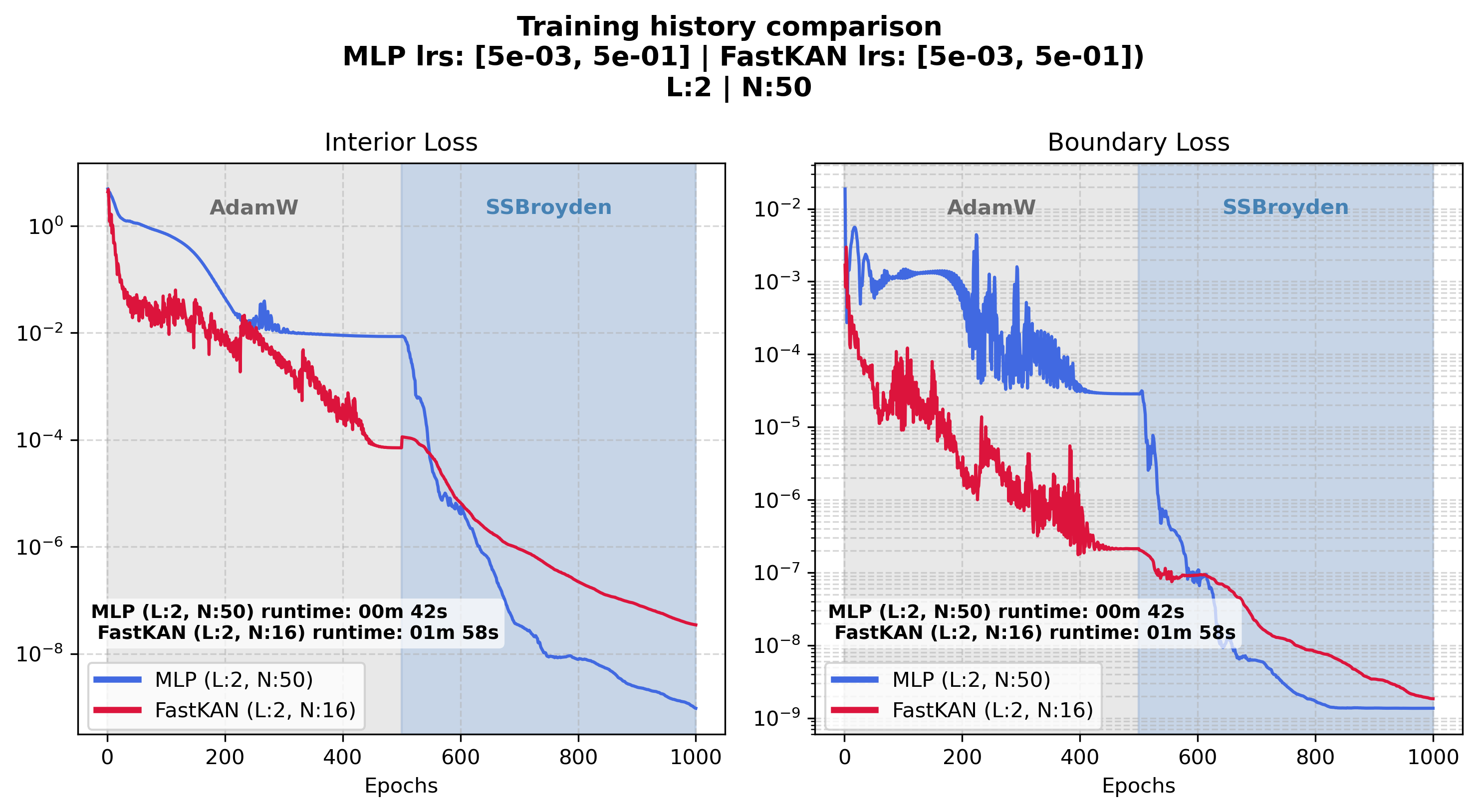}
\caption{Training loss dynamics during the AdamW and SS-Broyden optimization phases for the FastKAN (red) and $\text{MLP}_1$ (blue) models in the benchmark case.}
\label{fig:hist_BENCH}
\end{figure}

Figure~\ref{fig:hist_BENCH} illustrates the dynamic evolution of the loss terms during training, comparing the performance of the two models. Both loss terms, after 500 AdamW and 500 SSBroyden epochs, attain values below $10^{-7}$ with the FastKAN model. The wide MLP$_1$ achieves better accuracy during the SSBroyden phase in significantly less GPU time, corroborating our earlier remark that MLPs can be tuned to outperform KANs for linear equilibria. However, as demonstrated subsequently, this advantage does not hold for nonlinear equilibria.

\begin{figure}[!htb]
	\centering
	\includegraphics[scale=0.5]{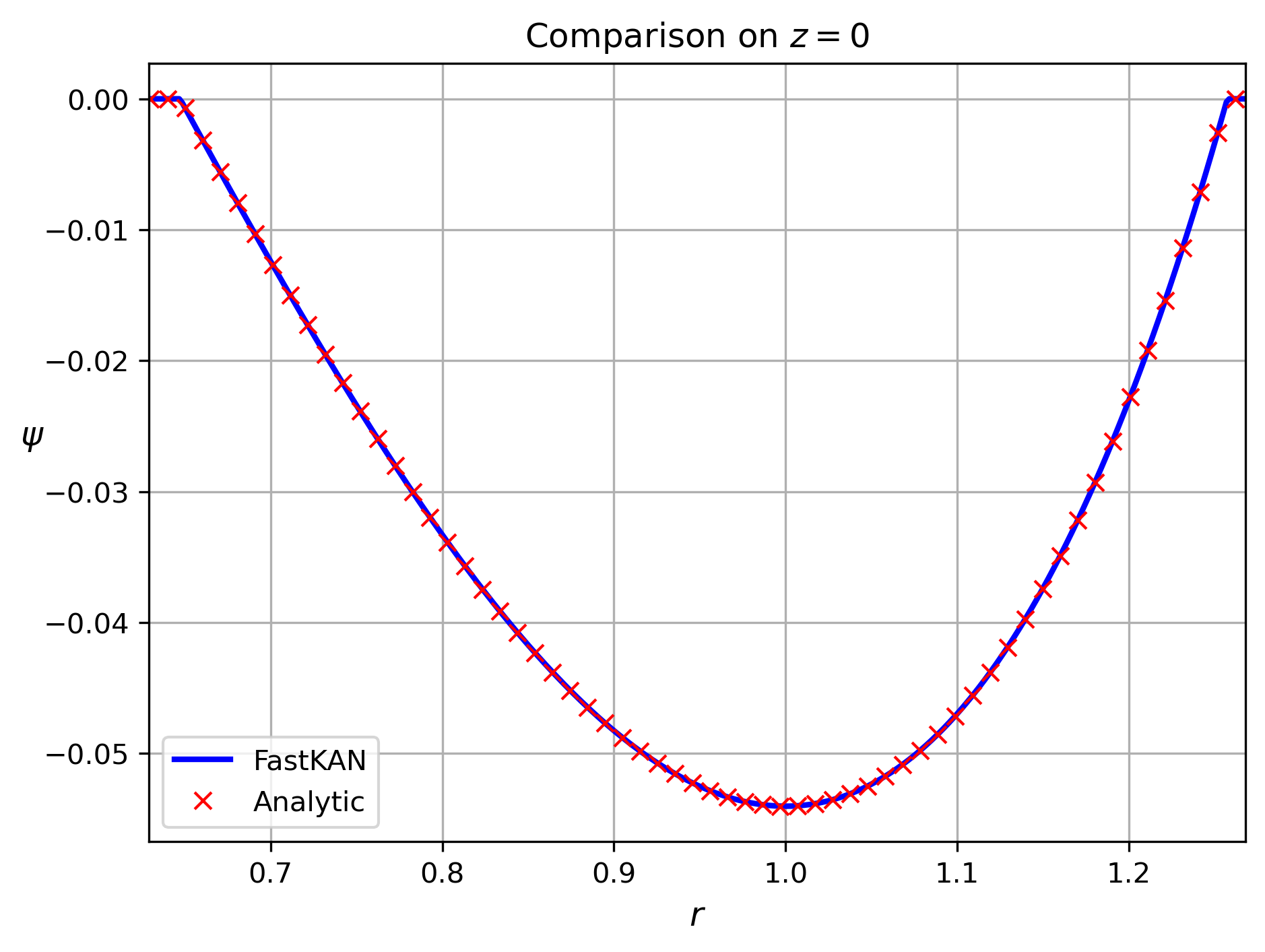}
	\caption{Comparison of the  poloidal magnetic flux function $\psi(R, z=0)$ along the midplane between the FastKAN approximation $\psi_{\text{KAN}}$ and the analytical Solov'ev solution $\tilde{\psi}$.}
	\label{fig:comparison_error}
\end{figure}

\begin{figure}[!htb]
	\centering
	\includegraphics[height=6.5cm]{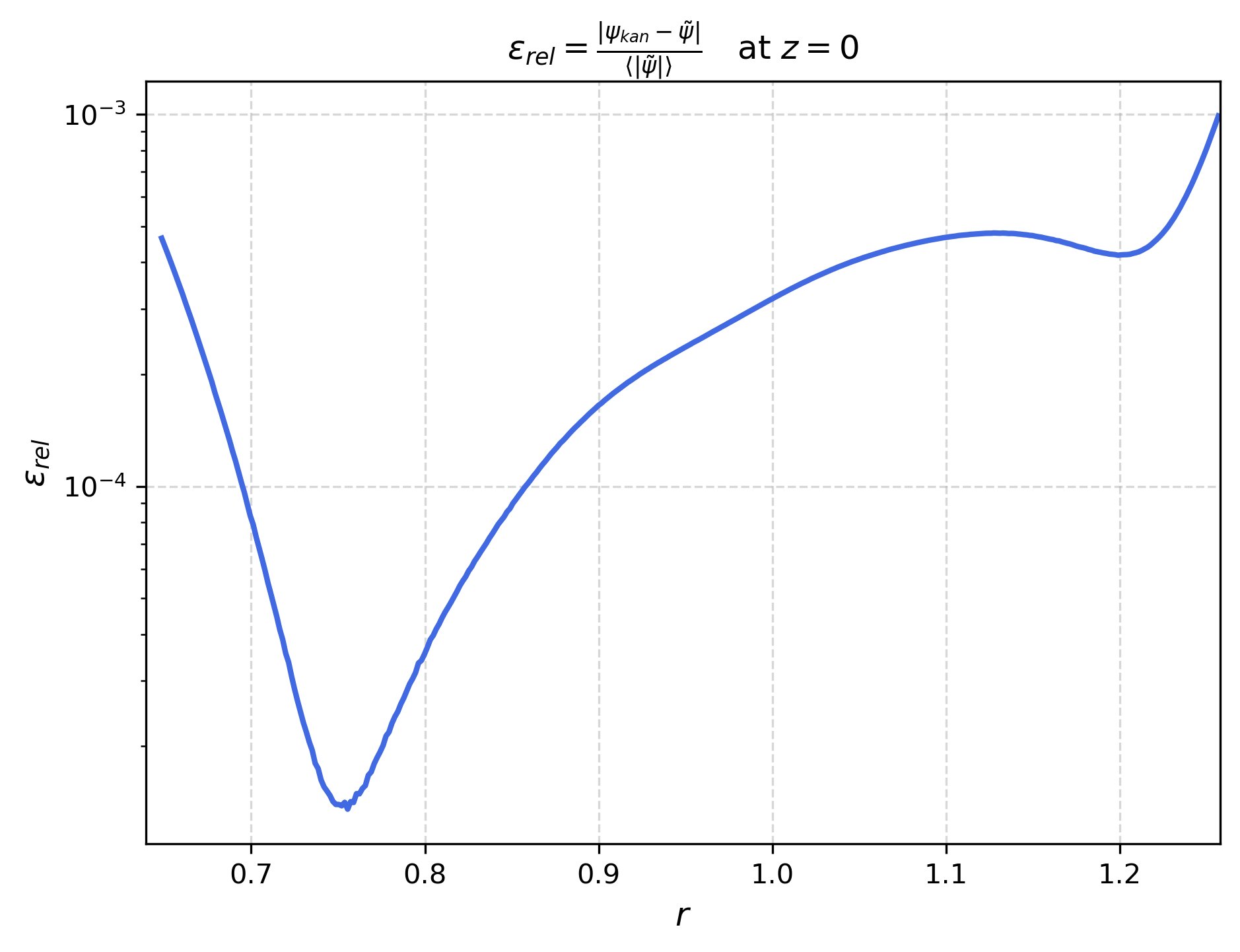}%
	\hspace{0.2cm} 
	\includegraphics[height=6.5cm]{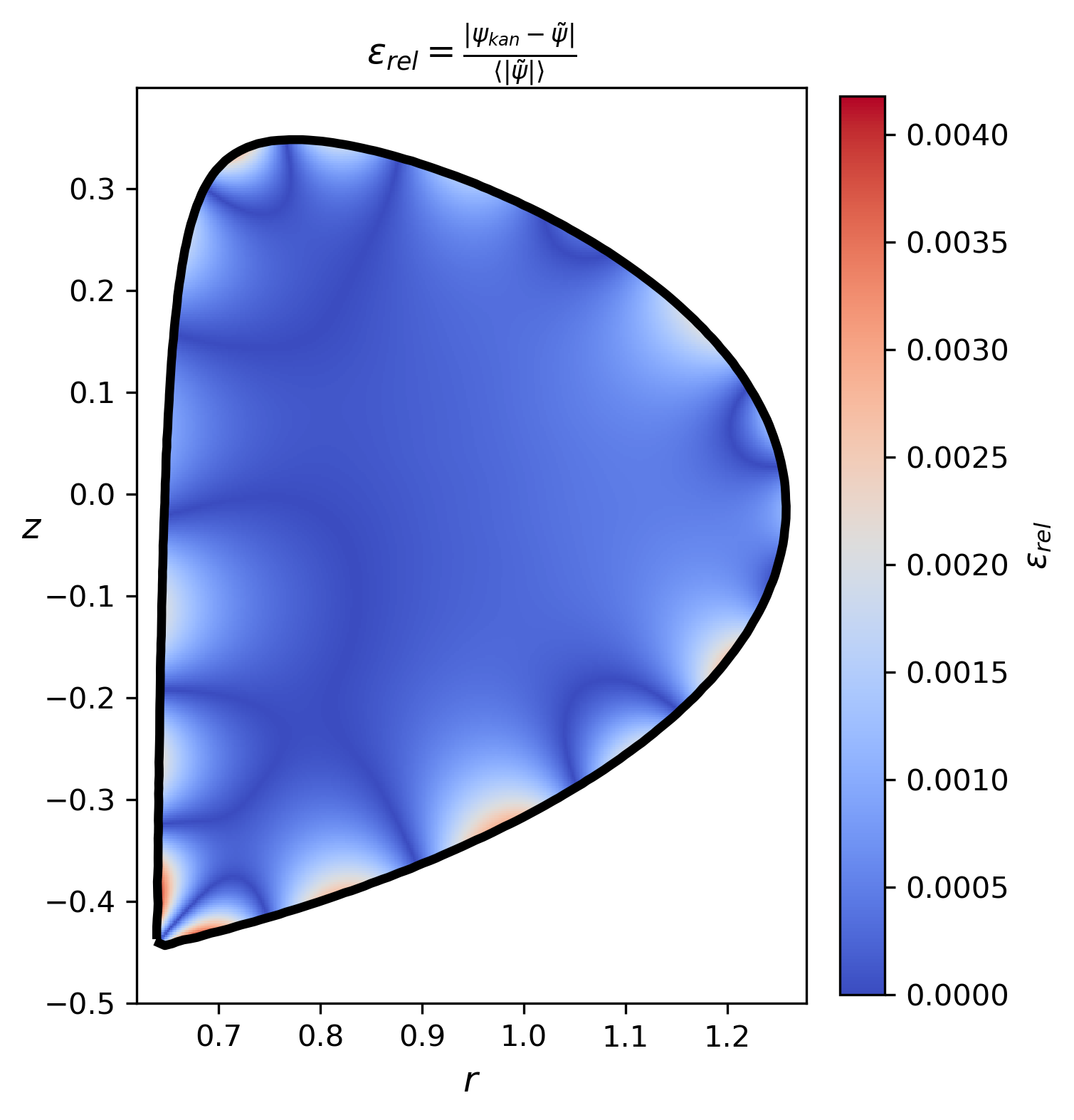}%
	\caption{(Left) Relative error profile along the midplane ($z = 0$). (Right) 2D spatial distribution of the relative error $\varepsilon_{\text{rel}}$ across the poloidal plane.}
	\label{fig:errors_bench}
\end{figure}

In Fig.~\ref{fig:comparison_error}, we provide a direct comparison between the KAN solution and the exact analytical solution along the midplane ($z=0$), showing excellent agreement. To perform a more rigorous benchmark, Fig.~\ref{fig:errors_bench} presents both 1D and 2D spatial distributions of the relative error between the two solutions, defined as:
\begin{equation}
	\varepsilon_{\text{rel}} = \frac{|\psi_{\text{\textsc{kan}}} - \tilde{\psi}|}{\langle |\tilde{\psi}| \rangle}\,,
\end{equation}
where $\tilde{\psi}$ denotes the exact analytical solution and $\langle \cdot \rangle$ represents the spatial mean value computed over the entire domain.

The relative error remains low ($\varepsilon_{\text{rel}} < 10^{-3}$) throughout the central region. Toward the plasma boundary, faint fringe-like structures emerge, leading to a slight increase in error ($\sim 10^{-3}$). The relative error attains its larger values near the boundary and the X-point. We conjecture that these fringe-like patterns mirror higher-order eigenmodes of the Shafranov operator ($\Delta^*$), strongly influenced by the prescribed boundary geometry. A detailed spectral decomposition to rigorously check this hypothesis, alongside targeted  mitigation strategies, remains an active direction for future work.

\subsection{Equilibrium with typical L-mode nonlinear profiles}
Let us  now investigate the performance of the lightweight KAN network in approximating equilibria constructed by solving the GS equation with non-linear profiles given by \eqref{freegs_ansatz} with $a_n \neq 0$ and $b_n \neq 0$. First, we examine the capability of the PIKAN algorithm for simultaneous network training and parameter discovery. Through several numerical experiments, we found that among the candidate architectures employed in the previous Solov'ev case, namely FastKAN, MLP$_1$, and MLP$_2$, only the FastKAN network was able to train successfully. The computational domain as well as the target $\beta_t$ and $I_t$ values are identical to the previous case (see Table~\ref{tab:geometric_target_parameters}). The profile parameters $P_1$ and $H_1$ are trainable (initialized as $P_1^{(0)} = 0.1$ and $H_1^{(0)} = 0.01$), while the profile shape parameters take the values $a_m = 0.8$, $a_n = 1.2$, $b_m = 1.2$, and $b_n = 0.8$. In this training setting, only the homotopy-based method combined with the FastKAN architecture achieved convergence. The network architectures, training strategies, and hyperparameter tuning are outlined in Table~\ref{tab:architectures_hyperparameters}.  As in the Solov'ev case, training was conducted in two phases using the same schedule. Architectures and training strategies that failed to converge were also tested across various learning rates; however, they yielded similar behavior and are omitted here for the sake of brevity.
\begin{figure}[!htb]
	\centering
	\includegraphics[scale=0.6]{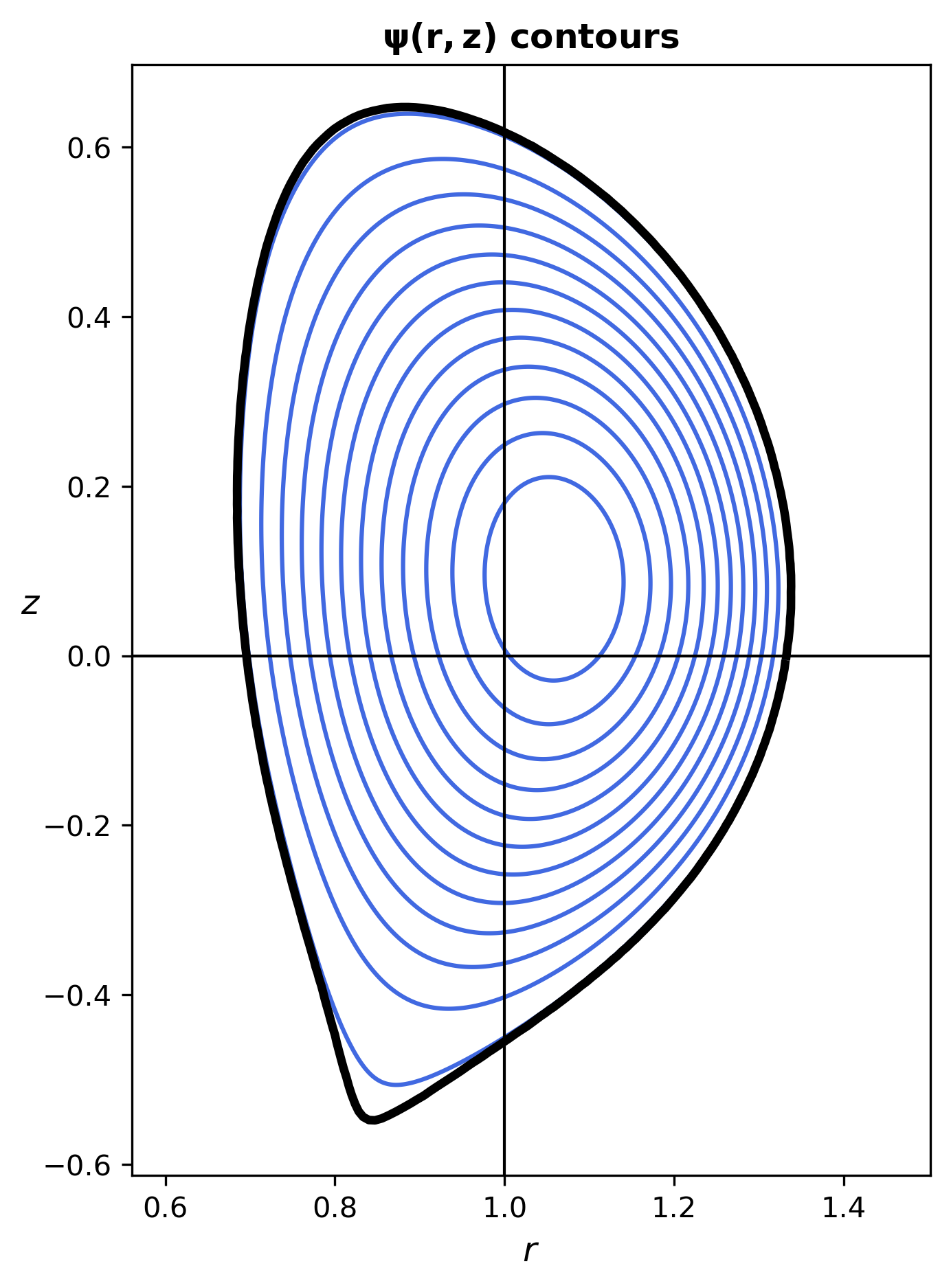}
		\includegraphics[scale=0.6]{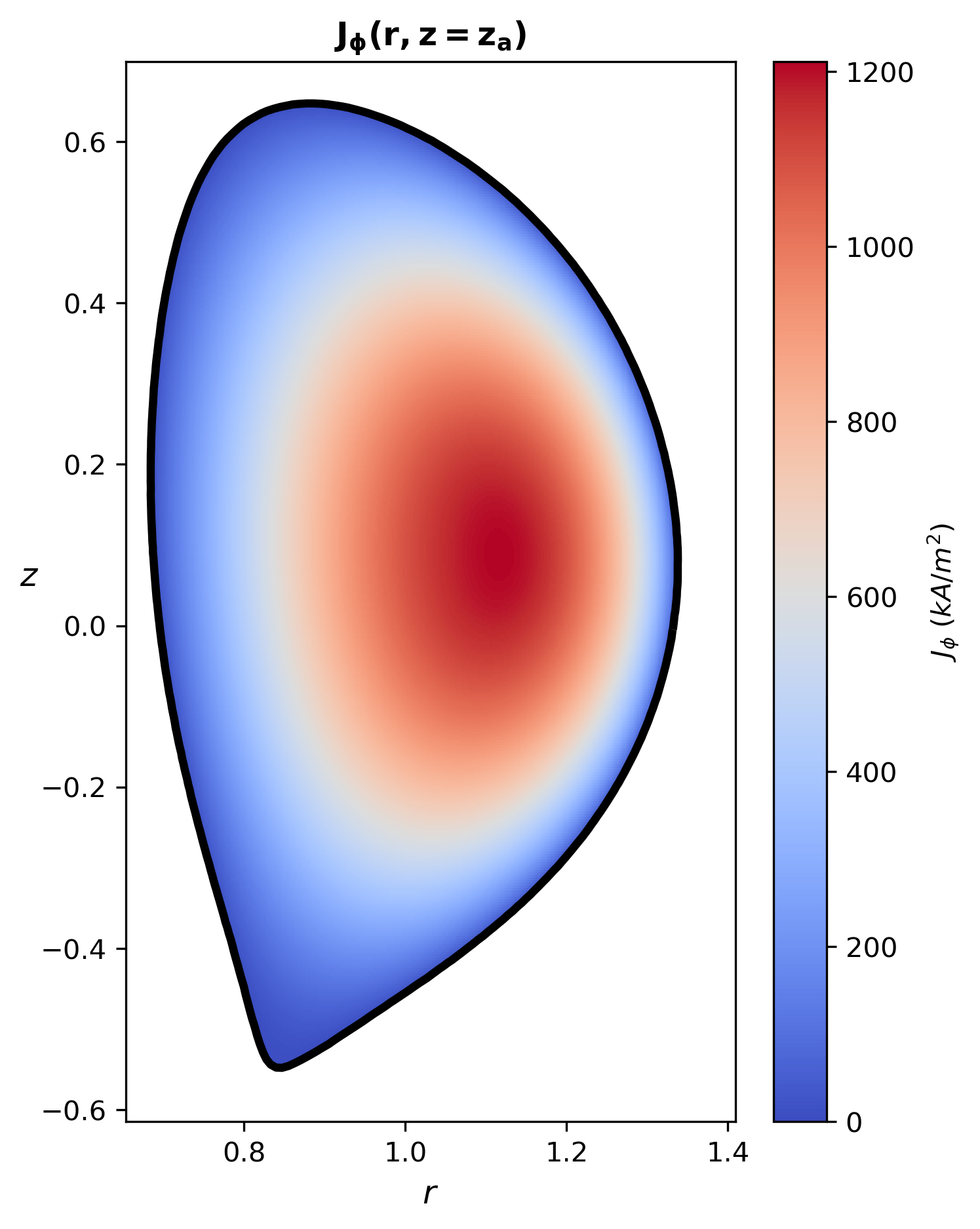}
	\caption{ (Left) Poloidal cross-section of the magnetic flux surfaces for the KAN equilibrium with nonlinear profiles associated with the ansatz \ref{freegs_ansatz}. (Right) The corresponding 2D toroidal current density distribution across the poloidal cross section.} \label{fig:freegs_contours_and_Jt}
\end{figure}
\begin{figure}[!htb]
	\centering
	\includegraphics[scale=0.5]{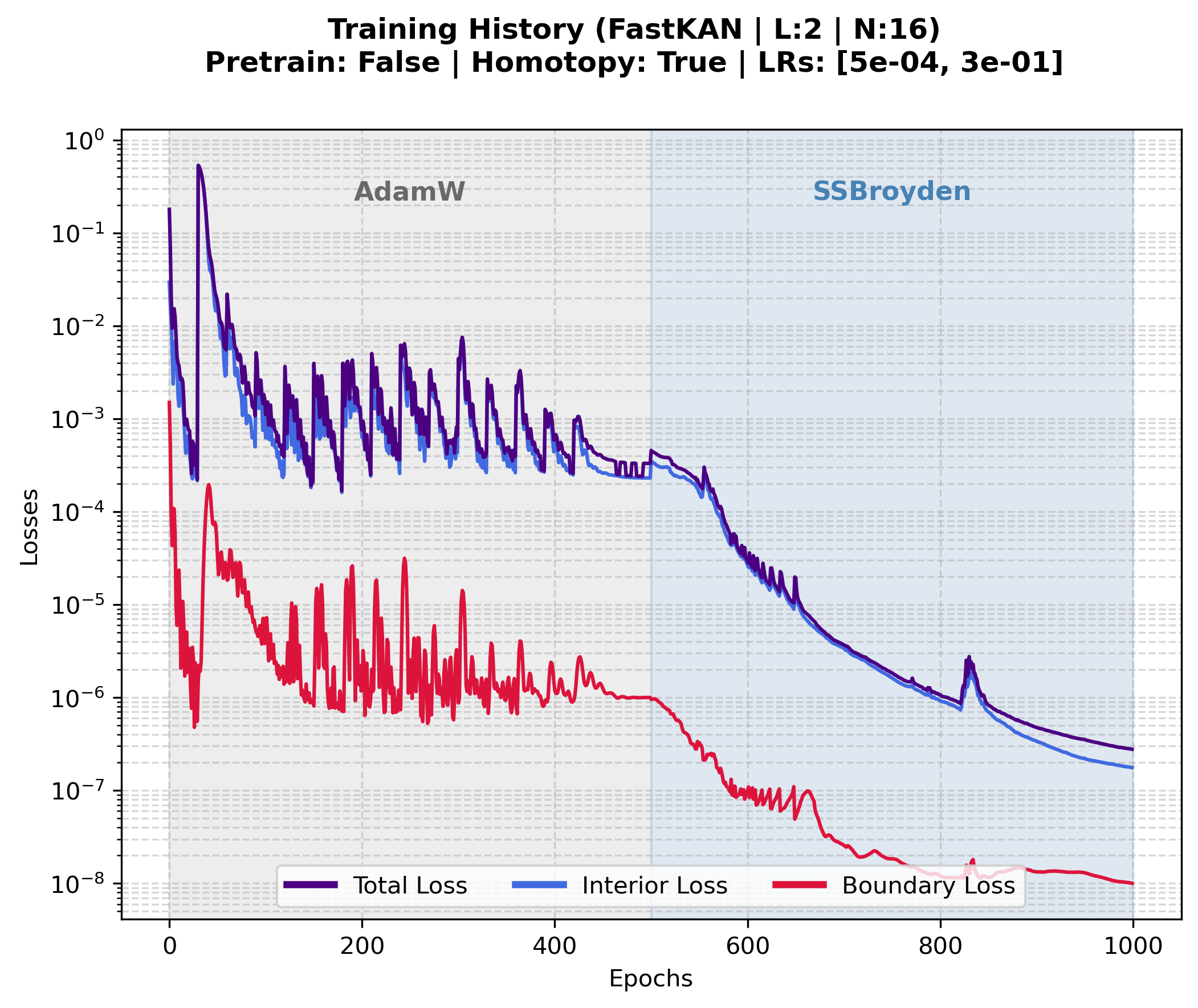}
 
	\caption{Training history for the typical L-mode nonlinear KAN equilibrium with profile parameter discovery during the first phase.}\label{fig:freegs_train_hist}
\end{figure}
\begin{figure}[!htb]
	\centering
	\includegraphics[scale=0.4]{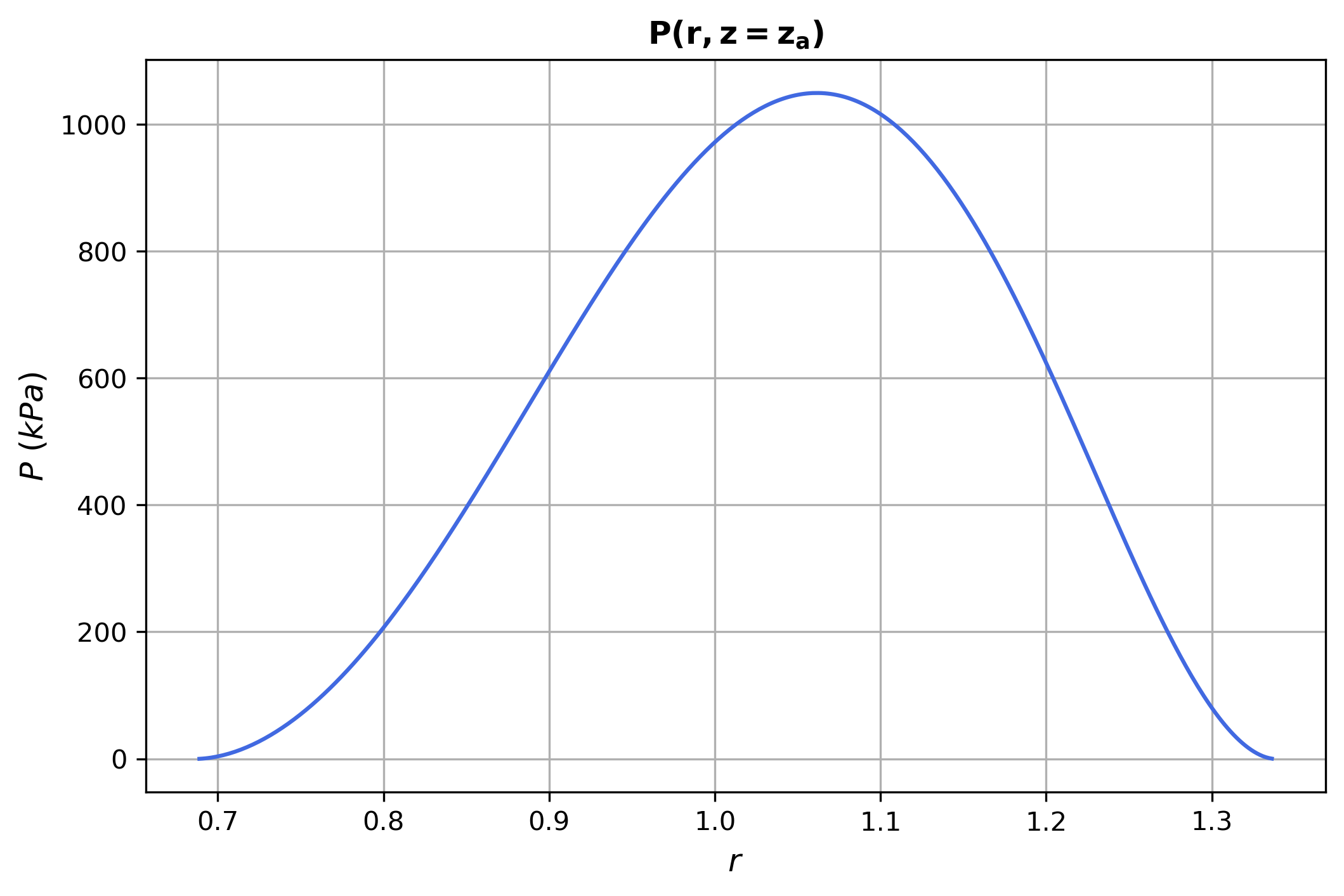 }	\includegraphics[scale=0.4]{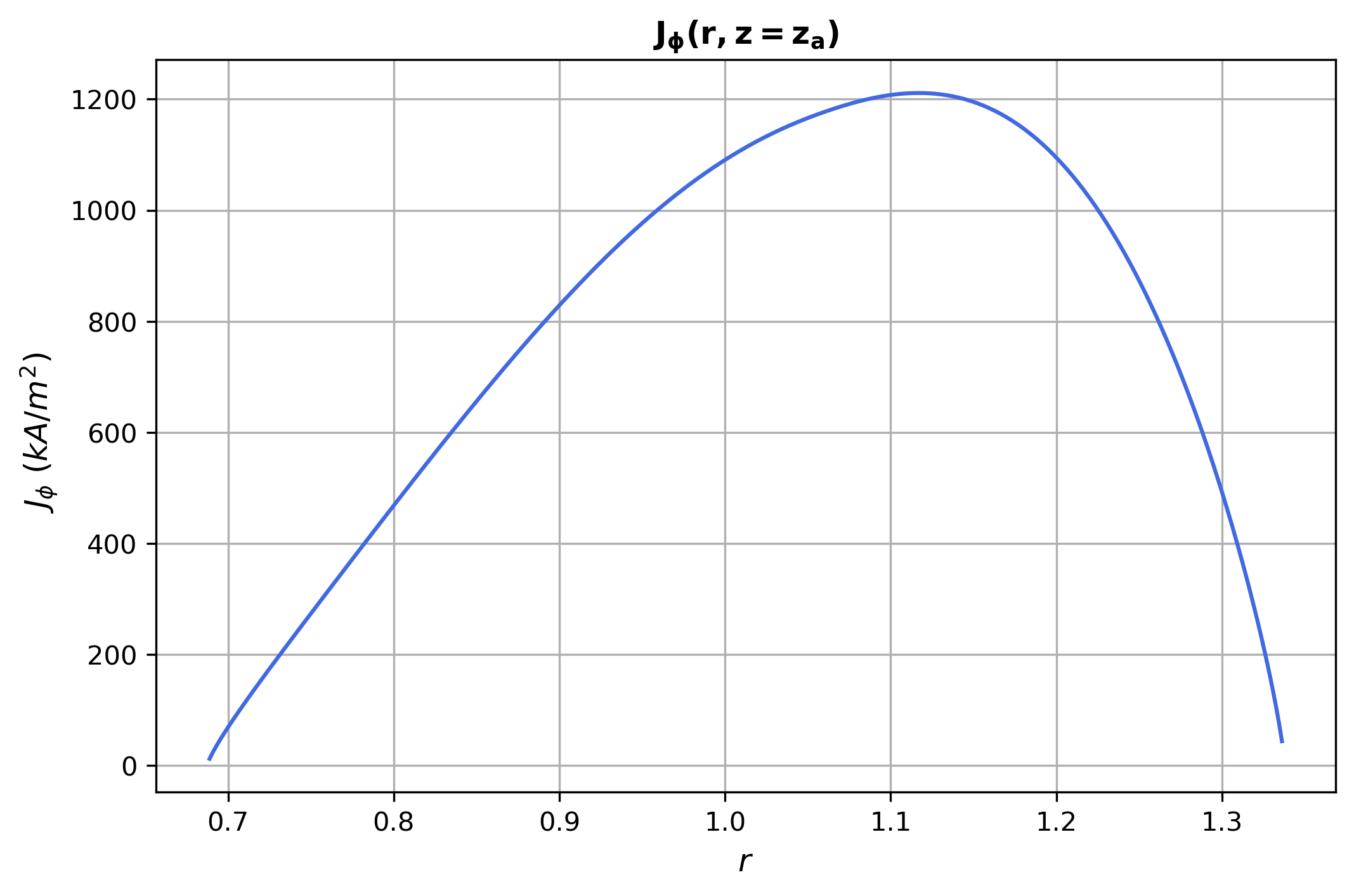 }
	
	\includegraphics[scale=0.4]{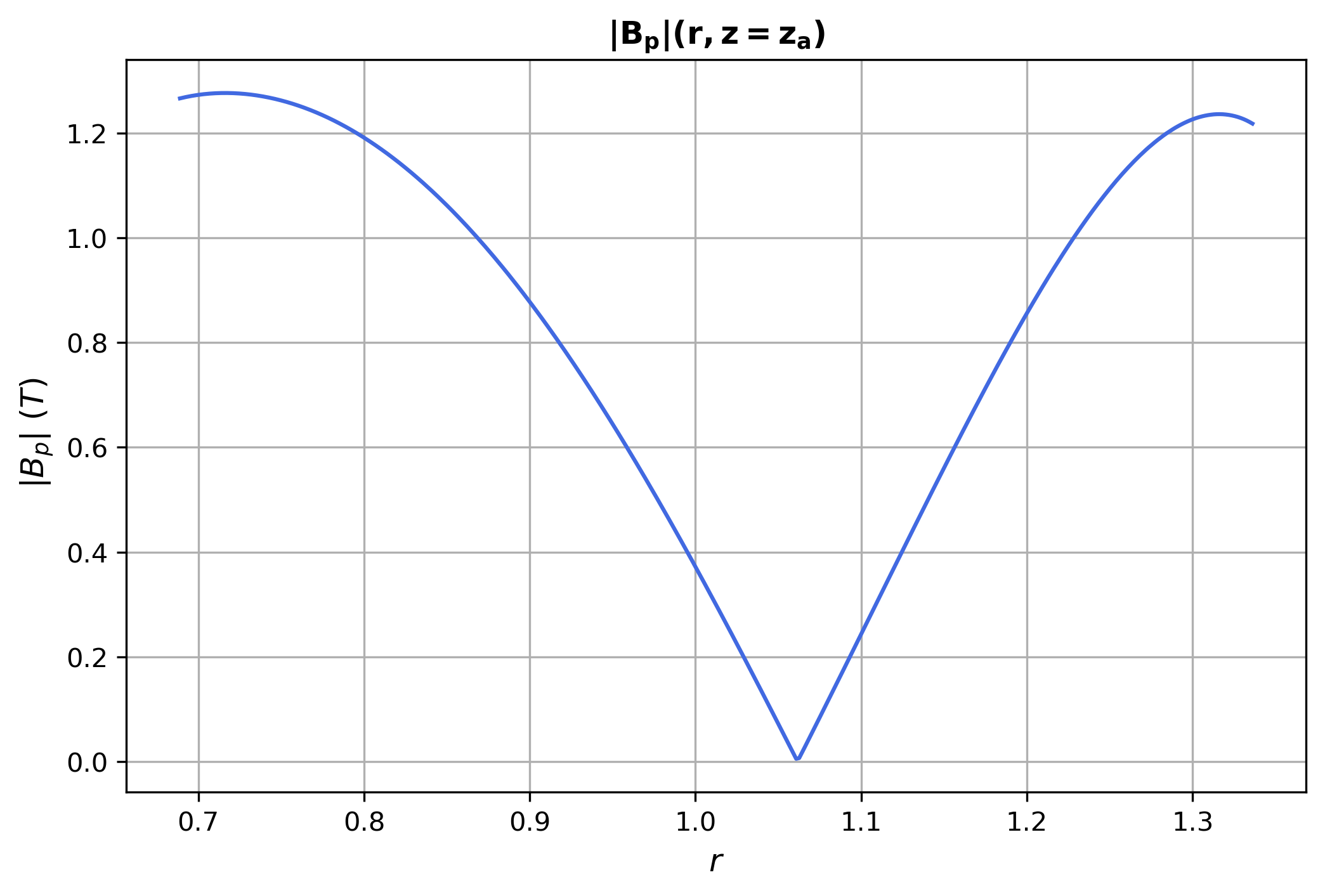}
	\includegraphics[scale=0.4]{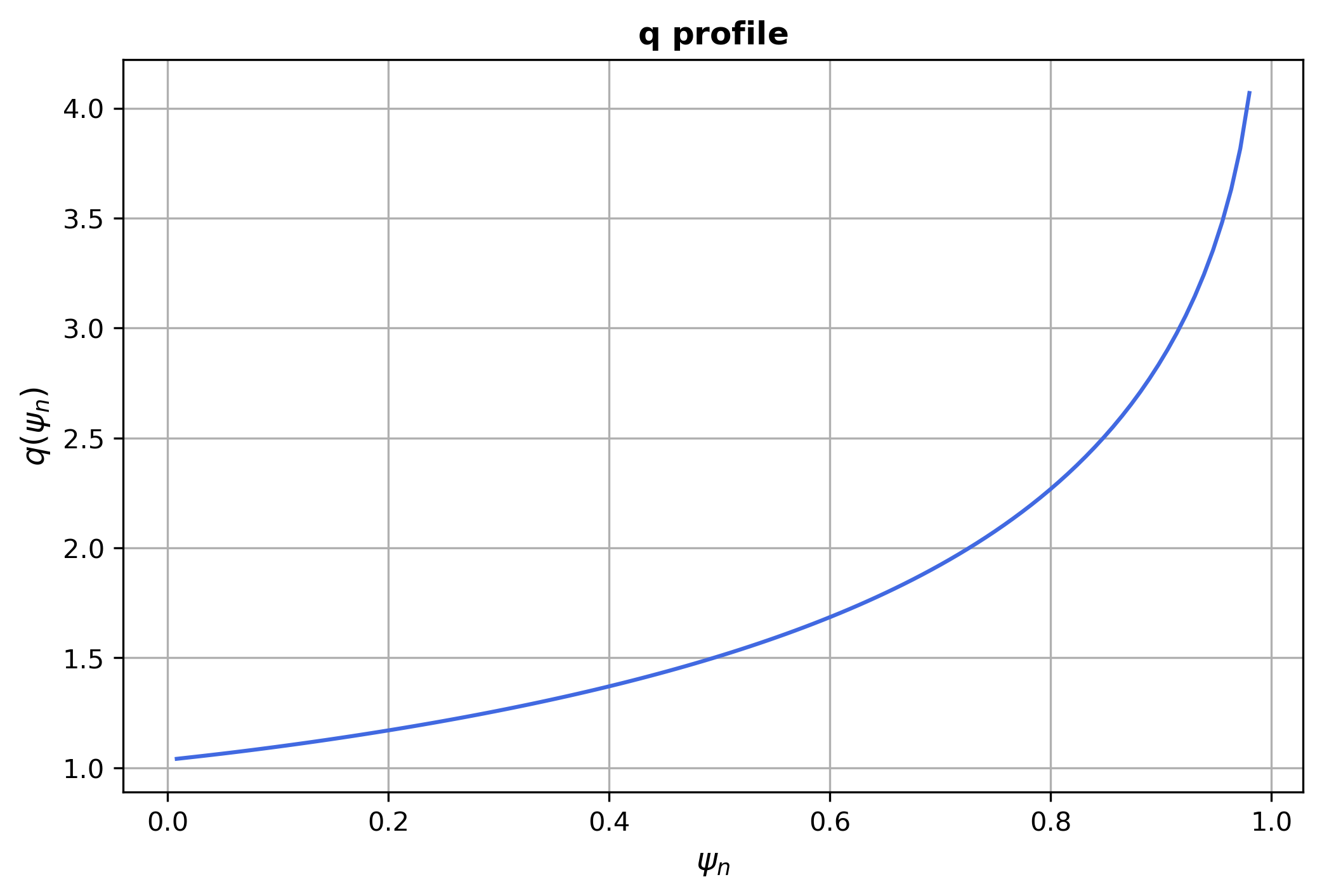}
	\caption{Profiles of the plasma pressure $P(r, z_a)$ (top-left), the toroidal current density $J_\phi(r, z_a)$ (top-right), and the modulus of the poloidal magnetic field $B_p(r, z_a)$ (bottom-left) on the midplane $z = z_a$, alongside the safety factor profile $q(\psi_n)$ as a function of the normalized poloidal flux $\psi_n$ (bottom-right) for the nonlinear equilibrium associated with the ansatz \eqref{freegs_ansatz}.}\label{fig:freegs_profiles}
\end{figure}
\begin{figure}[!htb]
	\centering
	\includegraphics[scale=0.5]{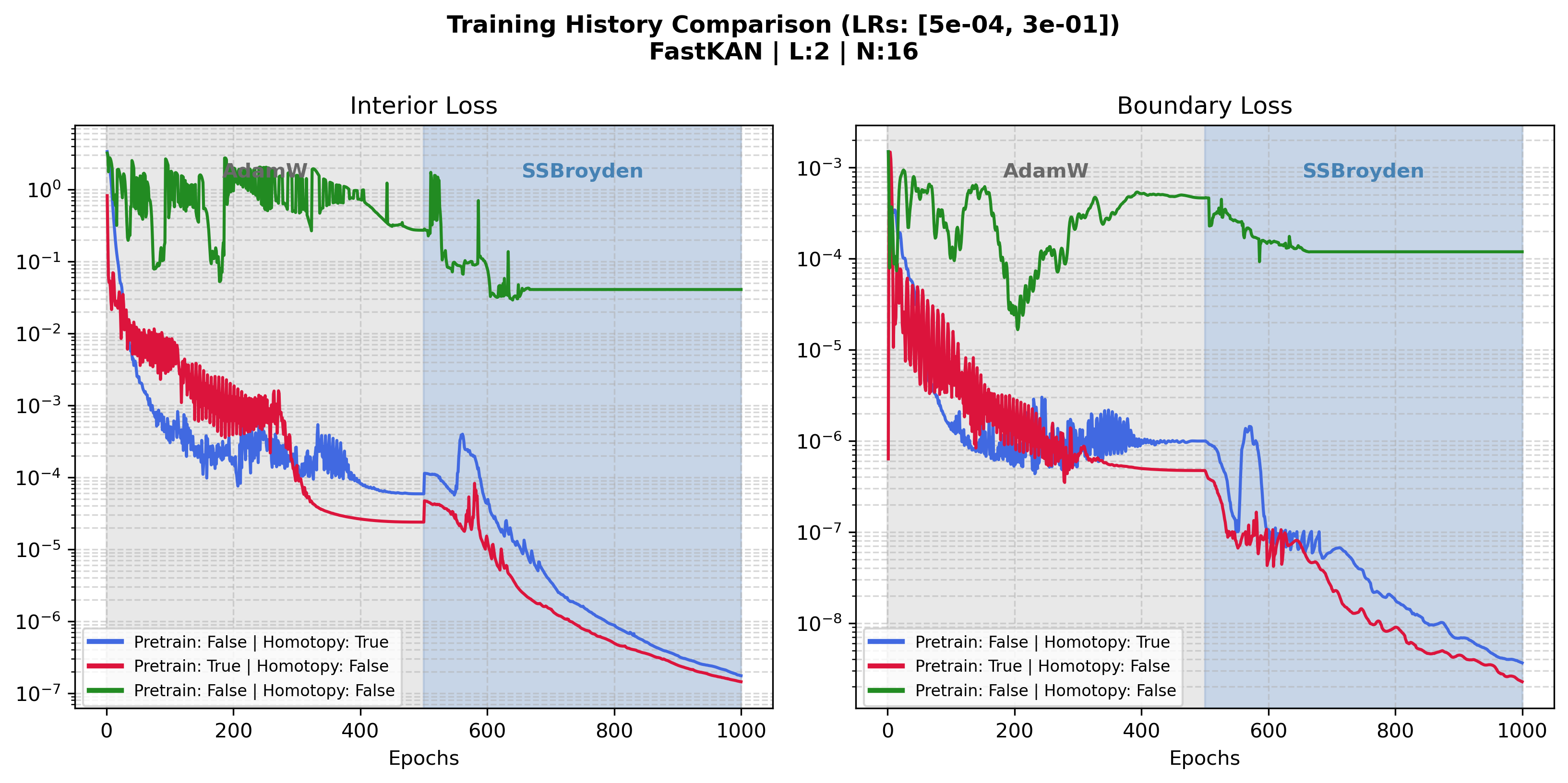}
\caption{Comparison of training loss dynamics for the typical nonlinear equilibrium under different convergence strategies:  Blue denotes the homotopy continuation method, red represents transfer learning via pretrained networks, and green corresponds to unguided optimization. (Left) interior loss term and (right) boundary loss term.}
	\label{fig:method_comparison_freegs}
\end{figure}

Figure~\ref{fig:freegs_contours_and_Jt} depicts the poloidal cross-section of the magnetic flux surfaces for the non-linear equilibrium (left panel) alongside the 2D spatial distribution of the toroidal current density given by Eq.~\eqref{J_phi} (right panel).  Figure~\ref{fig:freegs_train_hist} presents the training history of the loss terms for the scenario with discoverable parameters $(P_1, H_1)$ and Figure~\ref{fig:freegs_profiles} depicts the profiles of several equilibrium quantities on the plane $z=z_a$ as in the Solov'ev cases. Table~\ref{tab:freegs_comparison} summarizes key characteristics of the converged equilibrium, namely the reconstructed profile parameter values, the achieved toroidal beta, the total toroidal plasma current, the safety factor on axis, the position of the magnetic axis, the minimum values of the loss terms, and the GPU runtime. 

Note that for the complete determination of the equilibrium quantities we need to reconstruct the profile functions $P(\psi)$ and $H(\psi)$. Direct integration of \eqref{freegs_ansatz} yields:
\begin{align}
	P(\psi) = P_0 - (P_1\psi_a\psi_n ) \,{{}_2F_1\left(\frac{1}{b_m},-b_n,1+\frac{1}{b_m},\psi_n^{b_m}\right)}\,, \label{P_freegs}\\
	H(\psi) = H_0 -  (H_1\psi_a \psi_n ) \,{{}_2F_1\left(\frac{1}{a_m},-a_n,1+\frac{1}{a_m},\psi_n^{a_m}\right)}\,, \label{H_freegs}
\end{align}
where ${}_2F_1$ denotes the ordinary hypergeometric function and 
\begin{align}
	P_0 = P_1\psi_a\frac{\Gamma(1+b_n)\Gamma\left(1+\frac{1}{b_m}\right)}{\Gamma\left(1+b_n +\frac{1}{b_m}\right)}\,,\\
	H_0 =\frac{1}{2}+H_1 \psi_a \frac{\Gamma\left(1 + \frac{1}{a_m}\right) \Gamma(1 + a_n)}{\Gamma\left(1 + \frac{1}{a_m} + a_n\right)} \,.
\end{align}
The integration constants $P_0$ and $H_0$ have been determined by the boundary requirements at the plasma edge, $P(\psi = \psi_a) = 0$ and $F(\psi = \psi_a) = \sqrt{2H(\psi = \psi_a)} = 1$, respectively. The profiles in Fig.~\ref{fig:freegs_profiles} depict a typical peaked on-axis L-mode pressure profile, alongside the current density, the poloidal magnetic field modulus and the safety factor profiles. The safety factor has a typical behavior with values ranging from $q_a \approx 1$ on-axis to $q_b > 4$ at the boundary.

Then, to evaluate the performance of each training strategy, we conduct an ablation study, where  the physical profile parameters are held fixed. Under this  setting, four distinct training approaches are investigated: (i) unguided optimization, (ii) transfer learning using a Solov'ev pretrained model, (iii) homotopy-based continuation, (iv) MLPs with homotopy continuation.

The unguided approach and the MLP architectures fail to achieve convergence entirely, underscoring that specialized training procedures are necessary when handling nonlinear equilibrium profiles. Conversely, both transfer learning and homotopy continuation successfully enable training convergence, yielding similar results (Fig.~\ref{fig:method_comparison_freegs}), with the transfer learning method being slightly more effective in minimizing the loss terms during the second SSBroyden optimization phase. However, the homotopy continuation method retains the advantage that it does not require stored pretrained networks to be applied.

We therefore conclude that the accurate computation of nonlinear equilibria via neural networks is enabled by the combination of three key elements: the utilization of KAN architectures, homotopy-based curriculum learning, and quasi-Newton second-order optimization, yielding grid-independent, continuous, and differentiable surrogate equilibrium models.

\begin{table}[!htb]
	\centering
\footnotesize
\setlength{\tabcolsep}{3pt}
	\begin{tabular}{l c c c c c c c c}
		\toprule
				\multicolumn{9}{c}{\textit{Case A: Discoverable Profile Parameters}} \\
		\midrule
		\makecell{\textbf{Model} \\ \textbf{Method}} & $(P_1, H_1)$ & $u_a$ & $\beta_t$ & $I_t \; [10^7\text{ A}]$ & $q_a$ & $(r_a, z_a)$ & \makecell{\textbf{Losses} \\ $\mathcal{L}_{\mathcal{D}} / \mathcal{L}_{\partial \mathcal{D}}$} & \textbf{Runtime} \\
		\midrule

		\makecell{\textbf{KAN} \\ \textbf{Homotopy}}   & $(1.660, 0.079)$ & $0.053$ & $0.029$ & $1.173$ & $1.037$ & $(1.062, 0.091)$ & \makecell{$1.8 \times 10^{-7}$ \\ $1.0 \times 10^{-8}$} & 02m 09s \\
		\midrule
		\multicolumn{9}{c}{\textit{Case B: Frozen Profile Parameters}} \\
		\midrule
				\makecell{\textbf{Model} \\ \textbf{Method}} & $(P_1, H_1)$ & $u_a$ & $\beta_t$ & $I_t \; [10^7\text{ A}]$ & $q_a$ & $(r_a, z_a)$ & \makecell{\textbf{Losses} \\ $\mathcal{L}_{\mathcal{D}} / \mathcal{L}_{\partial \mathcal{D}}$} & \textbf{Runtime} \\
		\midrule
		\makecell{\textbf{KAN} \\ \textbf{Homotopy}}   & $(1.660, 0.079)$ & $0.053$ & $0.029$ & $1.174$ & $1.037$ & $(1.062, 0.091)$ & \makecell{$1.7 \times 10^{-7}$ \\ $3.7 \times 10^{-9}$} & 02m 13s \\ \addlinespace
		\makecell{\textbf{KAN} \\ \textbf{Pretrained}} & $(1.660, 0.079)$ & $0.053$ & $0.029$ & $1.174$ & $1.037$ & $(1.062, 0.091)$ & \makecell{$1.4 \times 10^{-7}$ \\ $2.3 \times 10^{-9}$} & 02m 10s \\
		\bottomrule
	\end{tabular}
	\caption{Equilibrium characteristics and figures of merit along with performance metrics for typical nonlinear equilibrium profiles. Case A presents the results when the profile parameters $(P_1, H_1)$ are optimized simultaneously with the network parameters to satisfy the equilibrium constraints on $\beta_t$ and $I_t$. In this setting, only the homotopy method enables convergence. Case B corresponds to frozen profile parameters, where the homotopy and transfer-learning methods demonstrate similar performance.}
	\label{tab:freegs_comparison}
\end{table}
\subsection{Equilibrium with pressure pedestal}
To construct equilibria with a pressure pedestal, we adopt a pressure profile function $P(\psi)$ of the form \eqref{pres_prof_ped}. The second free function $H(\psi)$ is selected according to \eqref{H_freegs}. As in the previous cases, the profile parameters $H_1$ and $P_1$ are determined by enforcing constraints on the total toroidal plasma current $I_t$ and toroidal beta $\beta_t$, while the remaining parameters are set to $P_2 = 0.03$, $P_3 = 45.0$, $\nu = 2 \times 10^{-5}$, $a_m = 0.8$, and $a_n = 1.2$.

The training procedure is again split into two phases utilizing AdamW and SSBroyden, with the specific hyperparameters detailed in Table~\ref{tab:architectures_hyperparameters}. The primary distinction in this scenario is that each optimization phase is extended to 1000 epochs, and network capacities are scaled up to enhance expressivity and accuracy. Specifically, the FastKAN depth is increased by adding an extra hidden layer, while the MLP sizes are expanded accordingly to maintain a comparable number of trainable parameters. Finally, we perform systematic comparisons among the homotopy-based, transfer learning via pretrained networks, and the unguided optimization baseline.
\begin{figure}[!htb]
	\centering
	\includegraphics[scale=0.6]{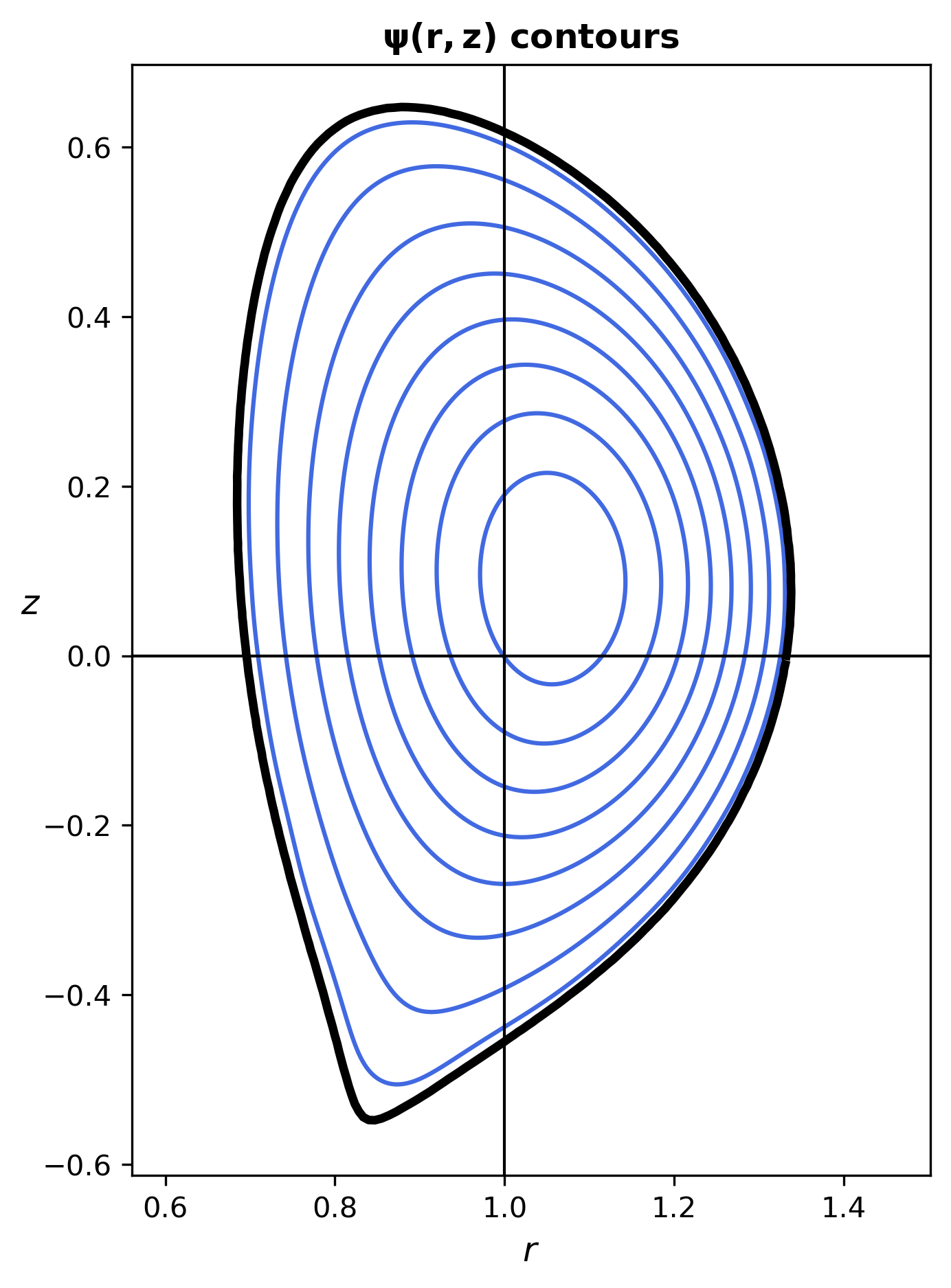}\includegraphics[scale=0.6]{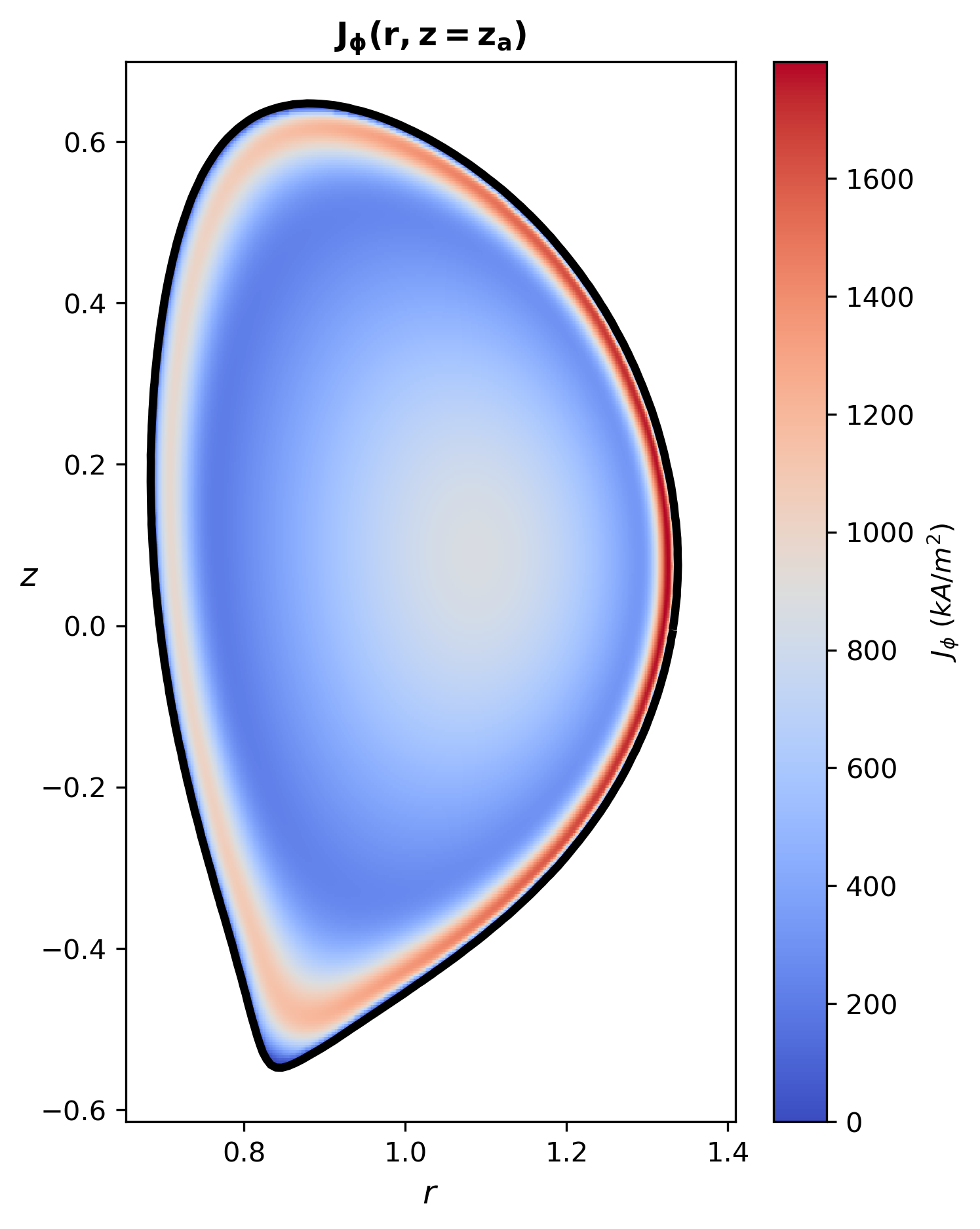}
\caption{(Left) Poloidal cross-section of the magnetic flux surfaces and (right) corresponding 2D toroidal current density distribution for the KAN equilibrium with pressure pedestal identified by the homotopy continuation method.} \label{fig:pedestal_contours_and_Jt}
\end{figure}
\begin{figure}[!htb]
	\centering
	\includegraphics[scale=0.4]{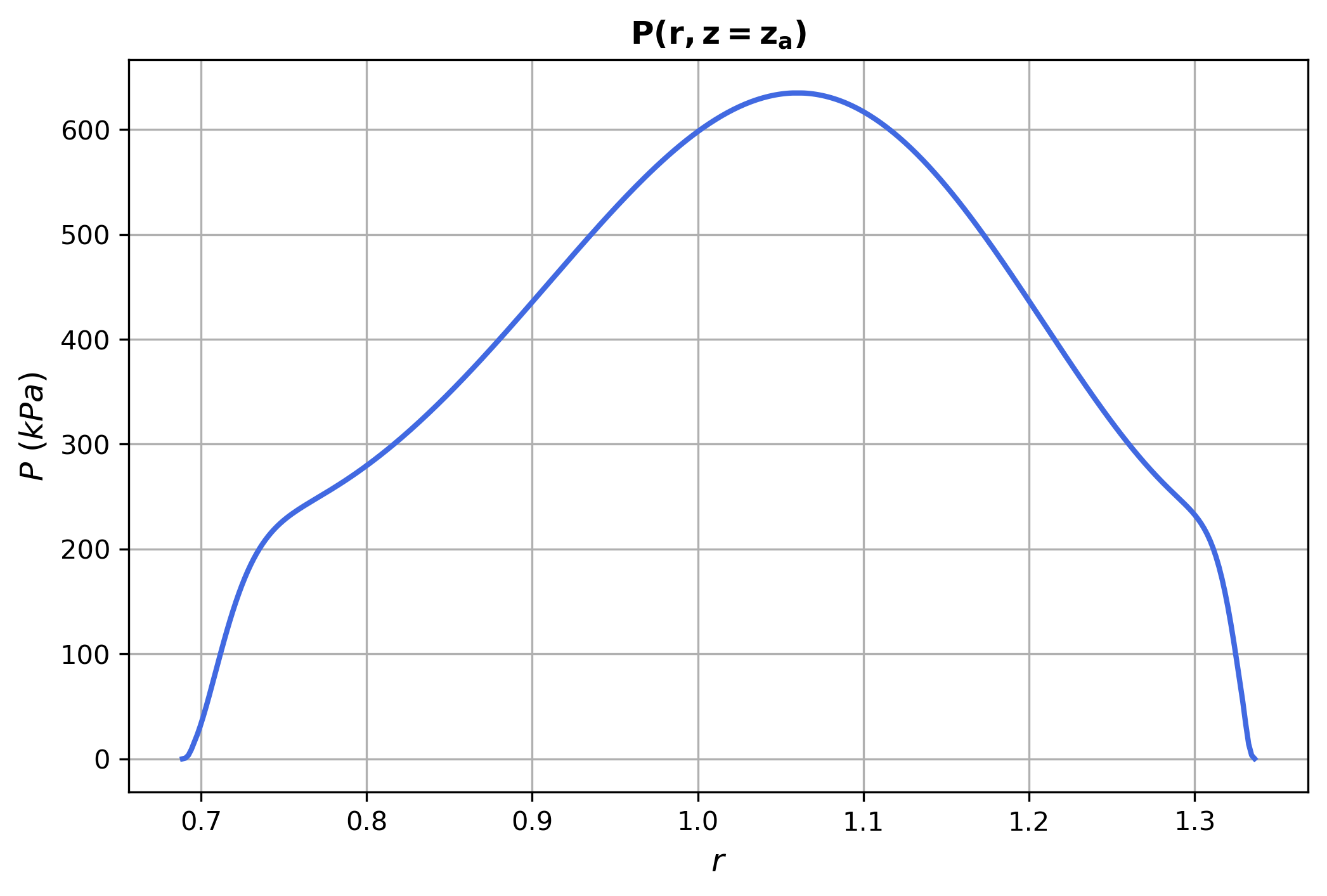}
	\includegraphics[scale=0.4]{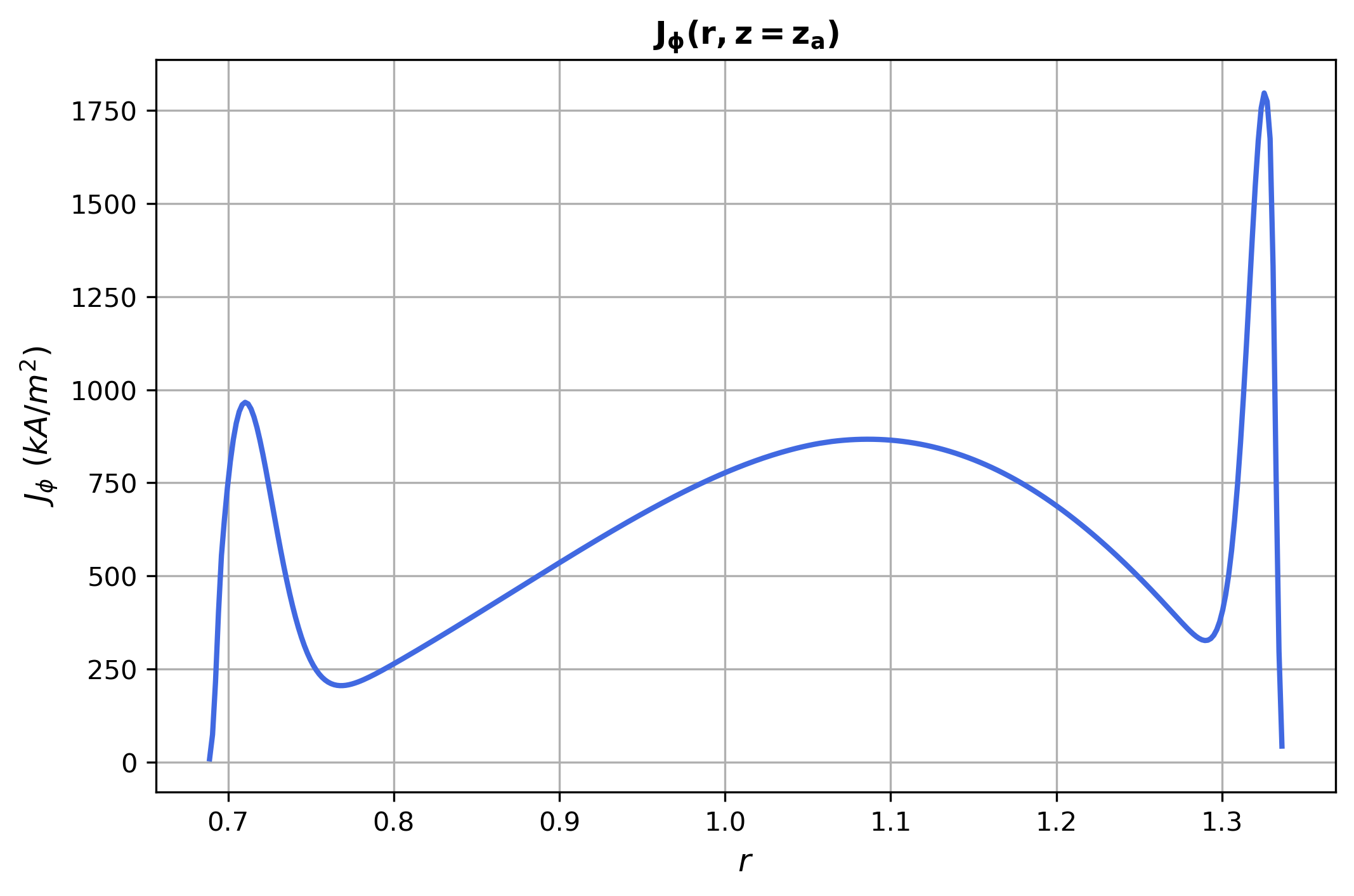}
	\includegraphics[scale=0.4]{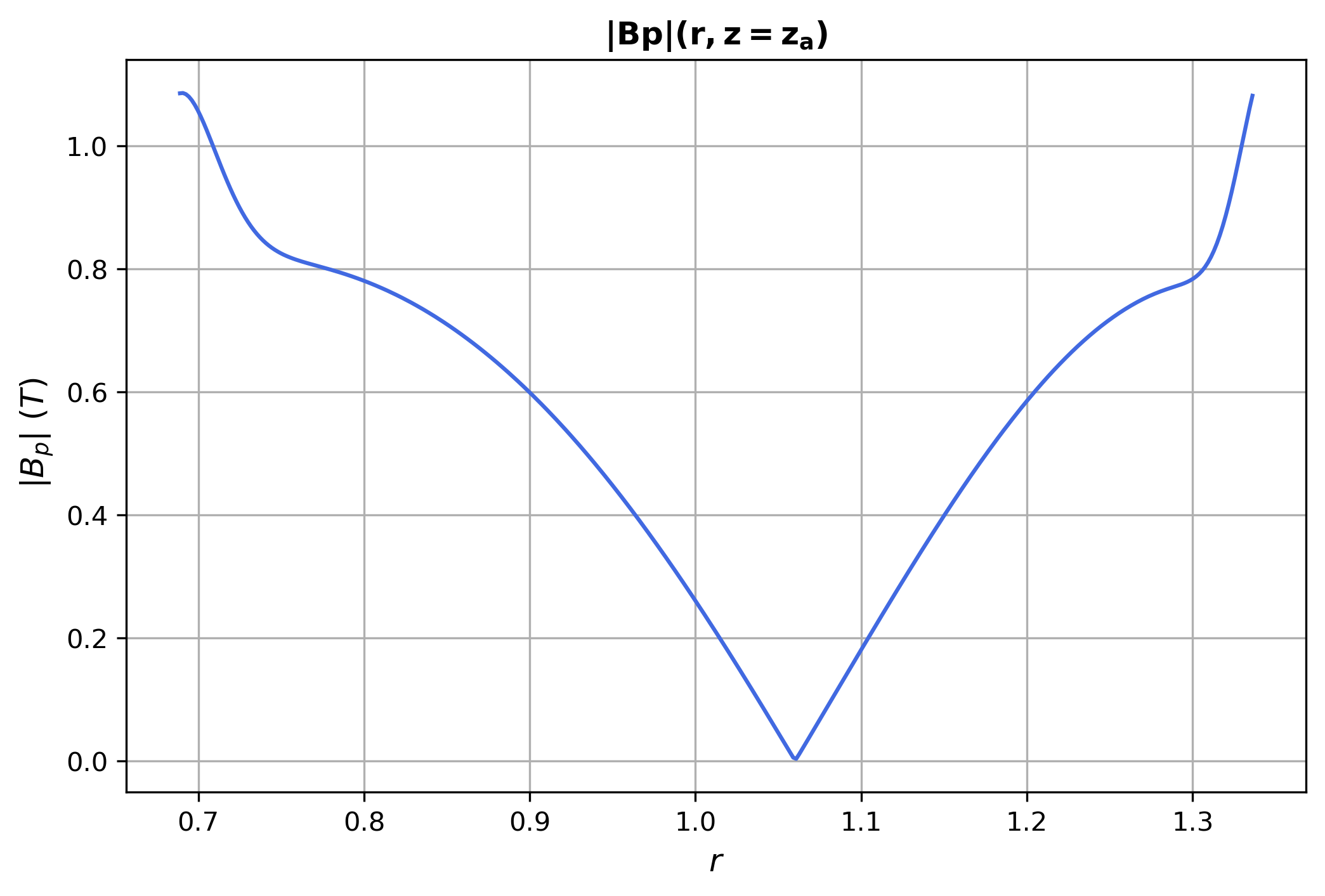}
	\includegraphics[scale=0.4]{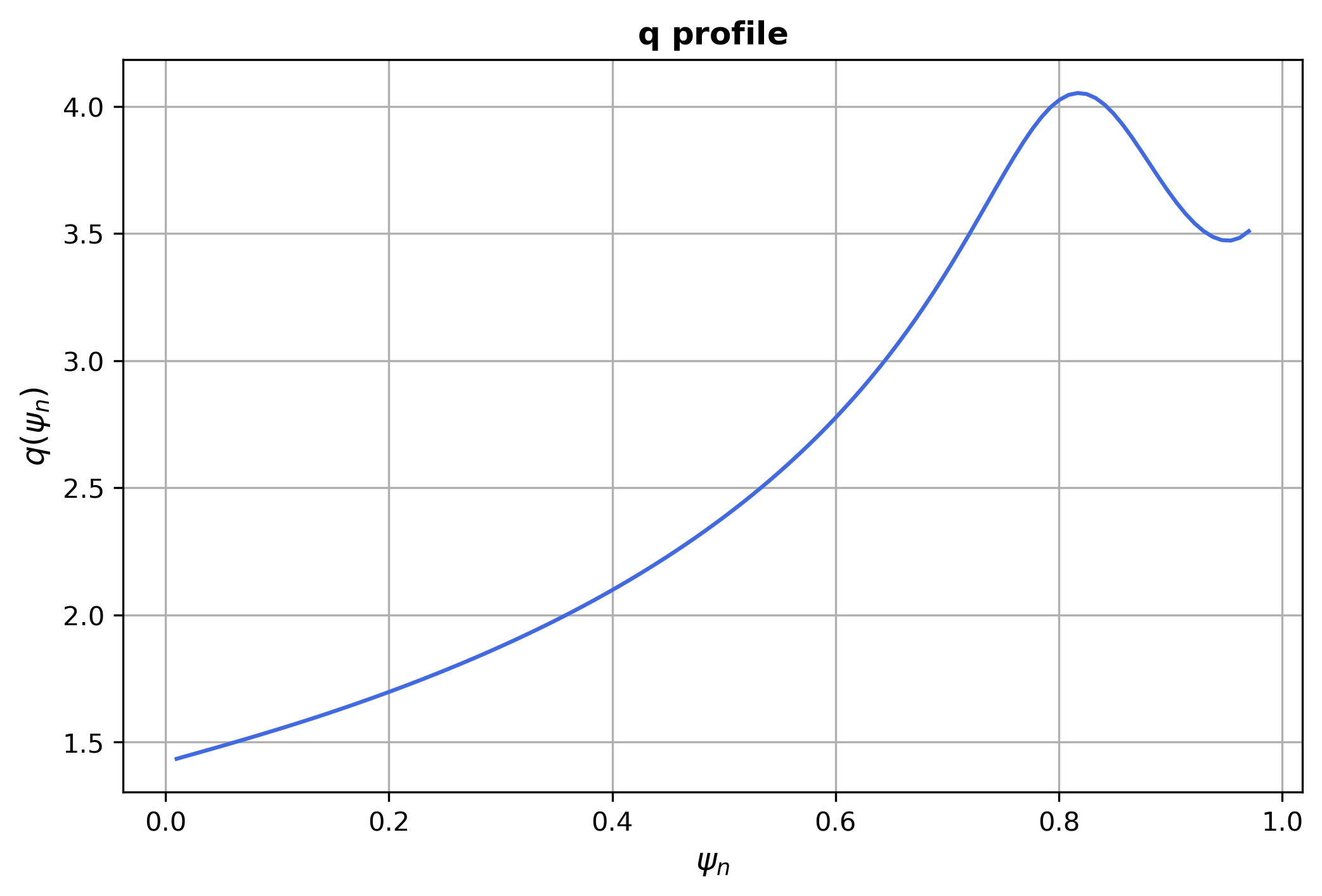}
	\caption{Midplane profiles ($z = z_a$) of the plasma pressure $P(R, z_a)$ (top-left), toroidal current density $J_\phi(R, z_a)$ (top-right), and modulus of the poloidal magnetic field $B_p(R, z_a)$ (bottom-left), alongside the safety factor profile $q(\psi_n)$ versus normalized poloidal flux $\psi_n$ (bottom-right) for the nonlinear equilibrium with pressure pedestal identified by the homotopy continuation method.} \label{fig:pedestal_profiles}
\end{figure}
\begin{figure}[!htb]
	\centering
	\includegraphics[scale=0.5]{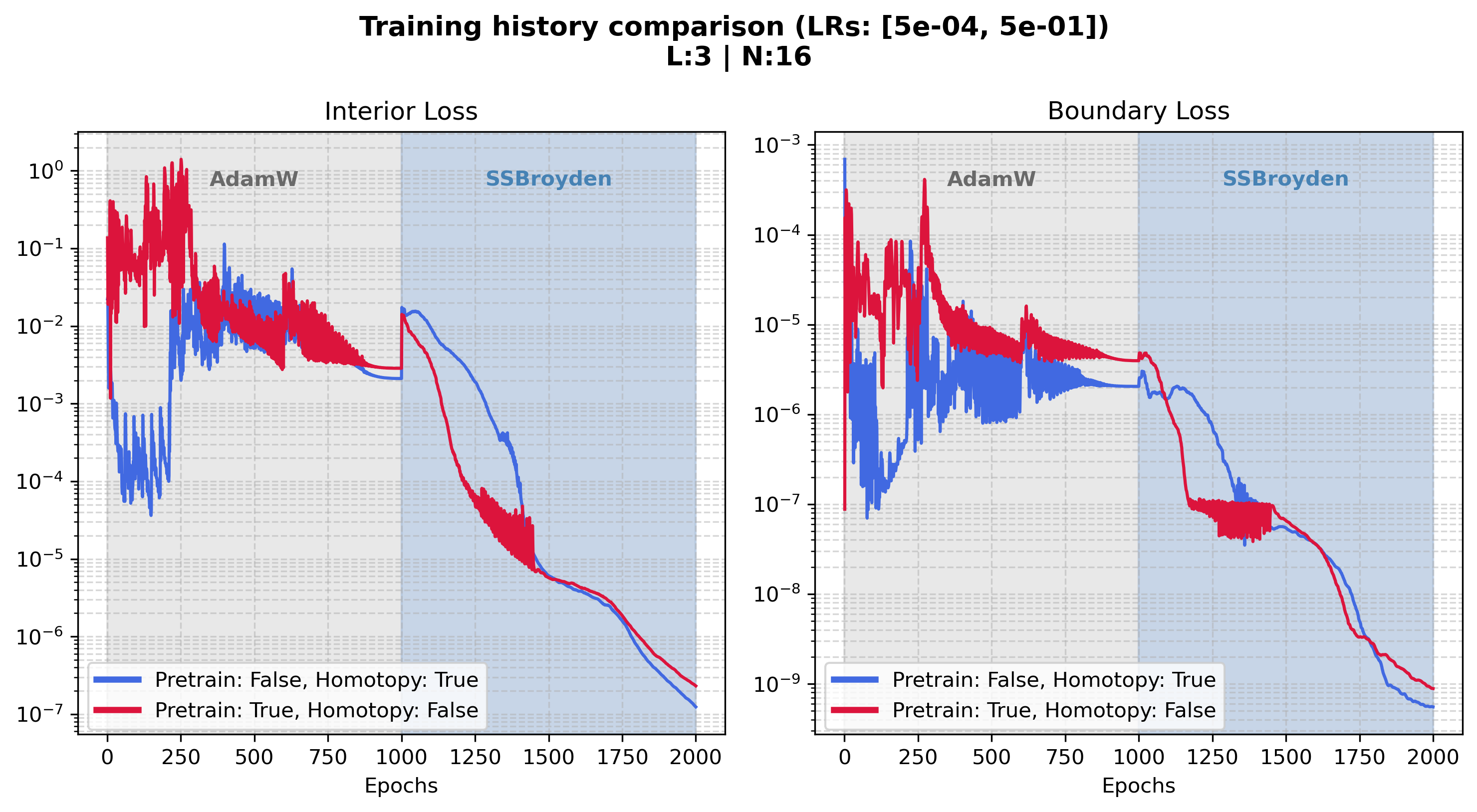}
\caption{Training history comparison between the homotopy continuation method (blue) and transfer learning via pretrained networks (red) for the pedestal equilibrium with profile parameter discovery: (left) evolution of the interior loss term and (right) boundary loss term.}
\label{fig:train_hist_pedestal_comparison_discovery}
\end{figure}
\begin{figure}[!htb]
	\centering
	\includegraphics[scale=0.45]{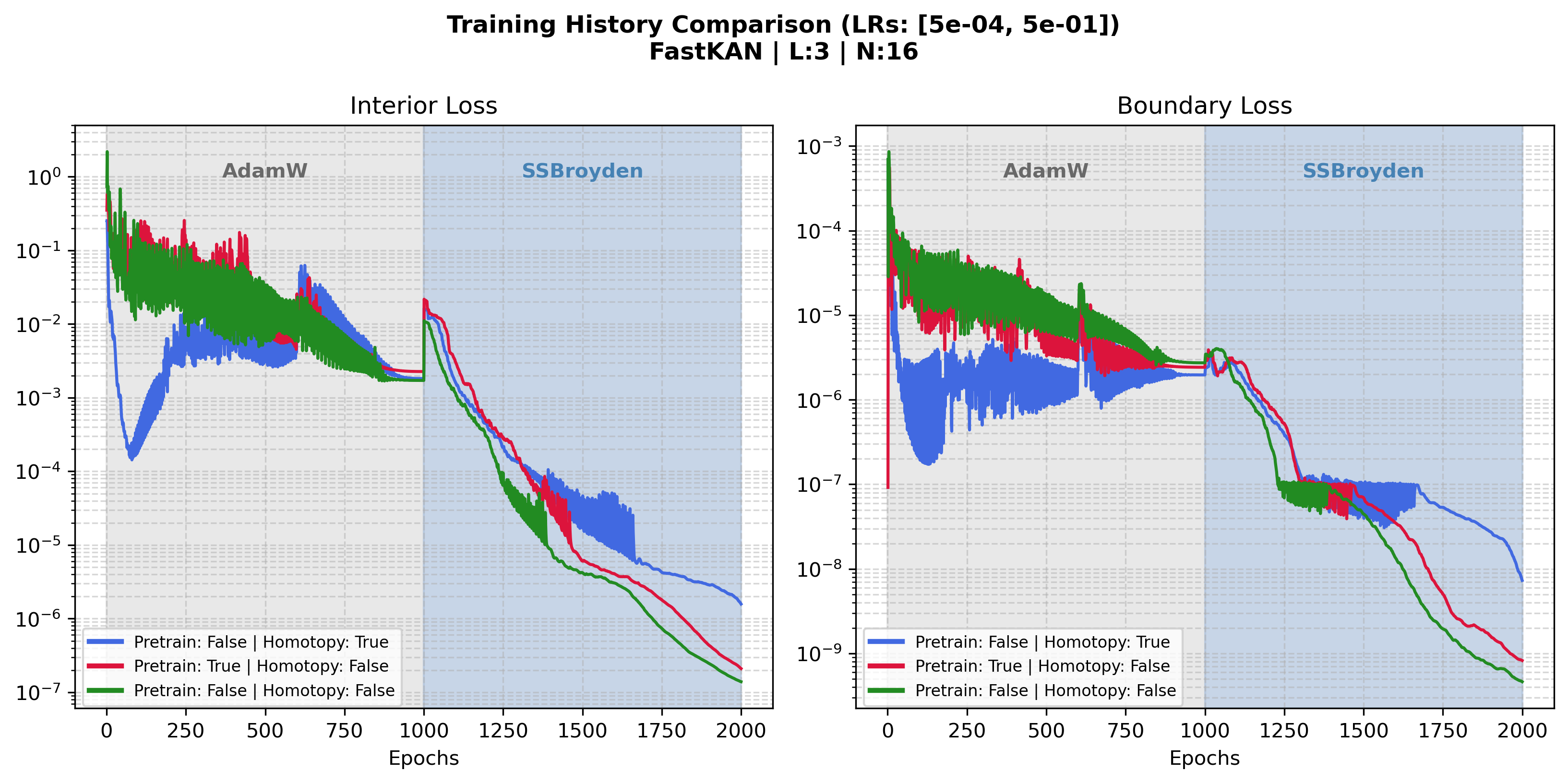}
	\caption{Training history comparison among the homotopy continuation method (blue), transfer learning via pretrained networks (red) and the unguided optimization (green) for the pedestal equilibrium with frozen profile parameters: (left) evolution of the interior loss term and (right) boundary loss term.}
	\label{fig:train_hist_pedestal_comparison_frozen}
\end{figure}
\begin{figure}[!htb]
	\centering
	\includegraphics[scale=0.6]{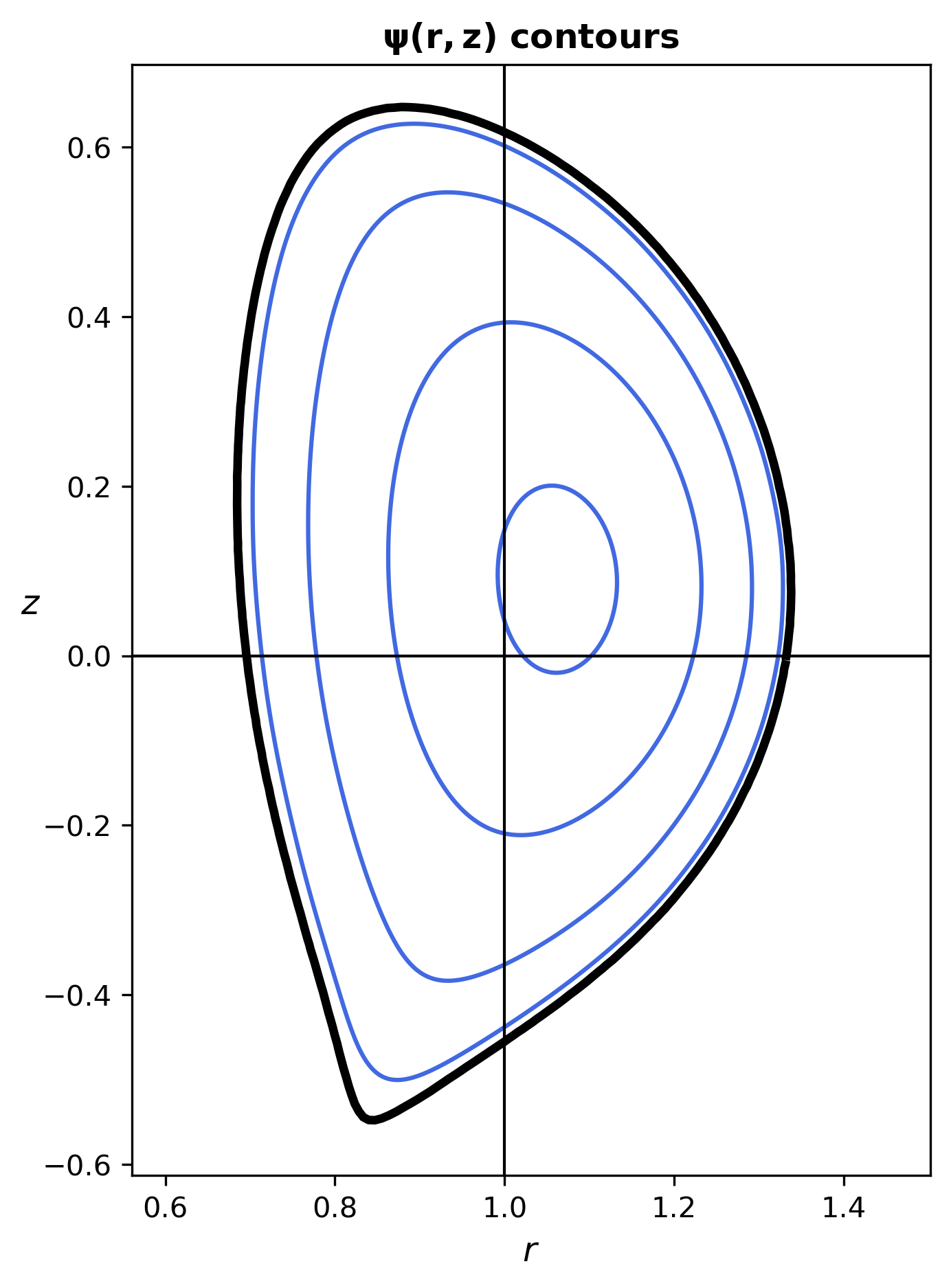}\includegraphics[scale=0.6]{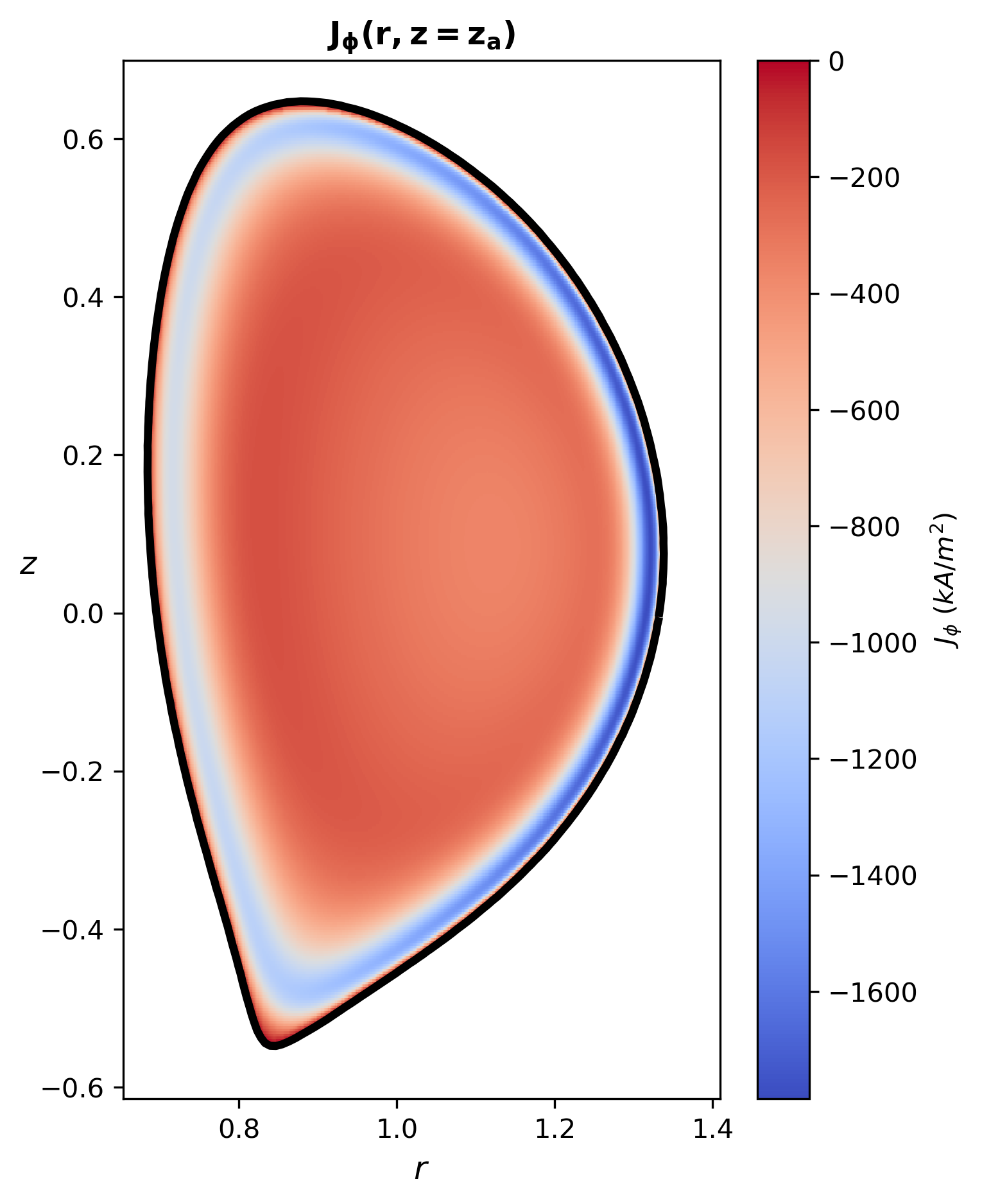}
	\caption{(Left) Poloidal cross-section of the magnetic flux surfaces and (right) corresponding 2D toroidal current density distribution for the KAN equilibrium with pressure pedestal identified by the  unguided method under frozen profile parameters.} \label{fig:pedestal_contours_and_Jt_II}
\end{figure}
\begin{figure}[!htb]
	\centering
	\includegraphics[scale=0.4]{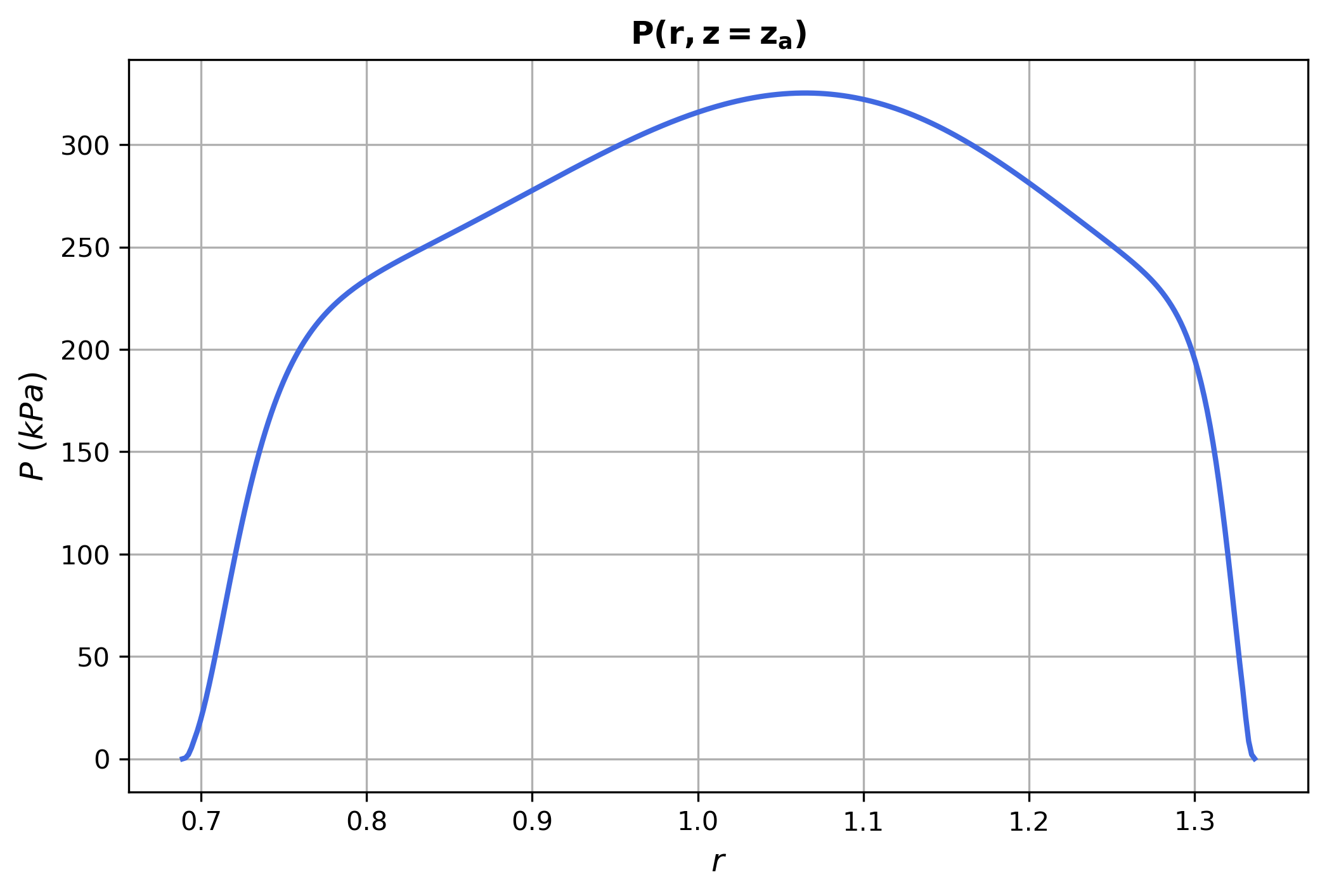}
	\includegraphics[scale=0.4]{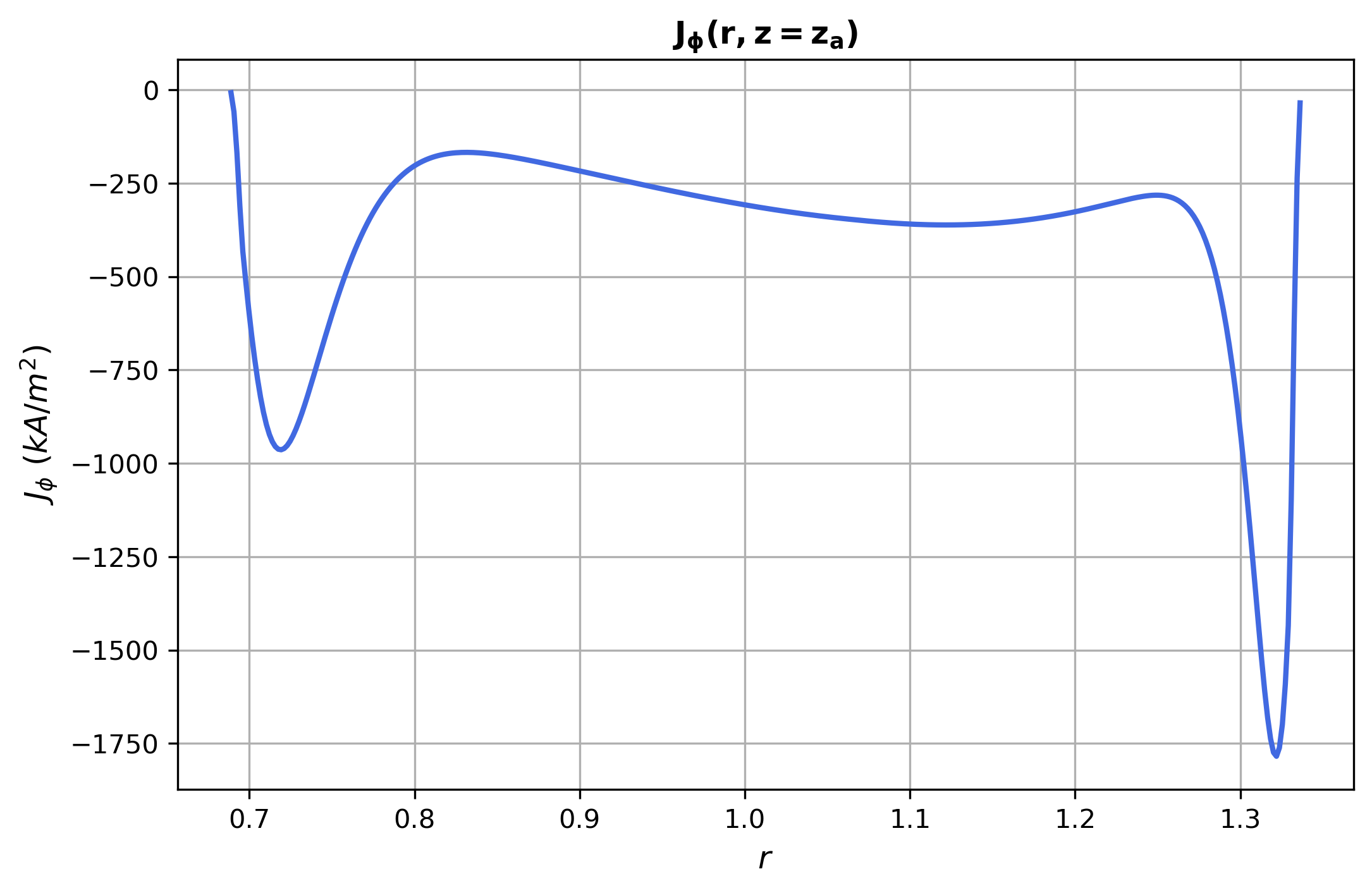}
	\includegraphics[scale=0.4]{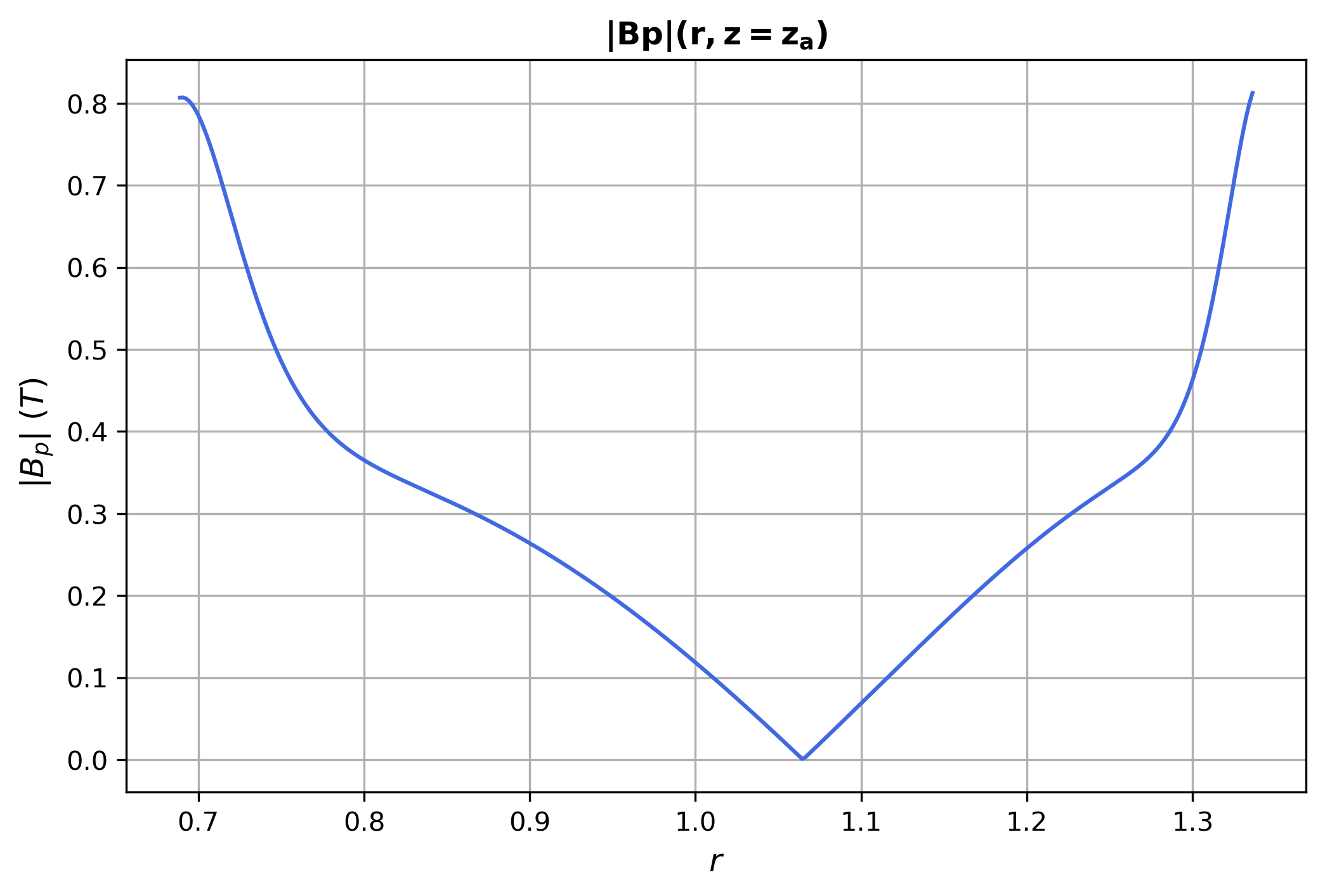}
	\includegraphics[scale=0.4]{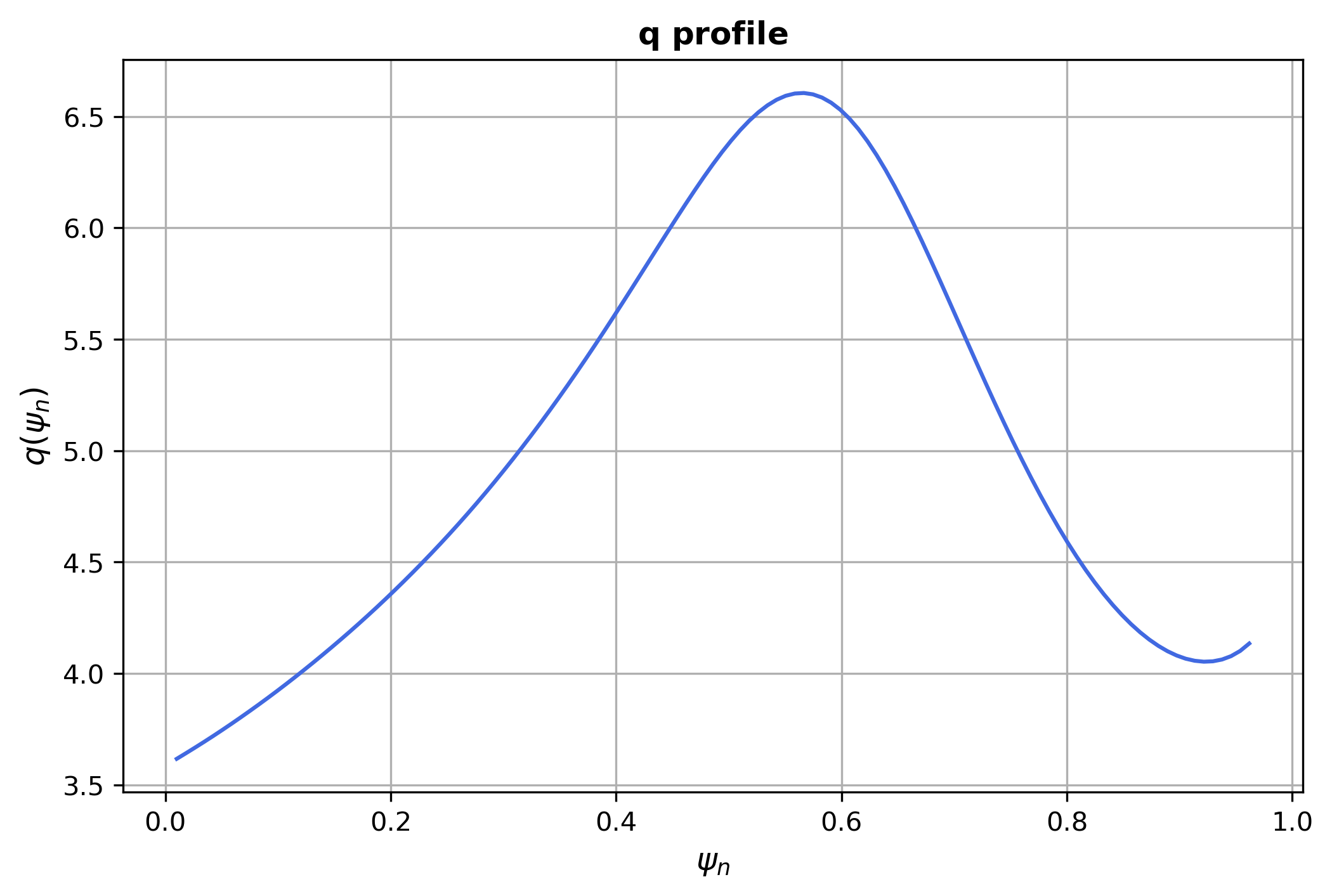}
		\caption{Midplane profiles ($z = z_a$) of the plasma pressure $P(R, z_a)$ (top-left), toroidal current density $J_\phi(R, z_a)$ (top-right), and modulus of the poloidal magnetic field $B_p(R, z_a)$ (bottom-left), alongside the safety factor profile $q(\psi_n)$ versus normalized poloidal flux $\psi_n$ (bottom-right) for the nonlinear equilibrium with pressure pedestal discovered by the unguided method.}
 \label{fig:pedestal_profiles_II}
\end{figure}
\begin{table}[!htb]
	\centering
	\footnotesize
	\setlength{\tabcolsep}{3pt} 
	\begin{tabular}{l c c c c c c c c}
		\toprule
				\multicolumn{9}{c}{\textit{Case A: Discoverable Profile Parameters}} \\
		\midrule
		\makecell{\textbf{Model} \\ \textbf{Method}} & $(P_1, H_1)$ & $u_a$ & $\beta_t$ & $I_t \; [10^7\text{ A}]$ & $q_a$ & $(r_a, z_a)$ & \makecell{\textbf{Losses} \\ $\mathcal{L}_{\mathcal{D}} / \mathcal{L}_{\partial \mathcal{D}}$} & \textbf{Runtime} \\
		\midrule

		\makecell{\textbf{KAN} \\ \textbf{Homotopy}}   & $(0.354, 0.109)$ & $0.036$  & $0.029$ & $1.274$ & $1.434$  & $(1.060, 0.091)$ & \makecell{$1.2 \times 10^{-7}$ \\ $5.5 \times 10^{-10}$} & 05m 23s \\ \addlinespace
		\makecell{\textbf{KAN} \\ \textbf{Pretrained}} & $(0.363, 0.030)$ & $0.032$  & $0.028$ & $1.235$ & $1.662$  & $(1.061, 0.091)$ & \makecell{$2.3 \times 10^{-7}$ \\ $8.9 \times 10^{-10}$} & 05m 33s \\ \addlinespace
		\makecell{\textbf{KAN} \\ \textbf{Unguided}}   & \text{---}       & \text{---} & \text{---} & \text{---} & \text{---} & \text{---}       & \textit{Did not converge} & \text{---} \\ \addlinespace
		\makecell{\textbf{MLP$_1$} \\ \textbf{Homotopy}} & $(0.220, 0.185)$ & $0.013$  & $0.013$ & $0.874$ & $5.271$  & $(1.075, 0.089)$ & \makecell{$1.2 \times 10^{-2}$ \\ $2.4 \times 10^{-6}$} & 03m 08s \\ \addlinespace
		\makecell{\textbf{MLP$_2$} \\ \textbf{Homotopy}} & $(0.172, 0.214)$ & $-0.009$ & $0.008$ & $0.611$ & $26.995$ & $(1.087, 0.083)$ & \makecell{$2.1 \times 10^{-4}$ \\ $4.7 \times 10^{-8}$} & 03m 24s \\
		\midrule
		\multicolumn{9}{c}{\textit{Case B: Frozen Profile Parameters}} \\
		\midrule
				\makecell{\textbf{Model} \\ \textbf{Method}} & $(P_1, H_1)$ & $u_a$ & $\beta_t$ & $I_t \; [10^7\text{ A}]$ & $q_a$ & $(r_a, z_a)$ & \makecell{\textbf{Losses} \\ $\mathcal{L}_{\mathcal{D}} / \mathcal{L}_{\partial \mathcal{D}}$} & \textbf{Runtime} \\
		\midrule
		\makecell{\textbf{KAN} \\ \textbf{Homotopy}}   & $(0.354, 0.109)$ & $0.037$  & $0.029$ & $1.282$ & $1.422$  & $(1.060, 0.091)$ & \makecell{$1.6 \times 10^{-6}$ \\ $7.3 \times 10^{-9}$} & 05m 17s \\ \addlinespace
		\makecell{\textbf{KAN} \\ \textbf{Pretrained}} & $(0.354, 0.109)$ & $0.037$  & $0.029$ & $1.277$ & $1.424$  & $(1.060, 0.091)$ & \makecell{$2.1 \times 10^{-7}$ \\ $8.3 \times 10^{-10}$} & 05m 15s \\ \addlinespace
		\makecell{\textbf{KAN} \\ \textbf{Unguided}}   & $(0.354, 0.109)$ & $-0.019$ & $0.021$ & $1.067$ & $3.589$  & $(1.065, 0.090)$ & \makecell{$1.4 \times 10^{-7}$ \\ $4.6 \times 10^{-10}$} & 05m 08s \\ \addlinespace
		\makecell{\textbf{MLP$_1$} \\ \textbf{Homotopy}} & $(0.354, 0.109)$ & $-0.020$ & $0.023$ & $1.064$ & $3.292$  & $(1.069, 0.083)$ & \makecell{$1.7 \times 10^{-3}$ \\ $2.0 \times 10^{-6}$} & 03m 07s \\ \addlinespace
		\makecell{\textbf{MLP$_2$} \\ \textbf{Homotopy}} & $(0.355, 0.110)$ & $-0.022$ & $0.023$ & $1.057$ & $3.068$  & $(1.071, 0.090)$ & \makecell{$2.1 \times 10^{-3}$ \\ $2.5 \times 10^{-6}$} & 03m 30s \\
		\bottomrule
	\end{tabular}
\caption{Equilibrium characteristics, figures of merit and performance metrics across different optimization strategies for the pedestal equilibrium profile. Case A presents the results when the profile parameters $(P_1, H_1)$ are optimized simultaneously with the network parameters to satisfy the equilibrium constraints on $\beta_t$ and $I_t$. Case B corresponds to frozen profile parameters, where the homotopy and transfer-learning strategies converge to the same equilibrium branch, whereas unguided runs converge to a different equilibrium branch.}
		\label{tab:pedestal_comparisons}
\end{table}

First, we train the KAN network with profile parameters ($P_1, H_1$) determined by the equilibrium constraints on $\beta_t$ and $I_t$, starting from initial values $P_1^{(0)} = 0.1$ and $H_1^{(0)} = 0.01$. In this scenario, both the homotopy-based and transfer-learning approaches converge to solutions with comparable accuracy and GPU computation times (see Table~\ref{tab:pedestal_comparisons} and Fig.~\ref{fig:train_hist_pedestal_comparison_discovery}), whereas the unguided method fails to converge. It should be noted, however, that the transfer-learning approach requires prior knowledge of a corresponding Solov'ev solution, necessitating the retraining of the Solov'ev network whenever the boundary conditions change. Both convergent methods yield similar values for $P_1$; however, this is not the case for $H_1$. This difference is also reflected in the safety factor values on the magnetic axis. We therefore conclude that the two methods converge to distinct solution branches that share similar $\beta_t$ and $I_t$ values but exhibit overall different characteristics. This provides initial evidence of solution multiplicity in the presence of pressure-pedestal non-linearity.  

This multiplicity becomes even more evident when training the KAN network with frozen profile parameters. For this experiment, parameters $P_1$ and $H_1$ are set to the values identified by the homotopy method. Under these conditions, all three methods converge, with the unguided and pretrained approaches minimizing more effectively the loss terms (Fig.~\ref{fig:train_hist_pedestal_comparison_frozen}). However, the unguided method does not converge to the same equilibrium solution. This is confirmed by the equilibrium characteristics provided in Table~\ref{tab:pedestal_comparisons}, as well as by comparing Figs.~\ref{fig:pedestal_contours_and_Jt}--\ref{fig:pedestal_profiles} with Figs.~\ref{fig:pedestal_contours_and_Jt_II}--\ref{fig:pedestal_profiles_II}, which depict the flux surface topology, current density distributions, and characteristic profile functions for the guided and unguided solutions, respectively, clearly showing that the two solutions have different equilibrium characteristics.

Both classes of equilibria are characterized by the pedestal structure of the pressure profile and edge current density peaks that emerge naturally due to the steep pressure gradient in the pedestal region, effectively capturing a prominent bootstrap current component. The edge region is also characterized by negative magnetic shear, as seen in the safety factor profiles. A primary distinction of the second equilibrium discovered by the unguided method is that the pressure attains overall smaller values, despite maintaining a similar pedestal height, resulting in a flatter pressure profile. Furthermore, the region of negative magnetic shear, marked by a decreasing safety factor toward the plasma edge, is noticeably more extended.

The MLP architectures (see Table~\ref{tab:architectures_hyperparameters}) also demonstrated improved performance in the pressure pedestal equilibrium than in the previous L-mode nonlinear equilibrium, exhibiting a  tendency to converge during parameter discovery and improved behavior under frozen parameters (see Table~\ref{tab:pedestal_comparisons}). Learning rate optimization could enable faster and more stable convergence for these MLP architectures but this was not pursued here. Interestingly, in the frozen-parameter scenario, the MLPs also tended to converge to the second solution branch discovered by the unguided method, despite being trained via the homotopy continuation approach.

Therefore, while convergence with the unguided method and homotopy-guided MLPs was unattainable for equilibria with the pressure profile given by \eqref{P_freegs} (even with frozen parameters), it is readily achieved for the pressure pedestal profile \eqref{pres_prof_ped}; nevertheless, unguided FastKAN settles on a distinct solution branch. Thus, the inherent multiplicity of solutions appears to be the mechanism that facilitates convergence in this case. The systematic isolation of distinct solution branches under non-linear profile functions and realistic tokamak geometries plays an important role in characterizing magnetic confinement regimes (e.g., \cite{Ham2024,Pentland2025}). Accordingly, a comprehensive investigation into the existence of multiple equilibrium branches and particularly those featuring H-mode characteristics such as pressure pedestals, is deferred to future work, as it represents a compelling direction for advancing our understanding of high-confinement regimes and the L-H transition.

\section{Conclusion}
\label{Sec_VI}
In this work, Physics-Informed Kolmogorov-Arnold Networks were utilized to compute Solov'ev and nonlinear tokamak equilibria with shaped fixed boundaries featuring a lower X-point. Equilibria with H-mode profile features, including pressure pedestals, edge current density peaks (qualitatively capturing bootstrap current), and negative magnetic shear at the plasma edge, were successfully constructed. The neural networks were trained in a physics-informed manner by minimizing the Grad-Shafranov residual while imposing Dirichlet boundary conditions as penalty terms. Additionally, a dedicated point-cloud generation method was introduced to construct training collocation points, ensuring relative spatial homogeneity across the computational domain. Furthermore, two of the GS profile parameters were dynamically determined by imposing physics-related constraints during training, and the SSBroyden method was employed to enhance convergence in all scenarios.

To attain nonlinear equilibria, KAN architectures were trained using guided methods that ensure robust convergence: (i) a homotopy-based curriculum learning scheme, where the networks continuously learn successively stiffer current density profiles starting from a baseline Solov'ev profile, and (ii) a transfer-learning approach where the KAN model is initialized with the weights of a network pretrained on a Solov'ev equilibrium sharing the same boundary conditions. The KAN models were benchmarked against MLP architectures with a comparable number of parameters.

Three main observations were made in this study: 1. While tuned MLPs may slightly outperform KANs in linear Solov'ev configurations with similar parameter counts, they fail to converge or converge very ineffectively when applied to highly nonlinear profiles. 2. For nonlinear equilibria involving profile parameter discovery, the homotopy-based method reliably ensures convergence, whereas the transfer-learning method enables convergence primarily in the pedestal case or when profile parameters are frozen. The homotopy approach offers the additional key advantage of eliminating the need for pretrained models. 3. Multiple equilibrium solutions exist for the pedestal configuration, complicating the optimization landscape with additional local minima. These  minima facilitate convergence even for the unguided method when profile parameters are frozen; however, different training strategies generally do not converge to the same equilibrium manifold.

In summary, the combination of three key elements, namely KAN architectures, guided training schemes, and second-order quasi-Newton SSBroyden method, enables the efficient, stable, and accurate computation of tokamak equilibria with highly nonlinear profiles, a task that is elusive or computationally inefficient when relying on standard MLPs and unguided training.

\section*{Acknowledgements}
One of the authors, J.L., acknowledges support from the National Magnetic Confinement Fusion Energy Research and Development Program of China (Nos. 2024YFE03020004).

\section*{Data and code availability}
The code used to generate the results will be made publicly available in a dedicated repository upon publication. In the interim, it is available from the corresponding author upon reasonable request.

\section*{Declaration of  AI-assisted tools}
The corresponding author acknowledges that during the preparation of this work, Generative AI tools were utilized to assist with improving manuscript language and overall readability, LaTeX layout formatting, and code script optimization (numerical and plotting routines). All generated code and text were reviewed, tested, and validated by the authors, who assume full responsibility for the scientific integrity and accuracy of the manuscript.

\begin{appendices}

\section{Point cloud generation method}
\label{appendix}
\subsection{Parametrization of boundaries prescribed by data}
In our previous work \cite{Kaltsas2022} we generated the computational domain and the collocations point data using explicit parametric equations that define the boundary. For a smooth D-shaped boundary we have used a set of two parametric equations determining the $r$ and $z$ coordinates of the boundary \cite{Turnbull1999} while for configurations with lower X-point that describe diverted tokamaks we used a set of suitable parametric equations described in \cite{Kuiroukidis2015}. The random distributions of the collocation points were created upon randomly discretizing the independent parameter $t$ in these parametric equations and a paramerer $s$ which is associated with the distance from the geometric center of the configuration. 

In this work, we introduce additional functions that enable the generation of 2D training point clouds with greater homogeneity from data-prescribed boundaries. The core principle of point cloud generation remains the same; namely, training points are generated by discretizing two continuous parameters corresponding to the poloidal angle and radial distance in parametric equations. The key distinction here is that these parametric equations are obtained by fitting boundary data using the discrete Fourier transform (DFT). This enables us to adjust the point density in both poloidal and radial directions according to the specific characteristics of the configuration. This is particularly useful for highly elongated configurations and geometries with X-points (boundary-aware point cloud generation), or equilibria with steep gradients in the physical profiles (physics-aware point cloud generation), as we can adjust the probability density function of the random number generator to ensure complete and uniform domain coverage while providing higher resolution in critical regions. 

To apply the algorithm, we first convert the boundary data into parametric functions given by trigonometric series, using the DFT to determine the amplitudes for each term. The domain is then described by equations of the form:
\begin{align}
	r(s,\theta) &= R_0 + \xi(s) \sum_{n} R_n \cos(n\theta + \varphi_n)\,, \\
	z(s,\theta) &= Z_0 + \xi(s) \sum_{n} Z_n \cos(n\theta + \phi_n)\,,
\end{align}
where $R_n, Z_n$ and $\varphi_n, \phi_n$ are the amplitudes and phase angles, respectively, obtained by applying the DFT to the boundary data given by coordinates $(R_d, Z_d)$. Here, $\xi(s)$ is a monotonic function satisfying $\xi(0)=0$ and $\xi(1)=1$, which determines the density of collocation points in the radial direction. Following our previous work \cite{Kaltsas2022}, we choose $\xi(s) = s^k$ with $k \in (0,1]$, and create the scattered point cloud within the domain $D$ by randomly sampling discrete values for the parameters $s$ and $\theta$ using appropriate distributions over $[0, 1)$ and $[0,2\pi)$, respectively.

\subsection{Geometry-aware point-cloud generation}
\begin{figure}[!htb]
	\centering
	\includegraphics[scale=0.6]{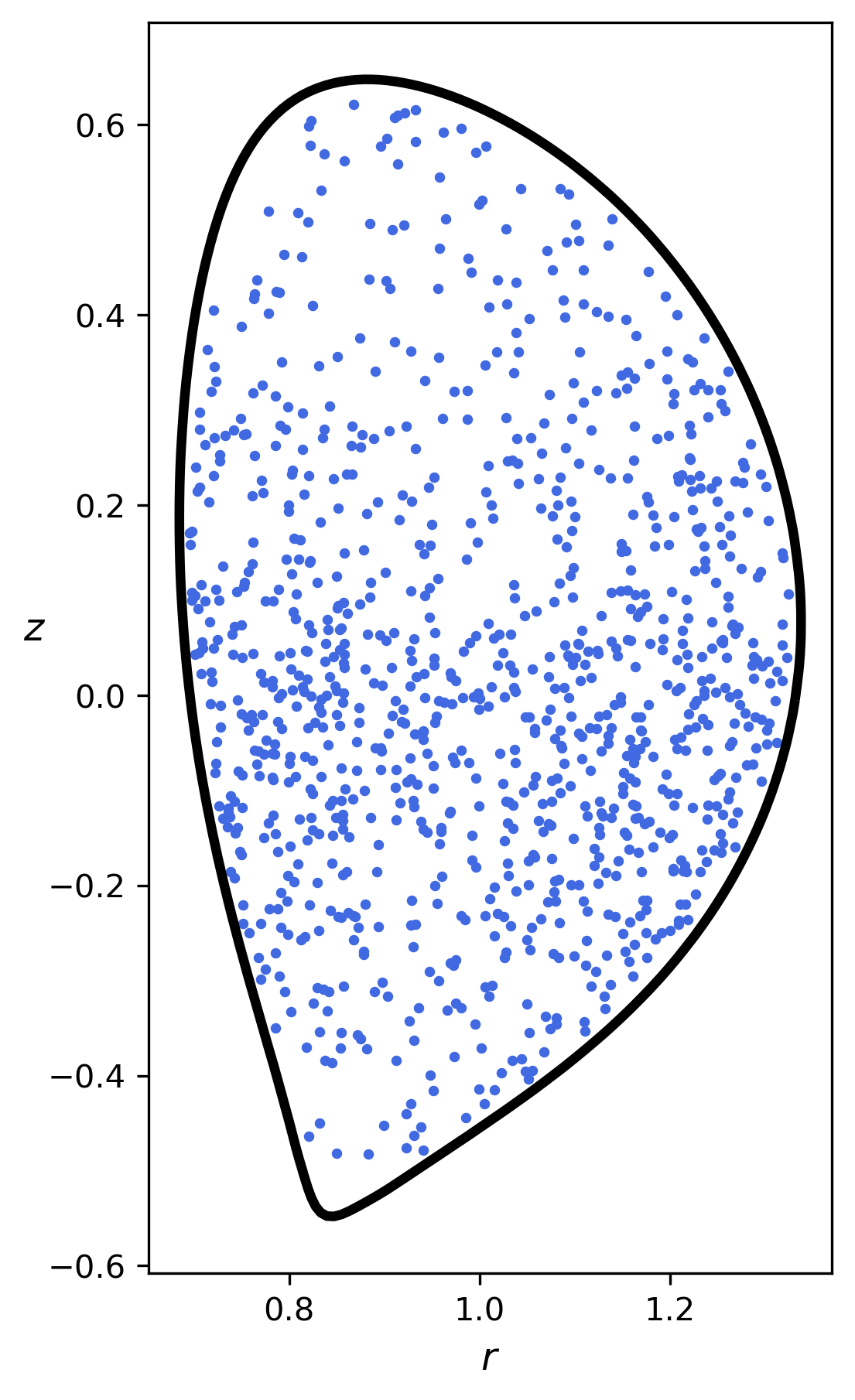}
	\includegraphics[scale=0.6]{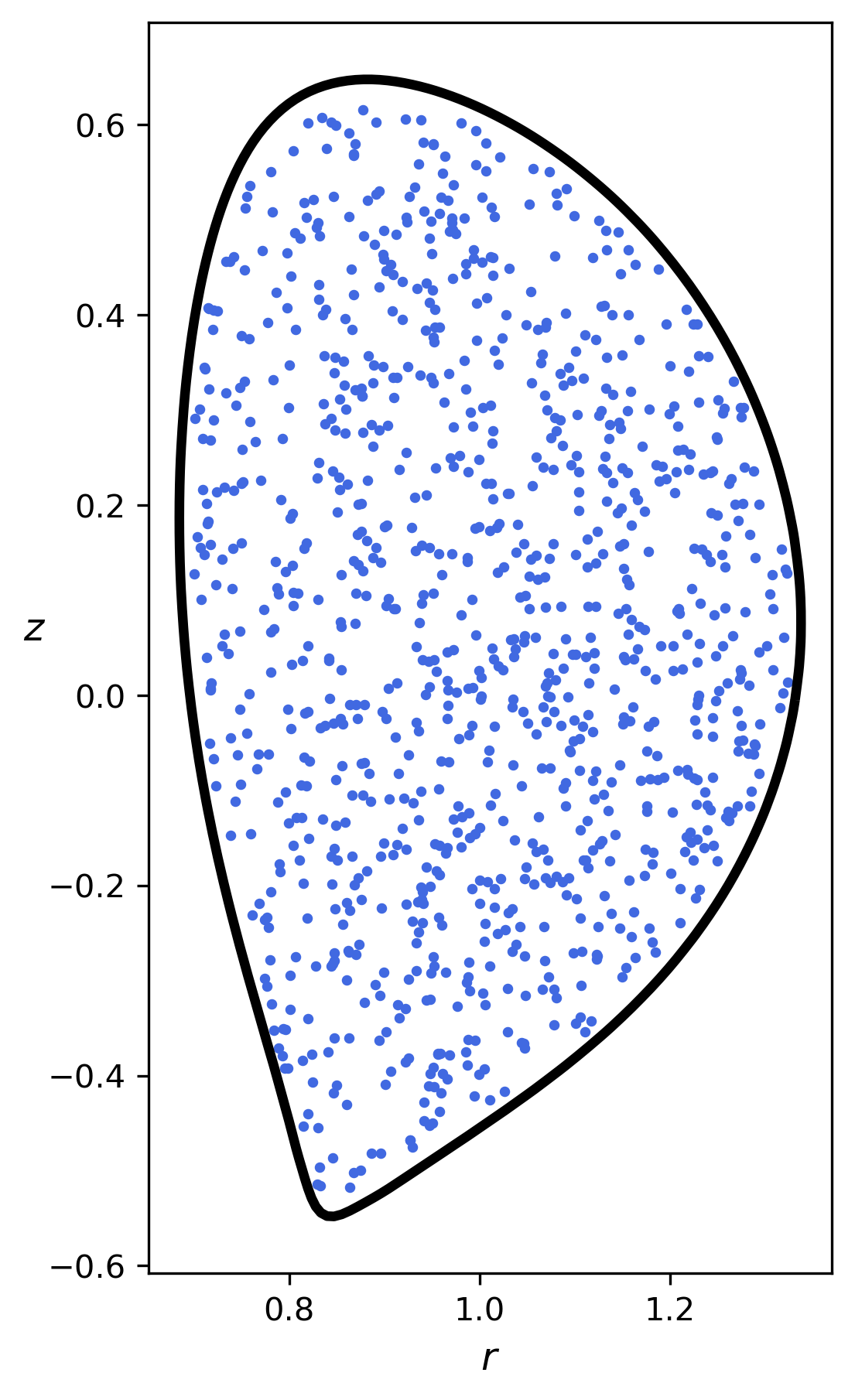}
\caption{(Left) Non-uniform point distribution generated by uniform sampling of parameters $s$ and $\theta$. (Right) Point distribution generated using geometry-aware probability density functions, thereby mitigating the effect of vertical elongation on point density.}
	\label{app:fig_1}
\end{figure}

The random numbers for the parameter $\theta$ are not drawn from a uniform distribution, but from a distribution that accounts for specific geometric characteristics of the domain. Specifically, if $\theta$ in the parametric equations is sampled uniformly, the resulting physical points $(r,z)$ are scattered in a non-uniform manner in highly elongated configurations; most points accumulate near the equatorial plane $z=Z_0$, leaving the furthermost regions along the vertical axis undersampled (see the left panel of Fig.~\ref{app:fig_1}). 

\begin{figure}[!htb]

	\centering
	\includegraphics[scale=0.5]{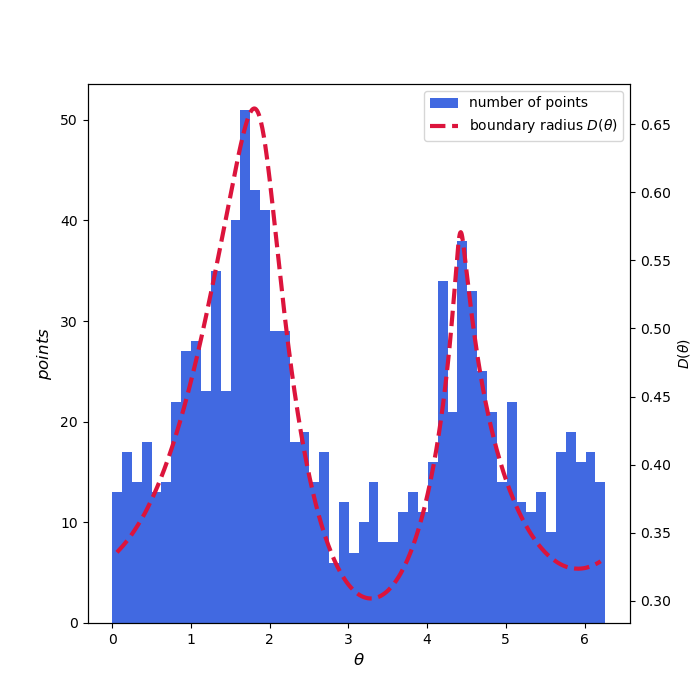} 
	\caption{Number of collocation points per poloidal angle interval generated via geometry-aware sampling (bars), plotted alongside the boundary radial distance (dashed line).}
	\label{app:fig_2}
\end{figure}

To circumvent this problem, we generate random samples for the poloidal angle $\theta$ using a custom probability density function $\Pc(\theta)$ that depends on the radial distance $D(\theta) = \sqrt{(r_b-R_0)^2+(z_b-Z_0)^2}$ of the boundary from the geometric center $(R_0,Z_0)$. The probability density function is defined as:
\begin{equation}
	\Pc(\theta) = \frac{D^{\ell}(\theta)}{\int_{0}^{2\pi} D^{\ell}(\theta)\,\mathrm{d}\theta}\,,
\end{equation}
where $D_b(\theta)$ interpolates the radial distance of the boundary points as a function of the poloidal angle $\theta$, and $\ell$ is a parameter that determines the dependence of the probability density on the radial distance. Here, we set $\ell=2$, yielding the angular distribution of points along $\theta$ shown in Fig.~\ref{app:fig_2}. This angular weighting produces a point cloud that covers the computational domain in a uniform manner, as shown in the right panel of Fig.~\ref{app:fig_1}.

\subsection{Point-cloud refinement: thinning and intelligent gap-filling}
\begin{figure}[!htb]
	\centering
	\includegraphics[scale=0.6]{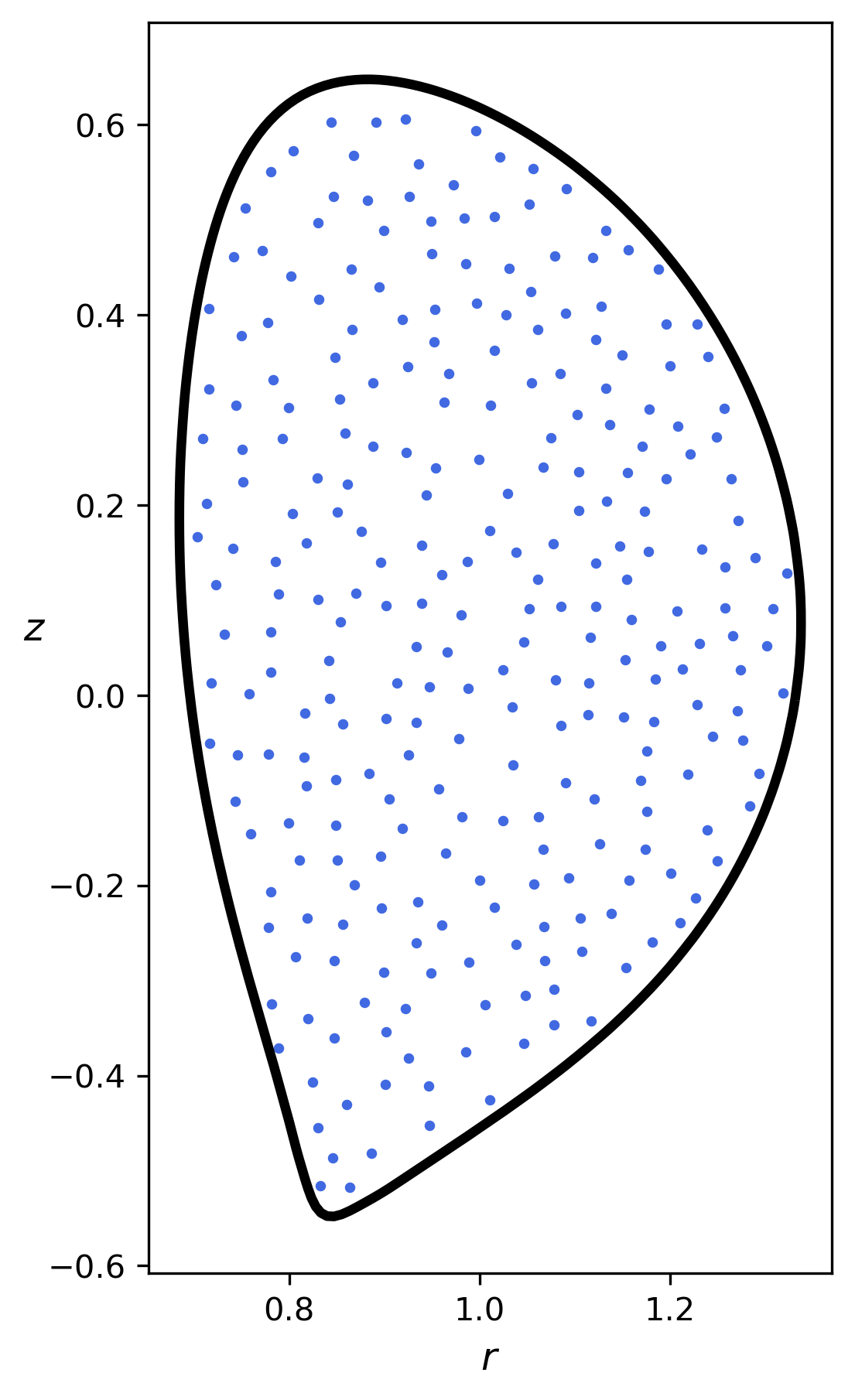}
	\includegraphics[scale=0.6]{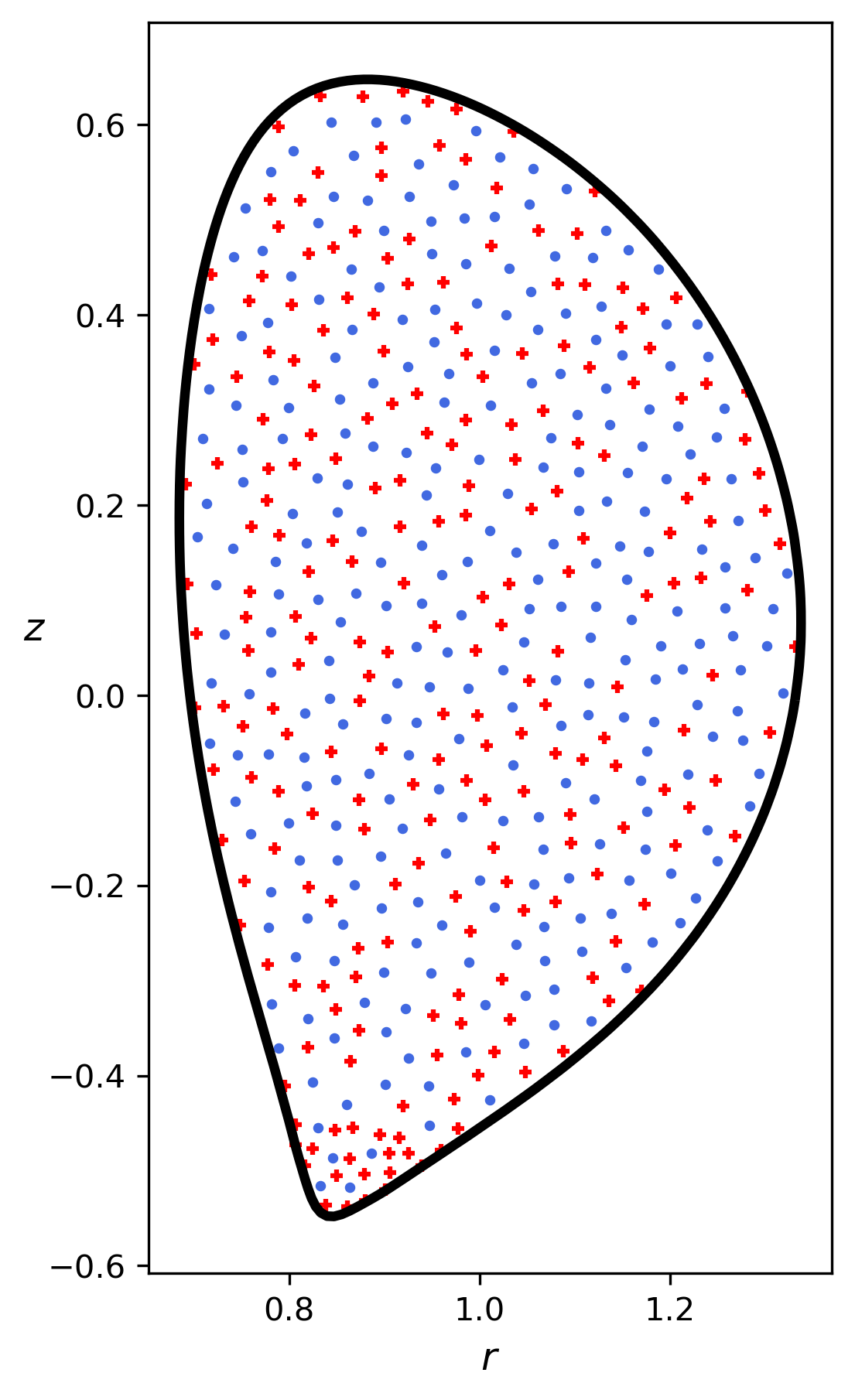}
\caption{(Left) Initial training point cloud filtered using the shortest-distance criterion. (Right) Refined training point cloud after energy-based point injection. Initial points are denoted by blue dots, and injected points by red crosses.}
	\label{app:fig_3}
\end{figure}

While the method described above achieves a globally uniform distribution, local point clouds may still contain dense clusters of clumped points or regions where the cloud is very sparse forming large gaps. To remedy this problem, we introduce two additional refinement steps. First, we establish a minimum separation radius $d_{\min}$ and iteratively eliminate all neighbors of each point that reside within a circle of radius $d_{\min}$ centered at that point. This is efficiently executed via $k$-d tree range queries rather than calculating the pairwise distance of all points.  This thinning  technique dissolves dense clusters (see left panel of Fig.~\ref{app:fig_3}a).

To fill remaining spatial gaps, we introduce additional points by evaluating a minimum ``potential energy'' criterion on the circumference of a circle of radius $d_{min}$ centered at the existing points. Each point in the cloud is modeled as a fictitious particle generating a repulsive potential $V(r) = r^{-1}$. For a given point $p$, we compute the total potential $\Sigma V$ evaluated on a circle of radius $d_{\min}$ centered at $p$, generated by all neighboring particles residing within a concentric ball of radius $d_{\max} = m_1 d_{\min}$ ($m_1 > 1$). A candidate point is placed at the location of minimum potential along this circumference. The candidate point is accepted provided that: (i) it resides strictly within the domain, (ii) it maintains a distance of at least $0.9 d_{\min}$ from all existing internal points, and (iii) it stays clear of the outer boundary by at least $m_2 D_{\max}$ ($m_2 \sim 10^{-3}$). This procedure is iteratively repeated over both original and newly generated points across multiple refinement passes ($p_{\mathrm{gen}}$) until gap filling is complete (see right panel of Fig.~\ref{app:fig_3}). The cloud can be refined further upon defining the minimal distance $d_{min}$ as a function of the radial distance $D$ and the boundary curvature $\kappa_b$. A recent paper that reviews similar or variants of the above techniques for point cloud generation for mesh-free methods is \cite{Suchde2022}. 

Furthermore, localized point cloud enrichment is applied near sharp geometric features of the boundary, such as divertor X-points or high-curvature boundary corners. The local boundary curvature $\kappa_b$ is computed as: $	\kappa_b = |\dot{r}\ddot{z} - \ddot{r}\dot{z}|/(\dot{r}^2 + \dot{z}^2)^{3/2}\,.$
At the location of maximum curvature, $(r_c, z_c) = \arg\max(\kappa_b)$, we determine the inward unit normal vector $\mathbf{n}_{\text{in}}$. Additional collocation points are then injected along and around $\mathbf{n}_{\text{in}}$ inside the domain, applying a relaxed minimum separation threshold ($m_3 d_{\min}$ where $m_3 < 1$). Candidate points are subjected to a sampling filter, requiring them to maintain a minimum boundary clearance of $m_4 D_{\max}$ ($m_4 \sim 10^{-3}$). This increases the resolution near the X-point (see the lower corner of the right panel in Fig.~\ref{app:fig_3}).


\end{appendices}

	\printbibliography
	
	\end{document}